\documentclass[amsmath,amssymb,notitlepage,nofootinbib]{revtex4-1}
\usepackage{graphicx}
\usepackage{mathtools}
\usepackage{dcolumn}
\usepackage{bm}
\begin{document}

\title{A General Treatment of Coupled Cavity Resonators}

\author{Osamu Kamigaito}
\email{kamigait@riken.jp}
\affiliation{RIKEN Nishina Center for Accelerator-Based Science,\\
2-1 Hirosawa, Wako-shi, Saitama 351-0198, Japan}%

             
\begin{abstract}
Consider several cavity resonators whose resonant frequencies and electromagnetic fields are known, and suppose that they are coupled through apertures.
How can the resonant frequencies and resonant electromagnetic fields of the resulting coupled-cavity resonator be expressed in terms of the original resonant frequencies and electromagnetic fields?
This is a problem frequently encountered in radio-frequency systems for accelerators.
One example is a multicell cavity used in high-energy electron accelerators.
In this note, the resonant frequencies and resonant electromagnetic fields of such coupled cavities are examined as generally as possible, based only on Maxwell's equations and boundary conditions.
Specifically, the response function of a cavity as viewed from its coupling ports, namely, its admittance matrix, is derived, and a formulation is developed to express the coupled resonant electromagnetic fields using this matrix.
The resulting formulation shows that the coupled resonant modes can be described in terms of overlap integrals between the resonant modes of the cavities and the waveguide modes at the coupling apertures.
It also shows that the coupled resonant frequencies are given by the zeros of the determinant of a matrix constructed from the admittance matrices.
The results thus obtained are applied to simple systems to confirm the validity of the formulation.
Numerical calculations based on this formulation are also carried out for periodic systems in an attempt to derive dispersion relations.
\end{abstract}

\maketitle
\tableofcontents
\clearpage

\section{\label{intro}Introduction and overview of this note}

In accelerators, several cavity resonators are often connected and used as a single coupled resonator.
Typical examples include coupled-cavity linear accelerators \cite{bib:nis66, bib:nag67} and superconducting multicell cavities \cite{bib:tes01}.
Other interesting applications of coupled cavities include ARES cavities \cite{bib:shi94, bib:kag97} and flat-top resonators for AVF cyclotrons \cite{bib:koh04}.

How can the resonant frequencies and resonant electromagnetic fields of such coupled-cavity resonators be expressed in terms of the frequencies and electromagnetic fields of the original cavity resonators?
In this note, such coupled-cavity resonators are examined in a general manner, based only on Maxwell's equations and boundary conditions.

The contents of this note are summarized below.
This note deals with lossless cavity resonators.
First, in Sec.~\ref{sec:admitance}, we derive the input admittance of a cavity resonator, namely, the response function relating the ``magnetic field'' generated at the cavity aperture to the ``electric field'' applied there.
To obtain this function, complete sets consisting of eigenvector fields of the Laplace operator are introduced in the cavity interior and at the aperture under consideration.
These vector fields have properties convenient for expanding Maxwell's equations.
Maxwell's equations are then expanded using these complete sets, and the response function of the cavity resonator, namely, its admittance, is derived.
This response function takes the form of a matrix whose dimension depends on the number of eigenvector fields introduced at the aperture.
With this ``admittance matrix,'' the electromagnetic fields inside the cavity are expressed in terms of the electric field (input) at the aperture, while the magnetic field (output) excited at the aperture is also expressed in terms of the electric field (input) at the aperture.

Next, in Sec.~\ref{sec:single}, the expression for the cavity admittance matrix obtained in this way is examined by restricting the eigenvector fields at the aperture to a single mode.
Since this treatment corresponds to an ordinary admittance, namely, the relation between voltage and current in a circuit, it makes the following sections easier to understand intuitively.
There, after considering the general properties of the expansion coefficients, we examine how the weakening of coupling is reflected in the expansion coefficients, how ``capacitive coupling'' is represented, and how the coupled resonant frequencies are related to the original resonant frequencies.

In Sec.~\ref{sec:matrix}, the coupled resonant modes are expressed using the admittance matrices.
First, a method is formulated for obtaining the resonant electromagnetic fields of a coupled resonator consisting of two cavities.
Since the two cavities to be coupled have apertures of the same shape, the response functions of the two cavities are defined at those apertures.
The condition for coupled resonant electromagnetic fields to arise is that the cavity responses to the ``electric field'' at the aperture are consistent with each other, that is, the ``magnetic field'' is smoothly connected across the two sides of the aperture.
This condition is expressed by requiring a matrix constructed from the admittance matrices to have a zero eigenvalue, and the solutions of its characteristic equation give the resonant frequencies of the coupled-cavity resonator.
Moreover, the eigenvector corresponding to the zero eigenvalue gives the electric-field components at the aperture in the resonant state.
This is one of the main results of this note.
Next, this formulation is applied to a system of two coupled cavities with only one aperture mode, and the properties of the coupled resonant modes are considered.
The formulation is then extended to the general case of $N$ cavities.

In Sec.~\ref{sec:periodic}, this formulation is applied to periodic systems.
First, the general features of the solutions are examined using the symmetry associated with a periodic system.
The method is then applied to a coupled resonator consisting of coaxial resonators, and its validity is confirmed by comparison with an exact solution.
Next, dispersion relations in general periodic systems are examined.
Simple numerical calculations are performed to confirm that, as in the coupled-circuit model, two types of dispersion relations are obtained.
Finally, a general theory of periodic systems is developed, and the characteristics of the solutions when extended to higher dimensions are described.
It will be seen that the subtle properties of positive-real matrices, which are also closely related to causality, lead to the uniqueness of the solutions.

In Appendix~\ref{ap:var}, it is shown that the formulation described in Sec.~\ref{sec:admitance} is consistent with a variational representation.
In Appendix~\ref{ap:split}, the frequency splitting that occurs when two identical cavities are coupled is explained using the method described in Sec.~\ref{sec:matrix}.
Although these subjects have been left out of the main text, they are considered important.

The principal textbook on which this note is based is that of Ref.~\cite{bib:kur69}.
Reference~\cite{bib:dom92} was also consulted concerning the classification of solutions of the Helmholtz equation and the boundary conditions.
Basic concepts of network theory are also used, but the minimum necessary background and references are given in footnotes in the main text.
Intermediate steps in the calculations are presented with as few omissions as possible.

Although several textbooks deal with coupled-cavity problems \cite{bib:bev61,bib:bev64,bib:sob99}, this note takes a different approach.
Another commonly used method is to represent the resonant electromagnetic fields under consideration by combinations of lumped-element circuits \cite{bib:nis66,bib:nag67,bib:wan08}, and this note provides a basis for that treatment.
We believe that it is worthwhile to consider a consistent treatment of coupled cavities starting only from Maxwell's equations and boundary conditions.
We hope that readers interested in this subject will find the note useful.

\section{Electromagnetic response of cavity resonators}
\label{sec:admitance}

In this section, we introduce eigenvector fields of the Laplace operator in the cavity interior (a three-dimensional region) and on the junction surface (a two-dimensional region), and use them to calculate the response function of the resonator.
The main result is Eq.~\eqref{eq:kekka1}.
Since the derivation presented here closely follows Professor Kurokawa's textbook \cite{bib:kur69}, readers who already know the derivation of Eq.~\eqref{eq:kekka1} may skip this section.

\subsection{Cavity eigenmodes\label{sub:eigen3}}

\begin{figure}[htb]
   \centering
   \includegraphics*[width=30mm]{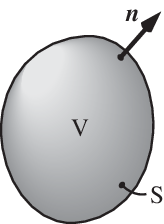}
   \caption{The three-dimensional region V, its boundary surface S, and the outward unit normal vector $\bm{n}$ on the boundary.
   V is enclosed by S, on which the perfect-conductor boundary conditions are imposed.
   The region V need not be simply connected.}
   \label{fig:single}
\end{figure}

Let $V$ denote the interior of the three-dimensional region shown in Fig.~\ref{fig:single}, $S$ its boundary surface, and $\bm{n}$ the outward unit normal vector on $S$.
We can define eigenvector fields of the Laplace operator in $V$ satisfying the following equations and boundary conditions:
\begin{eqnarray}
\label{eq:efun1}
\nabla\times\nabla\times\bm{E}-\nabla\nabla\cdot\bm{E}
-k^2\bm{E}&=&0\qquad \quad {\mbox {in V}}\,, \\
\label{eq:efun2}
\bm{n}\times\bm{E}&=&0\qquad \quad{\mbox {on S}}\,,\\
\label{eq:efun3}
\nabla\cdot\bm{E}&=&0\qquad \quad{\mbox {on S}}\,.
\end{eqnarray}
Equation~\eqref{eq:efun1} is the Helmholtz equation.
The eigenvector fields minimize the following functional:
\begin{eqnarray}
k^2\left(\bm{E}\right)=\frac{\int_{\tiny {\mbox V}}[\left(\nabla\times\bm{E}\right)^2+
\left(\nabla\cdot\bm{E}\right)^2]dv-2\int_{\tiny{\mbox S}}
\bm{n}\times\bm{E}\cdot\nabla\times\bm{E}dS}
{\int_{\tiny{\mbox V}}\bm{E}^2dv}\,.
\end{eqnarray}
In general, the eigenvalues $k^2$ form a discrete spectrum extending from zero to arbitrarily large values.

The eigenvector fields can be chosen to be orthonormal:
\begin{eqnarray}
\int_{\tiny {\mbox V}}\bm{E}_n \cdot \bm{E}_m dv = \delta_{nm}\,.
\end{eqnarray}
Here $\delta_{nm}$ is the Kronecker delta.
The eigenvector fields can be classified into three types \cite{bib:dom92}.
\begin{eqnarray}
\label{eq:classa_e}
{\mbox {(A)}}\qquad  \nabla\times\bm{E}_n=0,\qquad \quad \nabla\cdot\bm{E}_n=0\,,\\
\label{eq:classb_e}
{\mbox {(B)}}\qquad \nabla\times\bm{E}_n\neq0,\qquad \quad \nabla\cdot\bm{E}_n=0\,,\\
\label{eq:classc_e}
{\mbox {(C)}}\qquad \nabla\times\bm{E}_n=0,\qquad \quad \nabla\cdot\bm{E}_n\neq0\,.
\end{eqnarray}

A vector field $\bm{F}$ that is piecewise continuous and square-integrable in $V$ can be expanded in these eigenvector fields in the following sense:
\begin{eqnarray}
\lim_{N \rightarrow \infty}\int_{\tiny {\mbox V}}
|\bm{F}-\sum_{n=1}^N A_n \bm{E}_n|^2 dv=0\,,
\end{eqnarray}
where
\begin{eqnarray}
A_n=\int_{\tiny {\mbox V}}\bm{E}_n \cdot \bm{F}dv\,.
\end{eqnarray}

We can also define eigenvector fields satisfying a different set of boundary conditions:
\begin{eqnarray}
\label{eq:hfun1}
\nabla\times\nabla\times\bm{H}-\nabla\nabla\cdot\bm{H}
-k^2\bm{H}&=&0\qquad \quad {\mbox {in V}}\,,\\
\label{eq:hfun2}
\bm{n}\cdot\bm{H}&=&0\qquad \quad{\mbox {on S}}\,,\\
\label{eq:hfun3}
\bm{n}\times\nabla\times\bm{H}&=&0\qquad \quad{\mbox {on S}}\,.
\end{eqnarray}
The above eigenvector fields give the minimum value of the following functional.
\begin{eqnarray}
k^2\left(\bm{H}\right)=\frac{\int_{\tiny {\mbox V}}[\left(\nabla\times\bm{H}\right)^2+
\left(\nabla\cdot\bm{H}\right)^2]dv-2\int_{\tiny{\mbox S}}
(\bm{n}\cdot\bm{H})(\nabla\cdot\bm{H})dS}
{\int_{\tiny{\mbox V}}\bm{H}^2dv}\,.
\end{eqnarray}
In general, the eigenvalues $k^2$ form a discrete spectrum extending from zero to arbitrarily large values.

The eigenvector fields can be chosen to be orthonormal:
\begin{eqnarray}
\label{eq:hnorm}
\int_{\tiny {\mbox V}}\bm{H}_n \cdot \bm{H}_m dv = \delta_{nm}\,.
\end{eqnarray}
Here $\delta_{nm}$ is the Kronecker delta.
The eigenvector fields can be classified into three types \cite{bib:dom92}.
\begin{eqnarray}
\label{eq:classa_h}
{\mbox {(A)}}\qquad  \nabla\times\bm{H}_m=0,\qquad \quad \nabla\cdot\bm{H}_m=0\,,\\
\label{eq:classb_h}
{\mbox {(B)}}\qquad \nabla\times\bm{H}_m\neq0,\qquad \quad \nabla\cdot\bm{H}_m=0\,,\\
\label{eq:classc_h}
{\mbox {(C)}}\qquad \nabla\times\bm{H}_m=0,\qquad \quad \nabla\cdot\bm{H}_m\neq0\,.
\end{eqnarray}

A vector field $\bm{F}$ that is piecewise continuous and square-integrable in $V$ can be expanded in these eigenvector fields in the following sense:
\begin{eqnarray}
\lim_{M \rightarrow \infty}\int_{\tiny {\mbox V}}
|\bm{F}-\sum_{m=1}^M B_m \bm{H}_m|^2 dv=0\,,
\end{eqnarray}
where
\begin{eqnarray}
B_m=\int_{\tiny {\mbox V}}\bm{H}_m \cdot \bm{F}dv\,.
\end{eqnarray}

The type-(B) fields among $\bm{E}_n$ and $\bm{H}_m$ are related in a particular way.
Substitution into the variational expression shows that the eigenvalues $k_a^2$ of the type-(B) fields are nonzero.
Choose a type-(B) field $\bm{E}_a$ and define $\bm{H}_a$ by
\begin{eqnarray}
\nabla\times\bm{E}_a=k_a\bm{H}_a\,.
\label{eq:ea}
\end{eqnarray}
The resulting field $\bm{H}_a$ satisfies Eq.~\eqref{eq:classb_h}, the Helmholtz equation~\eqref{eq:hfun1}, and the boundary conditions~\eqref{eq:hfun2} and \eqref{eq:hfun3}%
\footnote{Let us verify only the boundary conditions.
First, take the scalar product of both sides of Eq.~\eqref{eq:ea} with $\bm{n}$.
\begin{eqnarray*}
\bm{n}\cdot\nabla\times\bm{E}_a=k_a\bm{n}\cdot\bm{H}_a\,.
\end{eqnarray*}
Take an arbitrary small region $A$ on the cavity surface $S$ and integrate both sides over $A$. By Stokes' theorem,
\begin{eqnarray*}
k_a\int_{\tiny{\mbox A}}\bm{n}\cdot\bm{H}_adS=\int_{\tiny{\mbox A}}\nabla\times\bm{E}_a\cdot\bm{n}dS=\int_{\tiny\partial{\mbox A}}\bm{E}_a\cdot\bm{t}dl\,.
\end{eqnarray*}
Here $\bm{t}$ is a vector tangent to $\partial{\mbox A}$.
The rightmost side of this equation vanishes by Eq.~\eqref{eq:efun2}.
Since $A$ is arbitrary, $k_a\bm{n}\cdot\bm{H}_a=0$.
As $k_a\ne0$, Eq.~\eqref{eq:hfun2} follows.
On the other hand, applying $\nabla\times$ to both sides of Eq.~\eqref{eq:ea} gives, from Eq.~\eqref{eq:hfun1},
\begin{eqnarray*}
k_a\nabla\times\bm{H}_a=\nabla\times\nabla\times\bm{E}_a={k_a}^2\bm{E}_a\,.
\end{eqnarray*}
Taking the vector product with $\bm{n}$ and using Eq.~\eqref{eq:efun2}, we obtain
\begin{eqnarray*}
k_a\bm{n}\times\nabla\times\bm{H}_a={k_a}^2\bm{n}\times\bm{E}_a=0\,.
\end{eqnarray*}
Again, since $k_a\ne0$, Eq.~\eqref{eq:hfun3} follows.
}.
Moreover, if $\bm{E}_a$ and $\bm{E}_b$ are orthonormal, the corresponding fields $\bm{H}_a$ and $\bm{H}_b$ are also orthonormal.

Conversely, given a type-(B) eigenvector field $\bm{H}_a$, let
\begin{eqnarray}
\nabla\times\bm{H}_a=k_a\bm{E}_a\,.
\label{eq:ha}
\end{eqnarray}
The resulting field $\bm{E}_a$ satisfies the Helmholtz equation~\eqref{eq:efun1} and the boundary conditions~\eqref{eq:efun2} and \eqref{eq:efun3}%
\footnote{Here again, let us verify only the boundary conditions.
Taking the vector product of both sides of Eq.~\eqref{eq:ha} with $\bm{n}$ gives
\begin{eqnarray*}
\bm{n}\times\nabla\times\bm{H}_a=k_a\bm{n}\times\bm{E}_a\,.
\end{eqnarray*}
The left-hand side vanishes by Eq.~\eqref{eq:hfun3}, and hence Eq.~\eqref{eq:efun2} follows because $k_a\ne0$.
Also, taking $\nabla\cdot$ of both sides of Eq.~\eqref{eq:ha} immediately gives Eq.~\eqref{eq:efun3}.
}.
Thus, without loss of completeness, the eigenvector fields belonging to (B) can be chosen to satisfy Eqs.~\eqref{eq:ea} and \eqref{eq:ha}.
In the following, we take the type-(B) eigenvector fields to be constructed in this way.

We use Greek indices for the type-(A) and type-(C) eigenvector fields to distinguish them from type-(B) fields. Thus,
\begin{eqnarray}
\nabla\times \bm{E}_{\nu}=0\qquad \quad {\mbox {in V}}\,,\\
\nabla\times \bm{H}_{\lambda}=0\qquad \quad {\mbox {in V}}\,.
\end{eqnarray}

The type-(A) and type-(B) eigenvector fields represent electromagnetic fields that can exist in the region $V$ enclosed by perfect conductors.

\subsection{Waveguide eigenmodes\label{sub:eigen2}}

\begin{figure}[htb]
   \centering
   \includegraphics*[width=50mm]{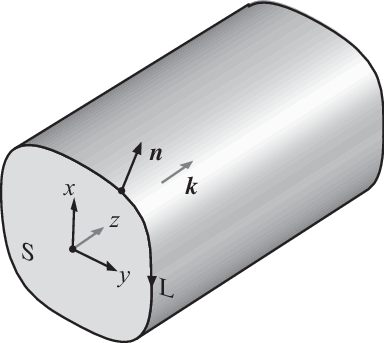}
   \caption{A waveguide with cross section S, its perimeter L, and the outward unit normal vector $\bm{n}$ on the boundary.}
   \label{fig:guide}
\end{figure}

Next, we introduce a complete set of vector fields on a two-dimensional plane in three-dimensional space.
Consider a waveguide with cross section S, as shown in Fig.~\ref{fig:guide}.
Let L denote the boundary of S and $\bm{n}$ the outward unit normal vector on this boundary.
We take the $(x,y)$ plane to coincide with S and the $z$ axis along the direction of propagation in the waveguide.
Eigenvector fields $\bm{E}_t=({E}_{tx}(x,y),{E}_{ty}(x,y),0)$ of the Laplace operator can then be defined in S by the following equations and boundary conditions:
\begin{eqnarray}
\nabla\times\nabla\times\bm{E}_t-\nabla\nabla\cdot\bm{E}_t
-k^2\bm{E}_t&=&0\qquad \quad {\mbox {in S}}\,, \\
\bm{n}\times\bm{E}_t&=&0\qquad \quad{\mbox {on L}}\,,\label{eq:e_bou_2dim}\\
\nabla\cdot\bm{E}_t&=&0\qquad \quad{\mbox {on L}}\,.
\end{eqnarray}
These eigenvector fields satisfy ``electric-type'' boundary conditions.
They minimize the following functional:
\begin{eqnarray}
k^2\left(\bm{E}_t\right)=\frac{\int_{\tiny {\mbox S}}[\left(\nabla\times\bm{E}_t\right)^2+
\left(\nabla\cdot\bm{E}_t\right)^2]dS-2\int_{\tiny{\mbox L}}
\bm{n}\times\bm{E}_t\cdot\nabla\times\bm{E}_tdl}
{\int_{\tiny{\mbox S}}\bm{E}_t^2dS}\,.
\end{eqnarray}
In general, the eigenvalues $k^2$ take discrete values ranging from zero to infinity.

The eigenvector fields can be chosen to be orthonormal:
\begin{eqnarray}
\label{eq:etnorm}
\int_{\tiny {\mbox S}}\bm{E}_{tn} \cdot \bm{E}_{tm} dS = \delta_{nm}\,.
\end{eqnarray}
Here $\delta$ is the Kronecker delta.
The eigenvector fields can also be classified into three types \cite{bib:dom92}:
\begin{eqnarray}
{\mbox {(A)}}\qquad  \nabla\times\bm{E}_{tn}=0,\qquad \quad \nabla\cdot\bm{E}_{tn}=0\,,\\
{\mbox {(B)}}\qquad \nabla\times\bm{E}_{tn}\neq0,\qquad \quad \nabla\cdot\bm{E}_{tn}=0\,,\\
{\mbox {(C)}}\qquad \nabla\times\bm{E}_{tn}=0,\qquad \quad \nabla\cdot\bm{E}_{tn}\neq0\,.
\end{eqnarray}

A field $\bm{F}$ that is piecewise continuous and square-integrable in S can be expanded in terms of these eigenvector fields in the following sense:
\begin{eqnarray}
\lim_{N \rightarrow \infty}\int_{\tiny {\mbox S}}
|\bm{F}-\sum_{n=1}^N a_n \bm{E}_{tn}|^2 dS=0\,,
\end{eqnarray}
where
\begin{eqnarray}
a_n=\int_{\tiny {\mbox S}}\bm{E}_{tn} \cdot \bm{F}dS\,.
\end{eqnarray}

For a method of constructing electromagnetic fields propagating through a waveguide from the eigenvector fields $\bm{E}_{tn}$, see Ref.~\cite{bib:kur69}.

As in the three-dimensional case, we can also define eigenvector fields satisfying ``magnetic-type'' boundary conditions:
\begin{eqnarray}
\nabla\times\nabla\times\bm{H}_t-\nabla\nabla\cdot\bm{H}_t
-k^2\bm{H}_t&=&0\qquad \quad {\mbox {in S}}\,,\\
\bm{n}\cdot\bm{H}_t&=&0\qquad \quad{\mbox {on L}}\,,\label{eq:h_bou_2dim}\\
\bm{n}\times \nabla\times \bm{H}_t&=&0\qquad \quad{\mbox {on L}}\,.
\end{eqnarray}
These eigenvector fields minimize the following functional:
\begin{eqnarray}
k^2\left(\bm{H}_t\right)=\frac{\int_{\tiny {\mbox S}}[\left(\nabla\times\bm{H}_t\right)^2+
\left(\nabla\cdot\bm{H}_t\right)^2]dS-2\int_{\tiny{\mbox L}}
(\bm{n}\cdot\bm{H}_t)(\nabla\cdot\bm{H}_t)dl}
{\int_{\tiny{\mbox S}}\bm{H}_t^2dS}\,.
\end{eqnarray}
In general, the eigenvalues $k^2$ take discrete values ranging from zero to infinity.

The eigenvector fields can be chosen to be orthonormal:
\begin{eqnarray}
\int_{\tiny {\mbox S}}\bm{H}_{tn} \cdot \bm{H}_{tm} dS = \delta_{nm}\,.
\end{eqnarray}
Here $\delta$ is the Kronecker delta.
The eigenvector fields can also be classified into three types \cite{bib:dom92}:
\begin{eqnarray}
{\mbox {(A)}}\qquad  \nabla\times\bm{H}_{tm}=0,\qquad \quad \nabla\cdot\bm{H}_{tm}=0\,,\\
{\mbox {(B)}}\qquad \nabla\times\bm{H}_{tm}\neq0,\qquad \quad \nabla\cdot\bm{H}_{tm}=0\,,\\
{\mbox {(C)}}\qquad \nabla\times\bm{H}_{tm}=0,\qquad \quad \nabla\cdot\bm{H}_{tm}\neq0\,.
\end{eqnarray}

A field $\bm{F}$ that is piecewise continuous and square-integrable in S can be expanded in terms of these eigenvector fields in the following sense:
\begin{eqnarray}
\lim_{M \rightarrow \infty}\int_{\tiny {\mbox S}}
|\bm{F}-\sum_{m=1}^M b_m \bm{H}_{tm}|^2 dS=0\,,
\end{eqnarray}
where
\begin{eqnarray}
b_m=\int_{\tiny {\mbox S}}\bm{H}_{tm} \cdot \bm{F}dS\,.
\end{eqnarray}

As with $\bm{E}_{tn}$, electromagnetic fields propagating through a waveguide can also be constructed from the eigenvector fields $\bm{H}_{tn}$.

\subsection{Cavity response\label{sub:adm}}

\begin{figure}[htb]
   \centering
   \includegraphics*[width=40mm]{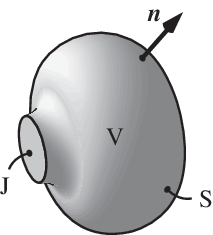}
   \caption{The cavity resonator for which the response function is calculated.
   The three-dimensional region V has a planar region J as part of its boundary.
   The remaining boundary, excluding J, is denoted by S.
   The outward unit normal vector on the boundary is $\bm{n}$.}
   \label{fig:junction}
\end{figure}

We now use the complete sets introduced above to derive the response function of a lossless cavity resonator.
Consider the cavity shown in Fig.~\ref{fig:junction}, whose boundary includes a planar region J. Let V denote the cavity interior and S the remainder of its boundary.
We seek to determine the magnetic field generated at J when this portion of the cavity wall is removed to form an aperture and an electric field is applied there.
In other words, we seek the input admittance of the cavity resonator.

Consider Maxwell's equations in V,
\begin{eqnarray}
\nabla\times\bm{H}&=&j\omega\epsilon\bm{E}\,,
\label{eq:edef}\\
\nabla\times\bm{E}&=&-j\omega\mu\bm{H}\,,
\label{eq:hdef}
\end{eqnarray}
under the boundary condition on S,
\begin{eqnarray}
\label{eq:ebound}
\bm{n}\times\bm{E}=0\qquad \quad{\mbox {on S}}\,.
\end{eqnarray}
Note that no boundary condition is imposed on J.
This corresponds to removing the portion J from the cavity wall.
For $\omega\neq0$, this boundary condition and Maxwell's equations imply that, on S,
\begin{eqnarray}
\label{eq:hbound}
\bm{n}\cdot\bm{H}=0\qquad \quad{\mbox {on S}}\,.
\end{eqnarray}
This follows from Stokes' theorem%
\footnote{Choose an arbitrary small region A on S, integrate both sides of Eq.~\eqref{eq:hdef} over A, and apply Stokes' theorem to obtain
\begin{eqnarray*}
-j\omega\mu\int_{\tiny{\mbox A}}\bm{H}\cdot\bm{n}dS=\int_{\tiny{\mbox A}}\nabla\times\bm{E}\cdot\bm{n}dS
=\int_{\tiny\partial{\mbox A}}\bm{E}\cdot\bm{t}dl\,.
\end{eqnarray*}
Here the rightmost side vanishes by Eq.~\eqref{eq:ebound}.
Since A is arbitrary, this means $-j\omega\mu\bm{n}\cdot\bm{H}=0$.
Thus, Eq.~\eqref{eq:hbound} is obtained for $\omega\ne0$.
}.
For $\omega\neq0$, Maxwell's equations also imply that
\begin{eqnarray}
\nabla\cdot\bm{E}&=&0\qquad \quad{\mbox {in V}}\,,\\
\nabla\cdot\bm{H}&=&0\qquad \quad{\mbox {in V}}\,.
\end{eqnarray}
In the following, we consider $\omega\neq0$, and hence the case of vanishing divergence.

As discussed in Sec.~\ref{sub:eigen3}, a complete set of vector fields can be introduced in the cavity interior V.
Since the fields are divergence-free, we can expand the solution of Maxwell's equations in terms of the type-(A) and type-(B) eigenvector fields:
\begin{eqnarray}
\label{eq:hexpand}
\bm{H}=\sum_{n}\bm{H}_n\int_{\tiny{\mbox V}}\bm{H}_n\cdot \bm{H} dv
+\sum_{\nu}\bm{H}_\nu\int_{\tiny{\mbox V}}\bm{H}_\nu\cdot \bm{H} dv\,.
\end{eqnarray}
Because only types (A) and (B) are retained, $\bm{H}_n$ represents a resonant electromagnetic field of the original cavity with J closed by a conducting lid.

We next determine the expansion coefficients in this equation.
For the first term, corresponding to the type-(B) fields, we transform the integral $\int_{\tiny{\mbox V}}\bm{E}_n\cdot \bm{E} dv$ in two different ways.
\begin{eqnarray}
\int_{\tiny{\mbox V}}\bm{E}_n\cdot \bm{E} dv
=\frac{1}{k_n}
\int_{\tiny{\mbox V}}\nabla\times\bm{H}_n\cdot\bm{E} dv\,.
\end{eqnarray}
Using the identity
\begin{eqnarray}
\nabla\cdot\left(\bm{H}_n\times \bm{E}\right)=\nabla\times\bm{H}_n\cdot\bm{E}
-\bm{H}_n\cdot\nabla\times\bm{E}\,,
\end{eqnarray}
gives
\begin{eqnarray}
\int_{\tiny{\mbox V}}\bm{E}_n\cdot \bm{E} dv
&=&\frac{1}{k_n}
\int_{\tiny{\mbox V}}\bm{H}_n\cdot\nabla\times\bm{E}dv+
\frac{1}{k_n}
\int_{\tiny{\mbox S}+\tiny{\mbox J}}\bm{H}_n\times\bm{E}\cdot\bm{n} dS
\nonumber \\
&=&-\frac{j \omega \mu}{k_n}
\int_{\tiny{\mbox V}}\bm{H}_n\cdot\bm{H}dv-
\frac{1}{k_n}
\int_{\tiny{\mbox J}}\bm{n}\times\bm{E}\cdot\bm{H}_n dS\,,
\end{eqnarray}
where the boundary condition~\eqref{eq:ebound} has been used in the last line.
On the other hand,
\begin{eqnarray}
\int_{\tiny{\mbox V}}\bm{E}_n\cdot \bm{E} dv
=\frac{1}{j\omega\epsilon}
\int_{\tiny{\mbox V}}\bm{E}_n\cdot\nabla\times\bm{H}dv\,,
\end{eqnarray}
and substitution of
\begin{eqnarray}
\nabla\cdot\left(\bm{H}\times \bm{E}_n\right)=\nabla\times\bm{H}\cdot\bm{E}_n
-\bm{H}\cdot\nabla\times\bm{E}_n\,,
\end{eqnarray}
gives
\begin{eqnarray}
\int_{\tiny{\mbox V}}\bm{E}_n\cdot \bm{E} dv
&=&\frac{1}{j \omega \epsilon}
\int_{\tiny{\mbox V}}\bm{H}\cdot\nabla\times\bm{E}_n dv+
\frac{1}{j \omega \epsilon}
\int_{\tiny{\mbox S}+\tiny{\mbox J}}\bm{H}\times\bm{E}_n\cdot\bm{n} dS
\nonumber \\
&=&\frac{k_n}{j \omega \epsilon}
\int_{\tiny{\mbox V}}\bm{H}_n\cdot\bm{H}dv\,,
\end{eqnarray}
In the last step, we used the boundary condition $\bm{n}\times\bm{E}_n=0$ on S+J.
Equating these two results gives
\begin{eqnarray}
\left(\frac{k_n}{j \omega \epsilon}+\frac{j \omega \mu}{k_n}\right)
\int_{\tiny{\mbox V}}\bm{H}_n\cdot\bm{H}dv
=-\frac{1}{k_n}
\int_{\tiny{\mbox J}}\bm{n}\times\bm{E}\cdot\bm{H}_n dS\,.
\end{eqnarray}
Using $\omega_n=c k_n$ and $\epsilon\mu=1/c^2$, and dividing both sides by $c\mu$, gives
\begin{eqnarray}
\label{eq:forced1}
\left(\frac{\omega_n}{j \omega}+\frac{j \omega}{\omega_n}\right)
\int_{\tiny{\mbox V}}\bm{H}_n\cdot\bm{H}dv
=-\frac{1}{\mu\omega_n}
\int_{\tiny{\mbox J}}\bm{n}\times\bm{E}\cdot\bm{H}_n dS\,.
\end{eqnarray}
Thus, the expansion coefficient for type (B) is
\begin{eqnarray}
\label{eq:var}
\int_{\tiny{\mbox V}}\bm{H}_n\cdot\bm{H}dv
&=&-\frac{1}{\left(\frac{\omega_n}{j \omega}+\frac{j \omega}{\omega_n}\right)}\cdot
\frac{1}{\mu\omega_n}
\int_{\tiny{\mbox J}}\bm{n}\times\bm{E}\cdot\bm{H}_n dS \nonumber \\
&=&-\frac{1}{j\mu\left(\omega-\frac{{\omega_n}^2}{\omega}\right)}
\int_{\tiny{\mbox J}}\bm{n}\times\bm{E}\cdot\bm{H}_n dS\,.
\end{eqnarray}

Next, the expansion coefficient $\int_{\tiny{\mbox V}}\bm{H}_\nu\cdot\bm{H}dv$ for type (A) is calculated.
This case differs slightly from the preceding one.
Recall that
\begin{eqnarray}
\nabla\times\bm{E}&=&-j\omega\mu\bm{H}\,.
\end{eqnarray}
Take the scalar product of both sides with $\bm{H}_\nu$ and integrate over V.
\begin{eqnarray}
\int_{\tiny{\mbox V}}\bm{H}_\nu\cdot\nabla\times\bm{E}dv
=-j\omega\mu\int_{\tiny{\mbox V}}\bm{H}_\nu\cdot\bm{H}dv\,.
\end{eqnarray}
To transform the left-hand side, we use
\begin{eqnarray}
\nabla\cdot\left(\bm{H}_\nu\times\bm{E}\right)
&=&\bm{E}\cdot\nabla\times\bm{H}_\nu
-\bm{H}_\nu\cdot\nabla\times\bm{E}\nonumber\\
&=&-\bm{H}_\nu\cdot\nabla\times\bm{E}\,.
\end{eqnarray}
Hence,
\begin{eqnarray}
\int_{\tiny{\mbox V}}\bm{H}_\nu\cdot\nabla\times\bm{E}dv
&=&\int_{\tiny{\mbox S}+\tiny{\mbox J}}\bm{E}\times\bm{H}_\nu\cdot\bm{n} dS
\nonumber\\
&=&\int_{\tiny{\mbox J}}\bm{n} \times\bm{E}\cdot\bm{H}_\nu dS\,.
\end{eqnarray}
Consequently,
\begin{eqnarray}
\label{eq:varnu}
\int_{\tiny{\mbox V}}\bm{H}_\nu\cdot\bm{H}dv
=-\frac{1}{j\omega\mu}
\int_{\tiny{\mbox J}}\bm{n}\times\bm{E}\cdot\bm{H}_\nu dS\,.
\end{eqnarray}
This gives the required expansion coefficient.
This result can be regarded as a special case ($\omega_n=0$) of the expansion coefficient~\eqref{eq:var} for type (B).

Let us now consider the expression obtained by transforming Eq.~\eqref{eq:forced1},
\begin{eqnarray}
\label{eq:forced2}
\left({\omega_n}^2-{\omega}^2\right)
\int_{\tiny{\mbox V}}\bm{H}_n\cdot\bm{H}dv
=-\frac{j\omega}{\mu}
\int_{\tiny{\mbox J}}\bm{n}\times\bm{E}\cdot\bm{H}_n dS\,.
\end{eqnarray}
As can be seen by replacing $j \omega$ with $d/dt$, this is an equation of forced oscillation and one of the most fundamental equations in considering resonance phenomena in cavities.
It should be noted that the driving term on the right-hand side is expressed by the ``electric field parallel to the boundary surface J.''
That is, the electric field on J induces the coefficient~\eqref{eq:var} through Maxwell's equations, and as a result a magnetic field represented by the expansion~\eqref{eq:hexpand} is produced {\bf inside V}.
Appendix~\ref{ap:var} shows that the same expression can also be obtained from a variational representation.

Let us examine more closely the magnetic field generated inside the cavity by the electric field at its boundary.
As discussed in Sec.~\ref{sub:eigen2}, a complete set of vector fields can be introduced on the two-dimensional region J.
We first expand the tangential component $\bm{E}_{\tiny{/\! /}}$ of $\bm{E}$ on J in terms of the complete set $\bm{E}_{tl}$ on J.
Since $\bm{E}_{\tiny{/\! /}}$ satisfies the boundary condition~\eqref{eq:e_bou_2dim} on the boundary of J, it is appropriate to use the complete set $\bm{E}_{tk}$ satisfying ``electric-type'' boundary conditions.
\begin{eqnarray}
\label{eq:coefv}
\bm{E}_{\tiny{/\! /}}
\coloneqq\sum_{l}V_{l} \bm{E}_{tl}\,.
\end{eqnarray}
Here $V_{l}$ is an expansion coefficient and may be regarded as an input parameter specifying the electric field on the boundary surface J.
Note that Maxwell's equations give $\nabla\cdot\bm{E}=0$, whereas the tangential component $\bm{E}_{\tiny{/\! /}}$ generally satisfies $\nabla\cdot\bm{E}_{\tiny{/\! /}}\neq0$ on J.
Thus, all three types (A) through (C) in two dimensions, considered in Sec.~\ref{sub:eigen2}, are required for the expansion.

Next, we expand the vector field $\bm{n} \times\bm{H}_n$ on J.
Since this vector field satisfies the ``electric-type'' boundary condition~\eqref{eq:e_bou_2dim} on the boundary of J, it is appropriate to expand it in terms of $\bm{E}_{tk}$.
\begin{eqnarray}
\label{eq:ntimesh}
\bm{n} \times\bm{H}_n
=\sum_{k}I_{nk}\bm{E}_{tk}\,,
\end{eqnarray}
where
\begin{eqnarray}
\label{eq:ink}
I_{nk}=\int_{\tiny{\mbox J}}\bm{E}_{tk}\cdot\left(\bm{n} \times\bm{H}_n\right)dS\,.
\end{eqnarray}
This coefficient is determined by the eigenvector fields of the cavity and the aperture.
As will be seen in the following sections, this expansion coefficient plays an essential role in considering resonant electromagnetic fields.

Taking the cross product of both sides of Eq.~\eqref{eq:ntimesh} with $-\bm{n}$ gives
\begin{eqnarray}
\bm{n}\times\left(\bm{n}\times\bm{H}_n\right)
=-\bm{H}_n,
\end{eqnarray}
and hence
\begin{eqnarray}
\label{eq:hna}
\bm{H}_n
=\sum_{k}I_{nk}\left(-\bm{n} \times\bm{E}_{tk}\right)\,.
\end{eqnarray}
Note that $-\bm{n} \times\bm{E}_{tl}$ is a vector field satisfying the ``magnetic-type'' boundary condition~\eqref{eq:h_bou_2dim} on the boundary of the two-dimensional plane J, as considered in Sec.~\ref{sub:eigen2}.
Similarly,
\begin{eqnarray}
\label{eq:hnu}
\bm{H}_\nu
=\sum_{k}I_{\nu k}\left(-\bm{n} \times\bm{E}_{tk}\right)\,,
\end{eqnarray}
where
\begin{eqnarray}
\label{eq:icoeff2}
I_{\nu k} =\int_{\tiny{\mbox J}}\bm{E}_{tk}\cdot\left(\bm{n} \times\bm{H}_\nu\right)dS\,.
\end{eqnarray}
Substituting expansions~\eqref{eq:coefv} and \eqref{eq:hna} into the surface integral over the aperture on the right-hand side of Eq.~\eqref{eq:var}, and using $\bm{n}\times\bm{E}=\bm{n}\times\bm{E}_{\tiny{/\! /}}$, gives
\begin{eqnarray}
-\int_{\tiny{\mbox J}}\bm{n}\times\bm{E}\cdot\bm{H}_n dS
&=&\int_{\tiny{\mbox J}}
\sum_{l}V_{l} \left(\bm{n} \times\bm{E}_{tl} \right)\cdot
\sum_{k}I_{nk}\left(\bm{n} \times\bm{E}_{tk}\right) 
dS \nonumber \\
&=&\sum_k V_k I_{nk}\,,
\end{eqnarray}
where the orthonormality condition for $\left\{\bm{E}_{tk}\right\}$ has been used.
Similarly, from Eqs.~\eqref{eq:coefv} and \eqref{eq:hnu}, the surface integral on the right-hand side of Eq.~\eqref{eq:varnu} becomes
\begin{eqnarray}
-\int_{\tiny{\mbox J}}\bm{n}\times\bm{E}\cdot\bm{H}_{\nu} dS
=\sum_k V_k I_{\nu k}\,.
\end{eqnarray}
Substituting the above results into Eq.~\eqref{eq:hexpand}, we obtain the following expansion of the magnetic field 
$\bm{H}$ generated {\bf inside V} when the voltages $\left\{ V_k \right\}$ are applied at the boundary surface:
\begin{eqnarray}
\label{eq:exp}
\bm{H}&=&
\sum_n{\frac{1}{j \left(\frac{\omega}{\omega_n}-\frac{\omega_n}{\omega}\right)}\cdot
\frac{1}{\mu\omega_n}
\sum_k V_k I_{nk}\bm{H}_n}
+\sum_\nu{\frac{1}{j \mu\omega}
\sum_k V_k I_{\nu k}\bm{H}_\nu}\,.
\end{eqnarray}

Consider the expansion~\eqref{eq:exp}.
First, $\omega$ is the frequency of the input voltage applied at the aperture.
If it is close to any one of the original resonant frequencies $\omega_n$, the corresponding field $\bm{H}_n$ is strongly excited.
However, even if an electric-field component $V_k$ is applied in a direction $\bm{E}_{tk}$ orthogonal to $\bm{n}\times\bm{H}_n$ on the junction surface J, the magnetic field is not excited, because $I_{nk}=0$.

What, then, is the magnetic field on the junction surface J?
Since expansions~\eqref{eq:exp} and \eqref{eq:hexpand} are defined as series converging in the mean-square sense in V, one might think that it is meaningless to discuss their values on J, which is the boundary of V.
In what follows, however, we assume that this
expansion can also be applied to the tangential
component of the magnetic field on $J$\footnote{
Mean-square convergence of the eigenmode expansion
in $V$ does not, by itself, guarantee convergence
of its tangential components on the boundary $J$.
In a rigorous treatment of Maxwell's equations,
tangential boundary values can be defined in a
generalized sense using trace operators on
$H(\mathrm{curl};V)$; see, e.g.,
Monk~\cite{bib:Mon03}, Chapter~3.
Convergence in $H(\mathrm{curl};V)$ implies convergence
of the corresponding tangential traces in an
appropriate boundary space.
Here we assume that the eigenmode expansion admits
such a boundary interpretation, without examining
the detailed convergence conditions.
}.
 Then, substituting the expansions~\eqref{eq:hna} and \eqref{eq:hnu} for $\bm{H}_n$ and $\bm{H}_\nu$ on the interface into Eq.~\eqref{eq:exp}, we obtain the following expression on J.
 \begin{eqnarray}
\label{eq:kekka1}
\bm{H}_{\tiny{/\! /}}&=&
\sum_n{\frac{1}{j \left(\frac{\omega}{\omega_n}-\frac{\omega_n}{\omega}\right)}\cdot
\frac{1}{\mu\omega_n}
\sum_k V_k I_{nk}\sum_{l}I_{nl}\left(-\bm{n} \times\bm{E}_{tl}\right)}\nonumber \\
&+&\sum_\nu{\frac{1}{j \mu\omega}
\sum_k V_k I_{\nu k}\sum_{l}I_{\nu l}\left(-\bm{n} \times\bm{E}_{tl}\right)}\nonumber \\
&=&\sum_{l}\left(-\bm{n} \times\bm{E}_{tl}\right)
\sum_k\left[
\sum_n\frac{I_{nl}I_{nk}}
{j \omega_n \mu \left(\frac{\omega}{\omega_n}-\frac{\omega_n}{\omega}\right)}
+\sum_\nu \frac{I_{\nu l}I_{\nu k}}{j \omega \mu}
\right]V_k\,.
\end{eqnarray}
That is, when the electric-field input~\eqref{eq:coefv} is applied at the boundary surface J, this equation expresses the magnetic field produced {\bf on J} through Maxwell's equations, and is therefore the most fundamental equation in considering the response of the cavity resonator.

\begin{figure}[htb]
   \centering
   \includegraphics*[width=120mm]{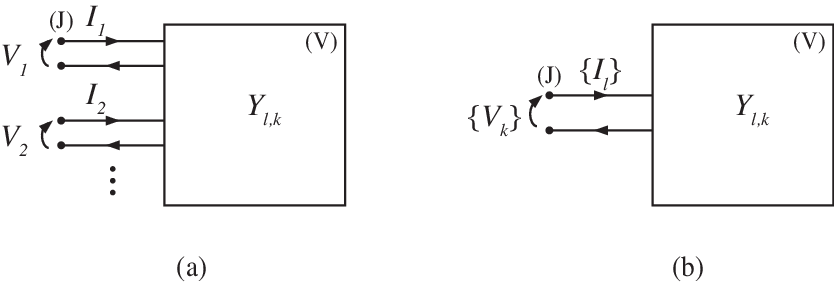}
   \caption{Schematic illustration of the input admittance of a cavity.
   As shown in (a), the cavity corresponds to a circuit having multiple ports according to the number of eigenvector fields on the boundary surface J.
   For simplicity, it is represented as in (b).}
   \label{fig:adm}
\end{figure}

For clarity, we rewrite Eq.~\eqref{eq:kekka1} for the magnetic field on J as
\begin{eqnarray}
\label{eq:kekka2}
\bm{H}_{\tiny{/\!/}}=\sum_{l}\left(-\bm{n} \times\bm{E}_{tl}\right) I_{l}\,.
\end{eqnarray}
That is,
\begin{eqnarray}
\label{eq:kekka3}
 I_{l}&=&\sum_k\left[\sum_n\frac{I_{nl}I_{nk}}
{j \omega_n \mu \left(\frac{\omega}{\omega_n}-\frac{\omega_n}{\omega}\right)}
+\sum_\nu \frac{I_{\nu l}I_{\nu k}}{j \omega \mu}\right]V_k \nonumber \\
&=&\sum_k Y_{lk}(j\omega)\cdot V_k\,,
\end{eqnarray}
where
\begin{eqnarray}
\label{eq:adm}
 Y_{lk}(j\omega)\coloneqq \sum_n\frac{I_{nl}I_{nk}}
{j \omega_n \mu \left(\frac{\omega}{\omega_n}-\frac{\omega_n}{\omega}\right)}
+\sum_\nu \frac{I_{\nu l}I_{\nu k}}{j \omega \mu}\,.
\end{eqnarray}
This is the response function, or admittance, of the cavity resonator, which gives the output ``current'' $I_l$ in response to the input ``voltage'' $V_k$.
Since the relation involves multiple components, $Y_{lk}$ is called the ``admittance matrix.''
The indices $l$ and $k$ refer to the eigenvector fields $\left\{\bm{E}_{tk}\right\}$ at the boundary surface J.
As mentioned above, the coefficients $I_{nk}$ and $I_{\nu k}$ are determined by the eigenvector fields of the cavity V and the boundary surface J, as can be seen from Eqs.~\eqref{eq:hna} and \eqref{eq:hnu}, and are intrinsic to the cavity, together with the eigenvalues $\omega_n$.
Figure~\ref{fig:adm} illustrates the cavity response represented by Eq.~\eqref{eq:kekka3}.
Unlike an ordinary one-port circuit, this description has as many ports as there are eigenvector fields on J (in principle, infinitely many).

\section{Properties of single-mode admittance}
\label{sec:single}

As shown by \eqref{eq:adm}, the input admittance of a cavity resonator is completely determined by the resonant frequencies $\{\omega_n\}$ and expansion coefficients $\{I_{nk}\}$ of the original cavity.
To make the discussion easier to follow, we consider the general properties of the input admittance by retaining only one eigenvector field at the aperture.

\subsection{Foster's form}

First, we set $l=k=1$ in \eqref{eq:adm} and consider
\begin{eqnarray}
\label{eq:adm1}
Y_{11}(j\omega)\coloneqq 
\sum_n\frac{{I_{n1}}^2}
{j \omega_n \mu \left(\frac{\omega}{\omega_n}-\frac{\omega_n}{\omega}\right)}
+\sum_\nu \frac{{I_{\nu 1}}^2}{j \omega \mu}\,.
\end{eqnarray}
This corresponds to retaining a single aperture mode.
Introducing the complex-frequency variable $s$ in place of $j\omega$, we obtain
\begin{eqnarray}
\label{eq:adm1s}
Y_{11}(s)= \sum_n\frac{{I_{n1}}^2 s}
{\mu(s^2+{\omega_n}^2)}
+\sum_\nu \frac{{I_{\nu 1}}^2}{\mu s}\,.
\end{eqnarray}
This is the typical form of a lossless positive-real function known as the Foster expansion\footnote{
A function $Z(s)$ of the complex frequency $s=\sigma+j\omega$ is called a positive-real function if it is analytic in the right half-plane $\mathrm{Re}\,s>0$, satisfies $\mathrm{Re}\,Z(s)\geq 0$, and also satisfies $Z(s^*)=Z(s)^*$.
The driving-point impedance of a passive one-port network is a positive-real function.
In particular, in the lossless case, $\mathrm{Re}\,Z(j\omega)=0$ except at poles on the imaginary axis, and the impedance is purely imaginary.
For a lossless real-rational positive-real function, all finite poles are simple poles on the imaginary axis, and their residues are positive real numbers.
This property forms the basis of the Foster expansion \cite{bib:van60}.
}.

\begin{figure}[htb]
   \centering
   \includegraphics*[width=160mm]{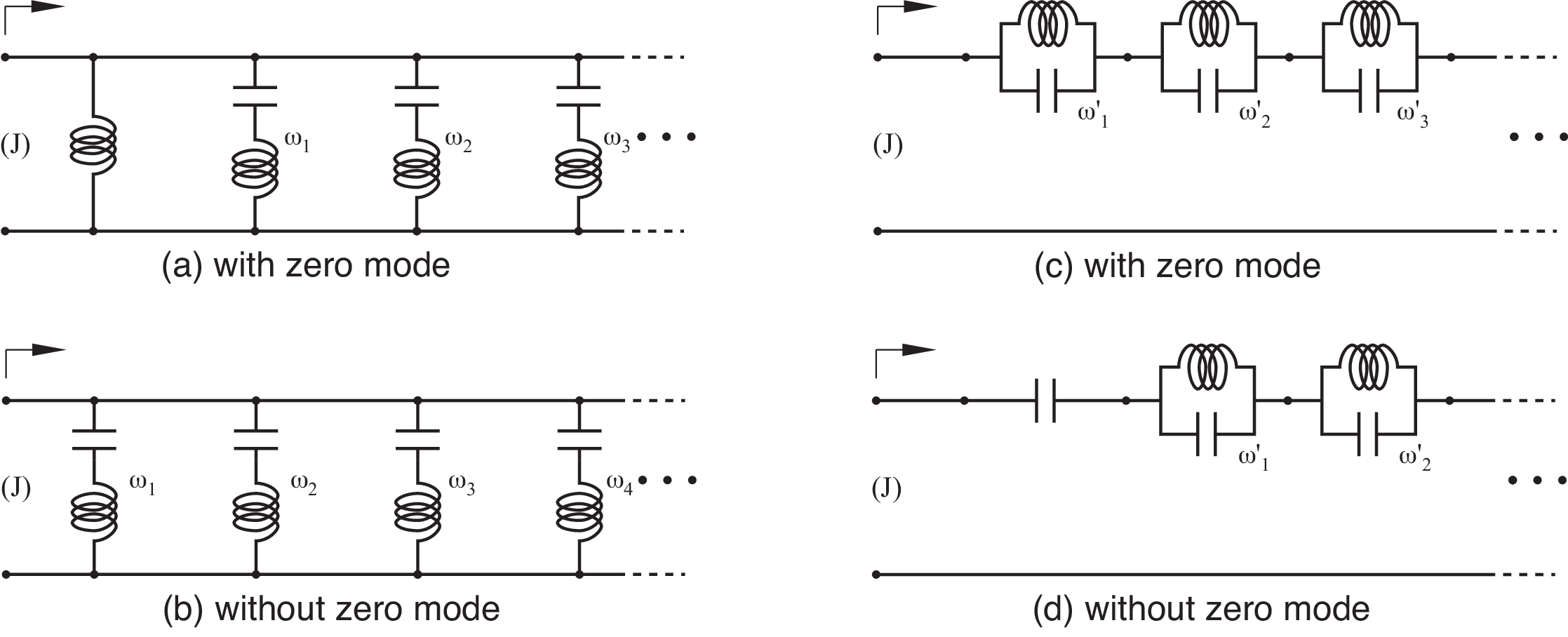}
   \caption{Equivalent circuits representing the input admittance of a cavity resonator. (a) corresponds to a cavity having a DC eigenvector field, while (b) corresponds to a cavity without a DC eigenvector field. These are Foster's first forms. (c) and (d) are Foster's second forms corresponding to (a) and (b), respectively.}
   \label{fig:foster_inf}
\end{figure}

The equivalent circuit for the input admittance in \eqref{eq:adm1}
is shown in Fig.~\ref{fig:foster_inf}(a) or (b).
Both are examples of Foster's first form.
Case (a) represents a cavity with a DC mode, whereas case (b) represents one without a DC mode.
The resonance frequencies of the individual series resonant circuits correspond to the poles of $Y_{11}(j\omega)$ in \eqref{eq:adm1}.
These frequencies correspond to the eigenvalues of the Laplacian associated with the eigenvector fields used in the expansion. They are the resonant frequencies allowed when the cavity is closed at $J$, that is, when $J$ is short-circuited.
The corresponding Foster second forms are shown in Fig.~\ref{fig:foster_inf}(c) and (d), respectively.
The resonant frequencies $\omega_1',~\omega_2',\cdots$ of the individual parallel resonant circuits are the zeros of $Y_{11}(j\omega)$.
These are the resonant frequencies allowed when $J$ is open-circuited.
The zeros and poles appear alternately\footnote{
It should be noted here that, for example in (a), the sequence of series resonant circuits connected in parallel continues infinitely.
This is a general property of the eigenvalues of the Laplacian \cite{bib:cou53}, and consequently the point at infinity ($s=\infty$) becomes an essential singularity.
Thus, for the input admittance of a cavity resonator, there are only two standard forms, according to whether $s=0$ is a pole or a zero, and these are exhausted by (a) and (b) in Fig.~\ref{fig:foster_inf}.
On the other hand, for a lossless circuit consisting of a finite number of elements, the partial-fraction expansion of the admittance contains only a finite number of terms, so that the point at infinity becomes either a simple pole or a simple zero.
Therefore, together with the behavior at $s=0$, there are four cases in all \cite{bib:wad06}.
The correspondence between the first and second forms also differs from that for a cavity.
}.

A simple example illustrating the correspondence between (a) and (c), and between (b) and (d), is the identity $1/\cot(x)=\tan(x)$.
Indeed, the partial-fraction expansions of $\cot$ and $\tan$ are
\begin{eqnarray}
\label{eq:cot}
\cot(x)&=&\frac{1}{x}+\sum_{n=1}^{\infty}\frac{2 x}{x^2-n^2\pi^2}\,,\\
\label{eq:tan}
\tan(x)&=&-\sum_{n=0}^{\infty}\frac{2 x}{x^2-(n+1/2)^2\pi^2}\,.
\end{eqnarray}
These expressions appear in the specific examples considered in the following subsection.

Finally, consider the low-frequency limit of \eqref{eq:adm1}.
For $\omega\ll\omega_n$, \eqref{eq:adm1} becomes approximately
\begin{eqnarray}
\label{eq:adm1_lf}
 Y_{11}(j\omega)\simeq j\omega C_0+\frac{1}{j\omega L_0}\,,
\end{eqnarray}
where $C_0$ and $L_0$ are defined by
\begin{eqnarray}
\label{eq:c0l0}
C_0\coloneqq \sum_n\frac{{I_{n1}}^2}{{\omega_n}^2\mu}, \qquad \frac{1}{L_0}
\coloneqq \sum_{\nu}\frac{{I_{\nu 1}}^2}{\mu}\,.
\end{eqnarray}
See also Ref.~\cite{bib:kur58}.
Thus, at low frequencies, the first term on the right-hand side of \eqref{eq:adm1} is capacitive, whereas the second is inductive.
This is also apparent from Fig.~\ref{fig:foster_inf}(a) or (b).
Note that each term in the sum defining the capacitance $C_0$ is proportional to ${I_{n1}}^2$.

\subsection{Expansion coefficients and input admittances}

\begin{figure}[htb]
   \centering
   \includegraphics*[width=40mm]{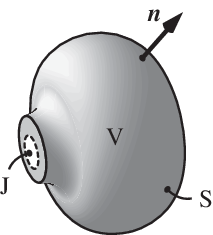}
   \caption{A situation in which the coupling aperture of a cavity is made progressively smaller. The region inside the dotted line is the junction surface $J$.}
   \label{fig:narrow}
\end{figure}

Next, we examine the properties of the expansion coefficients $\{I_{nk}\}$.
Consider how the coupling coefficients behave as the aperture of a cavity is made progressively smaller, as illustrated in Fig.~\ref{fig:narrow}.
Let $A_{\mbox {\tiny J}}$ denote the area of the aperture $J$.
The eigenvector fields $\{\bm{H}_n\}$ in $V$ and $\{\bm{E}_{tn}\}$ on $J$ are normalized according to \eqref{eq:hnorm} and \eqref{eq:etnorm}, respectively.
As the aperture becomes smaller, the local magnitude of $\{\bm{E}_{tn}\}$ therefore scales as ${A_{\mbox {\tiny J}}}^{-1/2}$.
By contrast, $\{\bm{H}_n\}$ remains unchanged.
Thus, from \eqref{eq:ink}, the order of magnitude of $I_{nk}$ is
\begin{eqnarray}
I_{nk}\propto {A_{\mbox {\tiny J}}}^{-1/2}\cdot A_{\mbox {\tiny J}}={A_{\mbox {\tiny J}}}^{1/2}\,.
\end{eqnarray}

\subsection{Examples of input admittances}

We now evaluate the expansion coefficients $\{I_{nk}\}$ for several examples and examine the corresponding input admittances.
\subsubsection{Coaxial resonator}

\begin{figure}[htb]
   \centering
   \includegraphics*[width=100mm]{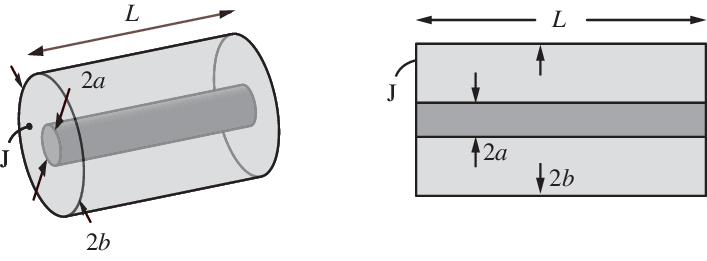}
   \caption{Coaxial resonator.}
   \label{fig:coax_sgl}
\end{figure}

We begin with a coaxial resonator.
Figure~\ref{fig:coax_sgl} shows a half-wavelength coaxial resonator of length $L$, with inner- and outer-conductor radii $a$ and $b$, respectively. We apply the preceding formulation to obtain its expansion coefficients and input admittance.
As before, we ask what magnetic field is produced on $J$ when the end wall is removed and an electric field is applied at the resulting aperture.

\begin{figure}[htb]
   \centering
   \includegraphics*[width=150mm]{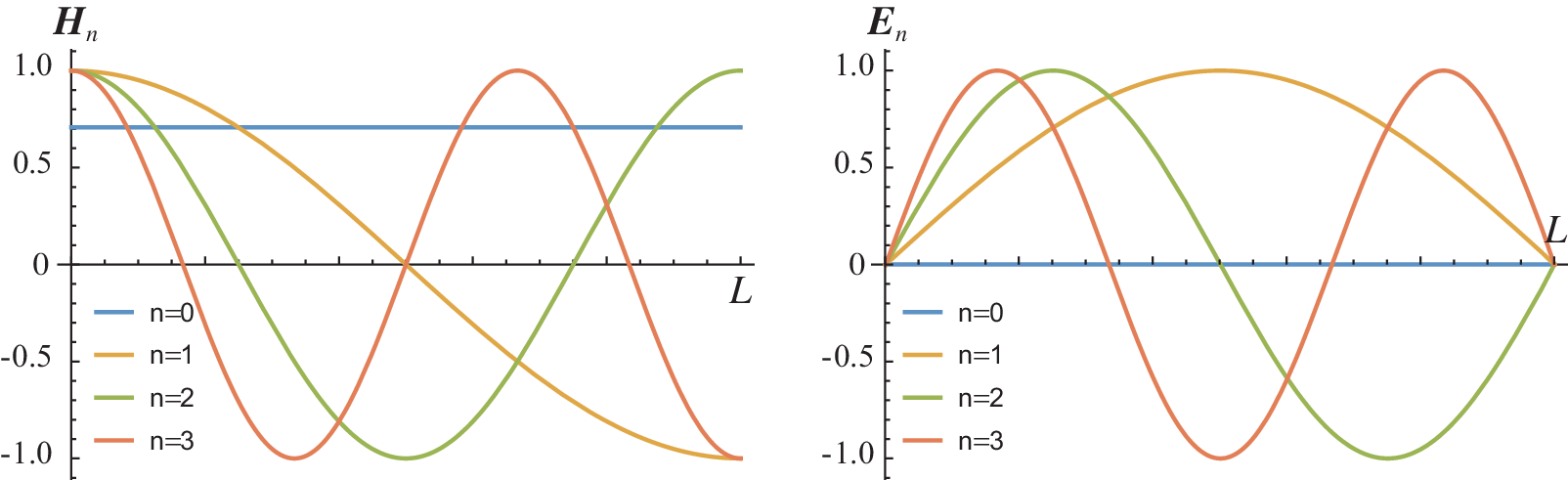}
   \caption{Eigenvector fields of a coaxial cavity resonator, shown for $n=0,1,2,3$.}
   \label{fig:unitcell}
\end{figure}

At low frequencies, it is sufficient to consider TEM modes as the eigenvector fields of the cavity.
With $n$ as the mode index, the TEM eigenvector fields can be written in cylindrical coordinates as
\begin{eqnarray}
\label{eq:hnpol}
\bm{H}_n&=&\frac{1}{\sqrt{\pi L\left(1+\delta_{n0}\right)\log\frac{b}{a}}}\cdot
\frac{1}{r}\cos\left(\frac{\omega_n}{c}z\right)\bm{e}_\theta
\qquad n=0,1,2,\cdots\,,\\
\bm{E}_n&=&\frac{1}{\sqrt{\pi L\log\frac{b}{a}}}\cdot
\frac{1}{r}\sin\left(\frac{\omega_n}{c}z\right)\bm{e}_r
\qquad n=1,2,\cdots\,,
\end{eqnarray}
and the eigenvalues $\omega_n$ are
\begin{eqnarray}
\label{eq:eigentem}
\omega_n=\frac{n c \pi}{L}
\qquad n=0,1,2,\cdots\,.
\end{eqnarray}
Figure~\ref{fig:unitcell} illustrates the lowest-order eigenvector fields.

We further retain only the lowest-order eigenvector field on the cross section $J$.
This field corresponds to a zero eigenvalue and is given by
\begin{eqnarray}
\label{eq:dammen}
\bm{E}_{t1}&=&\frac{1}{\sqrt{2 \pi \log\frac{b}{a}}}\cdot
\frac{1}{r}\bm{e}_r\,.
\end{eqnarray}
This is expected to be a good approximation when the frequencies under consideration are low.

Using these fields, we calculate the expansion coefficients at the junction from \eqref{eq:ink}:
\begin{eqnarray}
\label{eq:coeff_coax}
I_{n1}&=&\int_{z=0}\bm{E}_{t1}\cdot\left(-\bm{e}_z \times\bm{H}_n\right)dS
\nonumber\\
&=&\frac{2}{\sqrt{2L\left(1+\delta_{n0}\right)}}\,.
\end{eqnarray}
Substituting \eqref{eq:eigentem} and \eqref{eq:coeff_coax} into \eqref{eq:adm1} and using \eqref{eq:cot}, we obtain
\begin{eqnarray}
\label{eq:y11_coax}
Y_{11}(j\omega)=-\frac{j}{c\mu}\cot\left(\frac{L\omega}{c}\right)\,.
\end{eqnarray}
The difference from the coefficient in the usual circuit treatment arises from the normalization of the ``voltage'' and ``current''.
Using \eqref{eq:coefv} and \eqref{eq:kekka1} to calculate the voltage and current, we find that $Y_{11}$ must be multiplied by $2\pi/\log(b/a)$ to obtain the conventional circuit admittance.
Thus, the admittance in conventional circuit units is
\begin{eqnarray}
\label{eq:adm_coax}
-\frac{2\pi j}{c\mu\log(b/a)}\cot\left(\frac{L\omega}{c}\right)=\frac{1}{jZ_0\tan\left(\frac{L\omega}{c}\right)}\,,
\end{eqnarray}
where the characteristic impedance of the coaxial line is defined by
\begin{eqnarray}
\label{z_coax}
Z_0\coloneqq\frac{c\mu}{2\pi}\log(b/a)\,.
\end{eqnarray}

\begin{figure}[htb]
   \centering
   \includegraphics*[width=100mm]{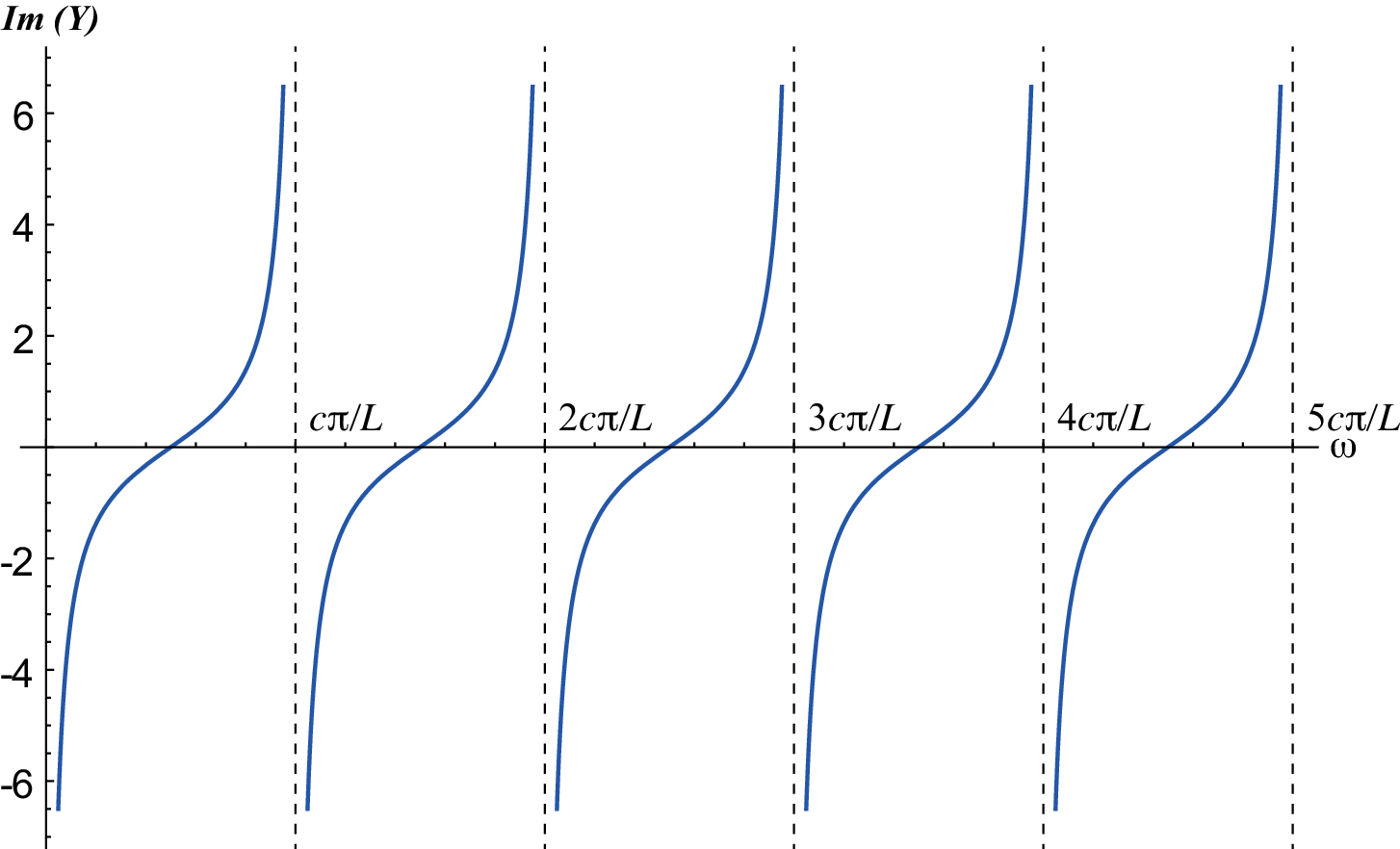}
   \caption{Imaginary part of $Y_{11}(j\omega)$ for a coaxial resonator. The dashed lines indicate the positions of the poles.}
   \label{fig:cot}
\end{figure}

Figure~\ref{fig:cot} shows the imaginary part of $Y_{11}(j\omega)$.
It has essentially the form of $-\cot(x)$, with a zero midway between each pair of successive poles.
This follows because the poles of $Y_{11}(s)$ are equally spaced along the imaginary axis and all the residues are equal (${I_{n1}}^2$ is constant).

\subsubsection{Quarter-wavelength resonator}

\begin{figure}[htb]
   \centering
   \includegraphics*[width=100mm]{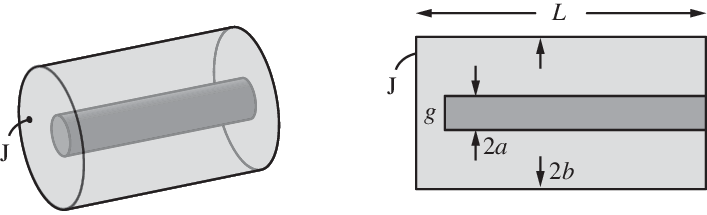}
   \caption{Quarter-wavelength resonator.}
   \label{fig:qwr}
\end{figure}

Next, consider the cavity shown in Fig.~\ref{fig:qwr}.
It is obtained by introducing a gap at one end of the resonator shown in Fig.~\ref{fig:coax_sgl}.
When $L$ is sufficiently larger than $b$, the lowest-frequency resonant electric field has a quarter-wavelength distribution, with the right end short-circuited and the left end open. This is why it is called a quarter-wavelength resonator.
In the following, we remove the end surface on the side with the gap, treat it as the coupling aperture $J$, and consider the cavity response as seen from this aperture.

Before evaluating the coupling coefficients, let us first consider the input impedance as seen from $J$.
An intuitive approximation is
\begin{eqnarray}
\label{eq:imp_qwr}
Z(j\omega)=\frac{1}{j\omega C_f}+jZ_0\tan\left(\frac{L\omega}{c}\right)\,.
\end{eqnarray}
Here $Z_0$ is the characteristic impedance of the coaxial section, and $C_f$ is the capacitance at the left end. The input impedance as seen from $J$ is approximated by their series combination.
When the coupling is strong, $C_f$ is large, whereas when the coupling is weak, $C_f$ is small.

\begin{figure}[htb]
   \centering
   \includegraphics*[width=100mm]{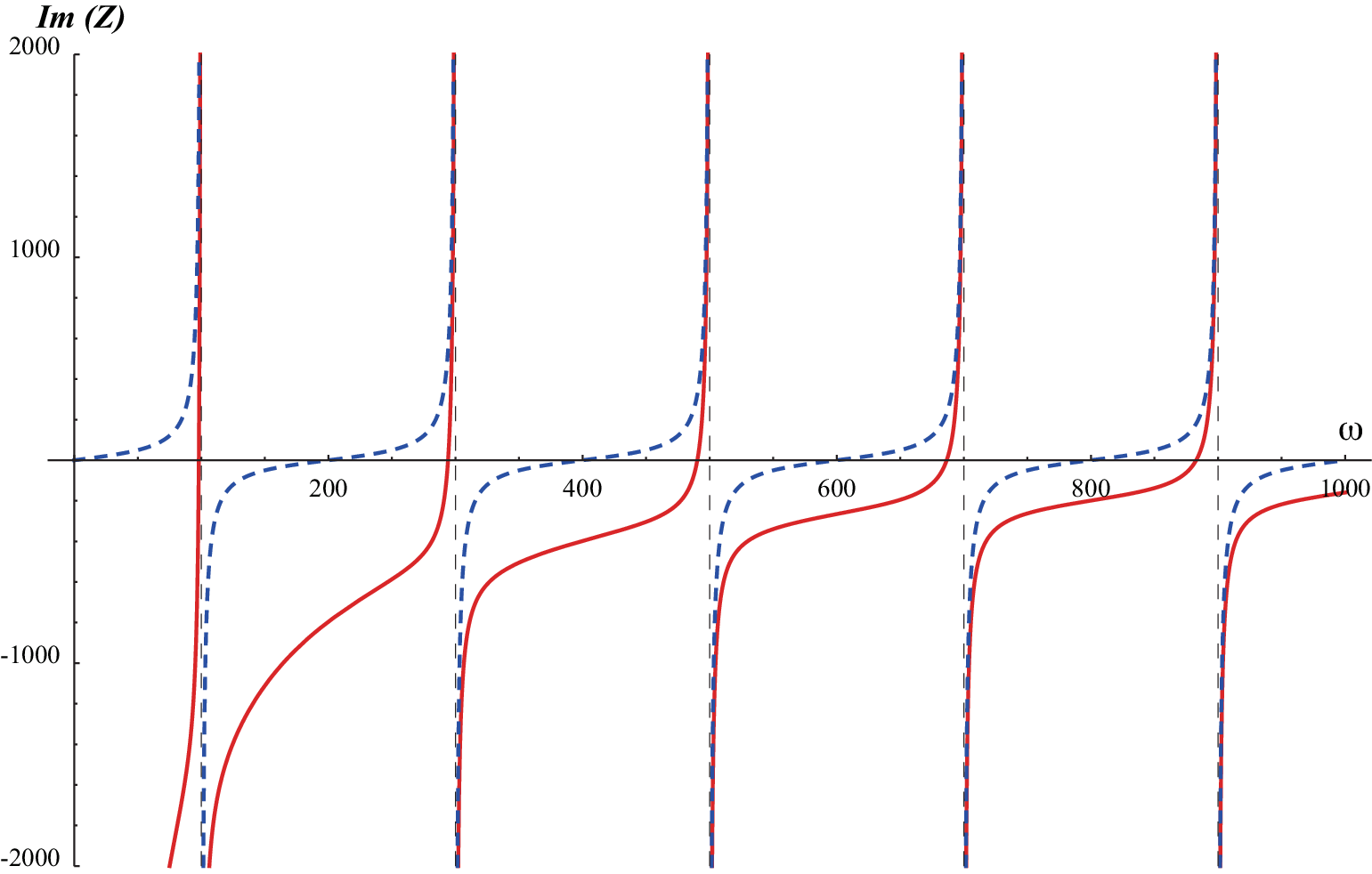}
   \caption{The red solid line shows the imaginary part of the circuit impedance of the quarter-wavelength resonator, given by \eqref{eq:imp_qwr}, as a function of frequency [MHz]. The parameters are $L=0.75 [{\mbox m}]$, $c=3.0\times10^8 [{\mbox {m/s}}]$, $Z_0=50 [\Omega]$, and $C_f=1.0\times 10^{-12} [{\mbox F}]$. The thin black dashed lines indicate the impedance poles. Note that the zeros and poles are close in the low-frequency region. The blue dashed line shows the result without the capacitance, corresponding to the second term on the right-hand side of \eqref{eq:imp_qwr}. The difference between the red solid and blue dashed lines is $-1/\omega C_f$, which causes the impedance zeros to approach the poles.}
   \label{fig:imp_qwr}
\end{figure}

The imaginary part of this impedance is plotted in Fig.~\ref{fig:imp_qwr}.
Unlike in Fig.~\ref{fig:cot}, the poles and zeros are close to each other, particularly in the low-frequency region.
This is because the series coupling capacitance shifts the graph of $\tan x$ downward.
As $C_f$ decreases, the graph shifts further downward, and the zero approaches the pole.
The zero also lies at a lower frequency than the pole.

\begin{figure}[htb]
   \centering
   \includegraphics*[width=100mm]{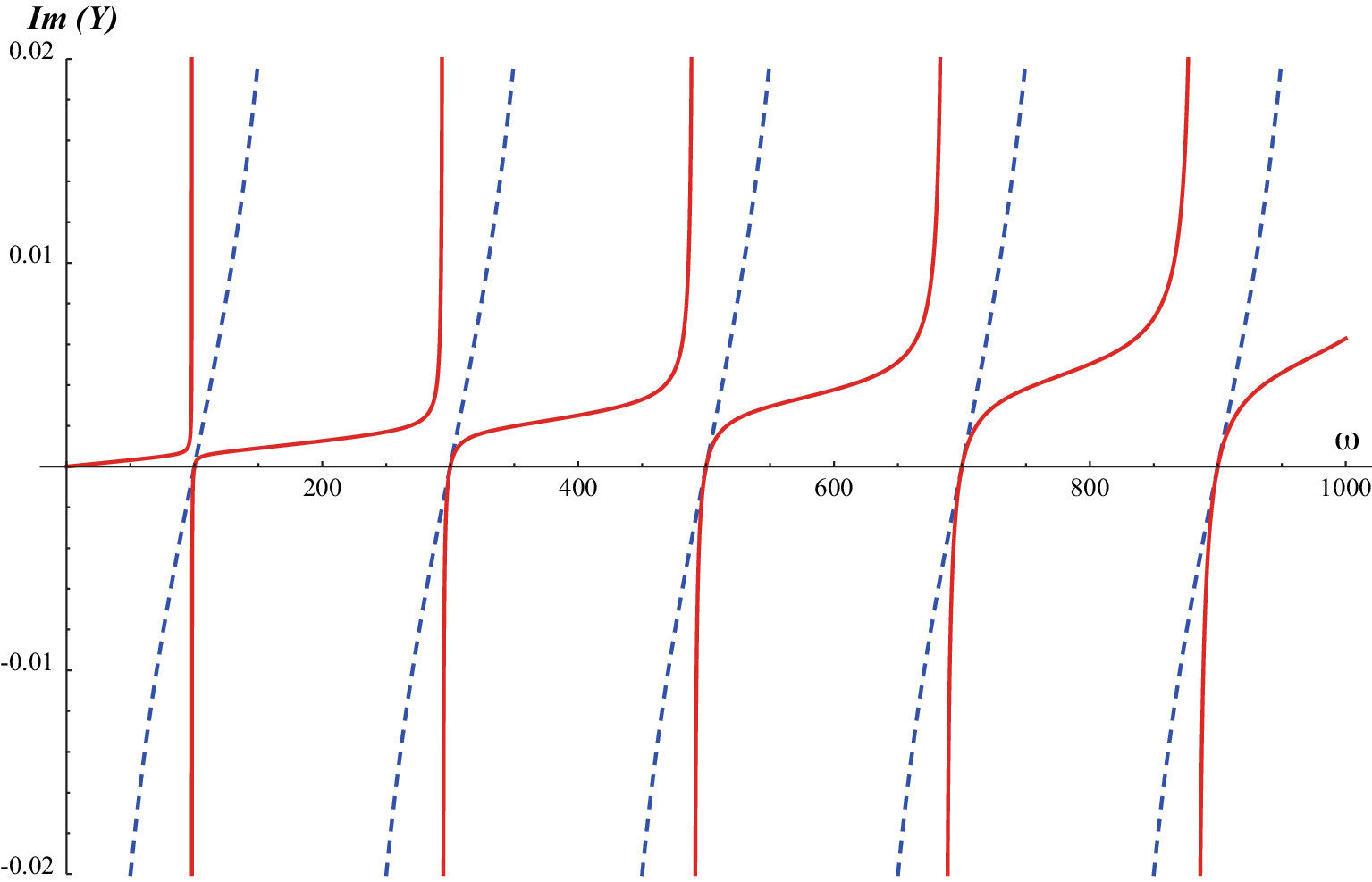}
   \caption{The red solid line shows the imaginary part of the circuit admittance of the quarter-wavelength resonator, obtained as the reciprocal of \eqref{eq:imp_qwr}, as a function of frequency [MHz]. The parameters are the same as in Fig.~\ref{fig:imp_qwr}. The blue dashed line is the imaginary part of the reciprocal of the impedance without the capacitance, namely \eqref{eq:adm_coax}.}
   \label{fig:adm_qwr}
\end{figure}

Figure~\ref{fig:adm_qwr} shows the imaginary part of the reciprocal of \eqref{eq:imp_qwr}, namely, the admittance.
Again, the poles and zeros are close to each other in the low-frequency region.
These observations suggest that, particularly for weak coupling, both the impedance and admittance can take arbitrarily large or small values in the vicinity of their respective zeros in the low-frequency region.
The meaning of this will become clear in the next section.

Equation~\eqref{eq:imp_qwr} corresponds to Fig.~\ref{fig:foster_inf}(d).
Therefore, its reciprocal, the admittance, can be expanded in partial fractions in the form shown in Fig.~\ref{fig:foster_inf}(b). The residue at each pole should then be obtainable from \eqref{eq:imp_qwr} by the procedure of network synthesis.
Here, however, rather than following that procedure, we consider the electromagnetic fields in the cavity and qualitatively examine how the expansion coefficients $I_{n1}$ in \eqref{eq:adm1} depend on $n$.

\begin{figure}[htb]
   \centering
   \includegraphics*[width=80mm]{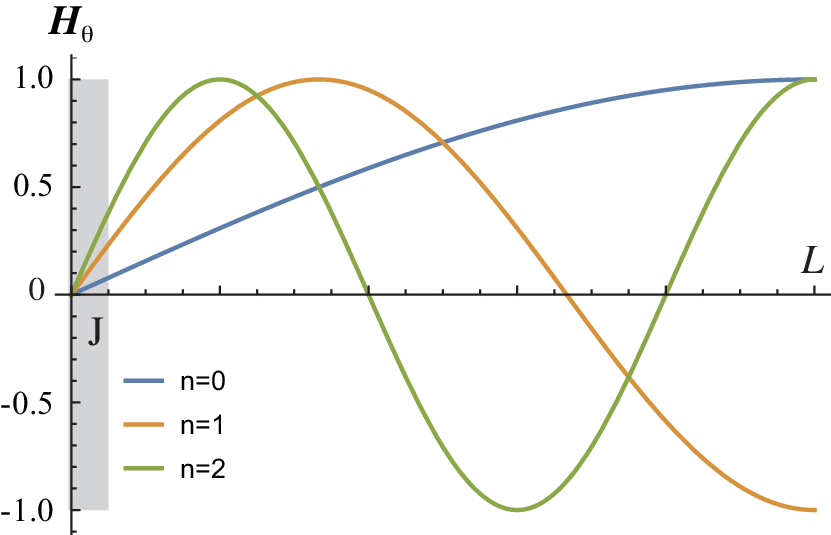}
   \caption{Resonant magnetic fields of a quarter-wavelength resonator. The right end is short-circuited and the left end is open. $n$ represents the number of nodes. The hatched region on the left is regarded as the junction surface $J$.}
   \label{fig:qwr_H}
\end{figure}

The resonant electromagnetic fields of the original cavity assumed in \eqref{eq:imp_qwr} are TEM-like $(2n+1)/4$-wavelength modes, where $n$ denotes the number of nodes.
The normalized eigenvector fields then have distributions such as those shown in Fig.~\ref{fig:qwr_H}.
Although the magnetic field is very weak near the open end, it gradually extends into this region as $n$ increases.
From Fig.~\ref{fig:qwr_H}, the magnitude of the magnetic field extending into the junction may be considered proportional to $2n+1$.
On the other hand, the two-dimensional eigenvector field on the junction that couples to these eigenvector fields is the same TM$_{01}$ mode.
Thus, from \eqref{eq:ink}, it is reasonable to expect $I_{n1}$ to increase with $n$.

\begin{figure}[htb]
   \centering
   \includegraphics*[width=100mm]{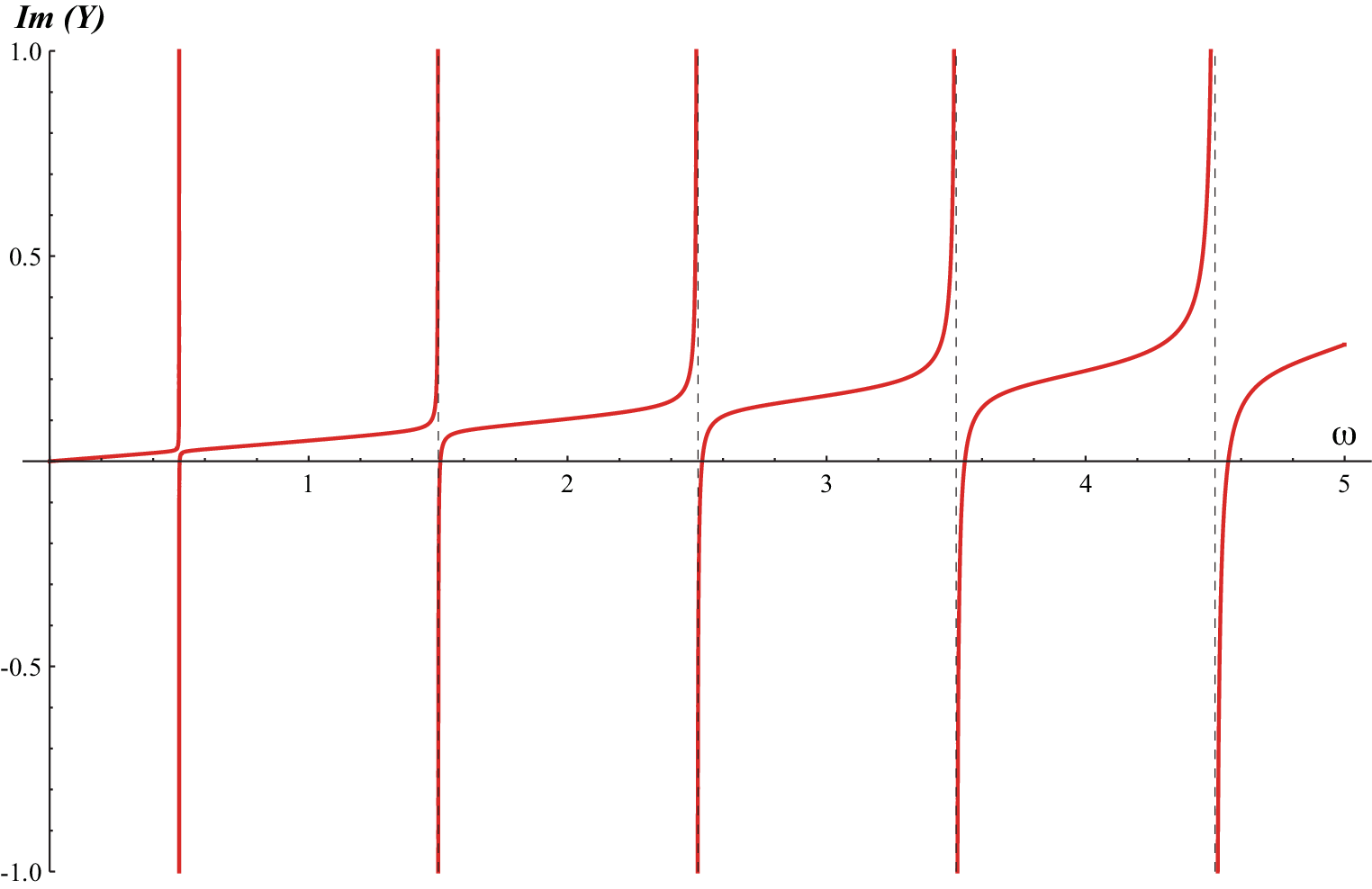}
   \caption{A quarter-wavelength-resonator-like admittance constructed by adjusting the coupling coefficients. The red solid line is the imaginary part of \eqref{eq:adm1} for $\mu=1$, $\omega_n=0.5\times (2n-1),~(n=1,\cdots,20)$, and $I_{n1}=0.015\times n,~(n=1,\cdots,20)$. The thin black dashed lines indicate the poles. Compare with Fig.~\ref{fig:adm_qwr}.}
   \label{fig:y11}
\end{figure}

\begin{figure}[htb]
   \centering
   \includegraphics*[width=100mm]{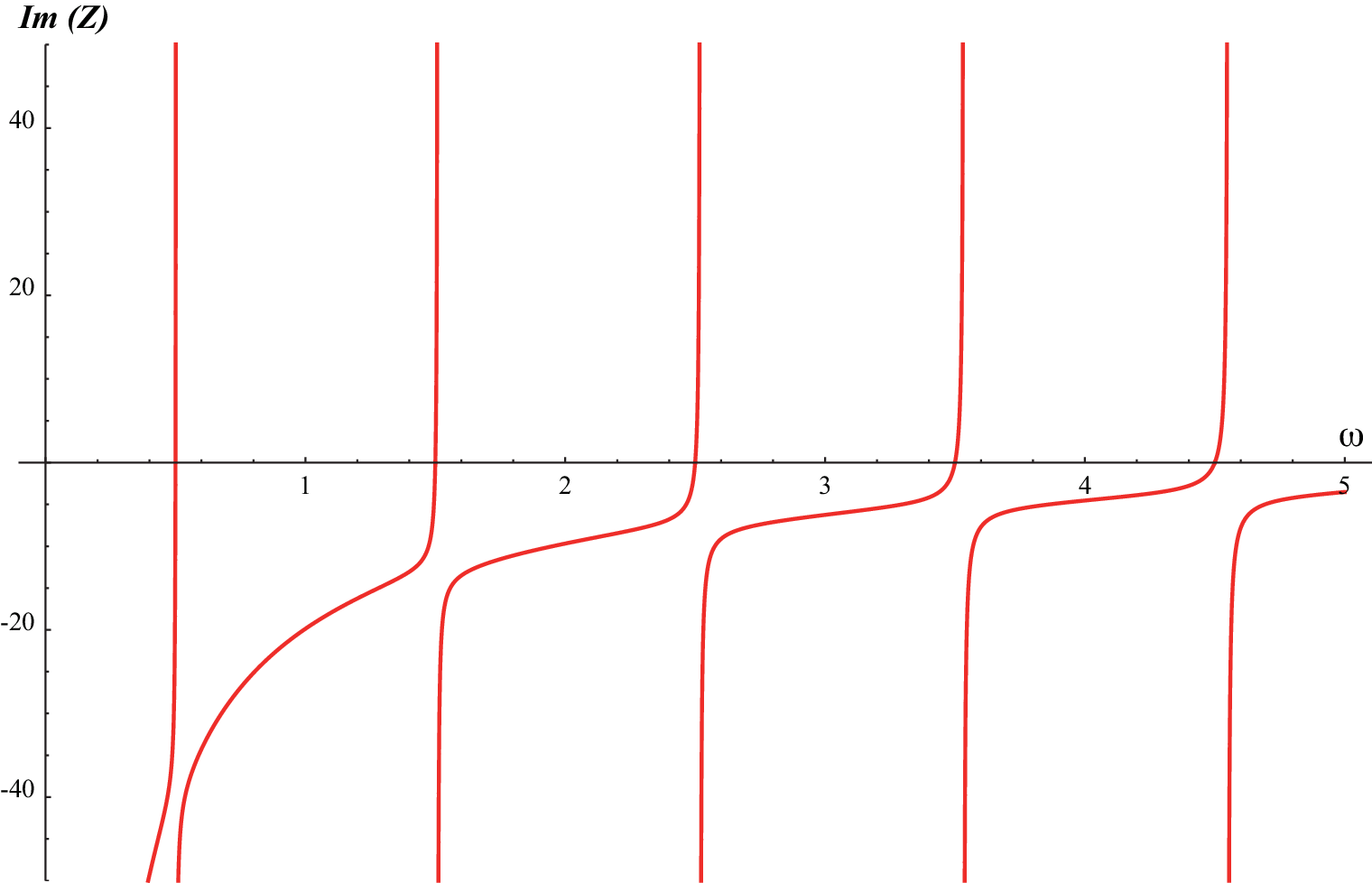}
   \caption{Imaginary part of the reciprocal of the admittance in Fig.~\ref{fig:y11}, namely, the impedance. Compare with Fig.~\ref{fig:imp_qwr}.}
   \label{fig:z11}
\end{figure}

Figure~\ref{fig:y11} shows the imaginary part of an input admittance calculated by choosing $I_{n1}$ proportional to $n$ to represent this behavior.
Figure~\ref{fig:z11} shows the imaginary part of the corresponding input impedance.
Note that their shapes resemble those in Figs.~\ref{fig:adm_qwr} and \ref{fig:imp_qwr}, respectively.
Thus, choosing $I_{n1}$ to increase with $n$ allows us to represent the admittance and impedance associated with capacitive coupling.


\subsubsection{Pillbox cavity with single port}

\begin{figure}[htb]
   \centering
   \includegraphics*[width=100mm]{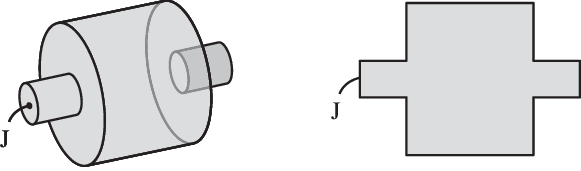}
   \caption{Schematic illustration of a pillbox cavity. The left end surface is taken as the junction surface.}
   \label{fig:pillbox}
\end{figure}

As a third example, let us consider qualitatively the input admittance of a pillbox cavity.
The junction surface $J$ is defined as shown in Fig.~\ref{fig:pillbox}.

We consider the TM$_{01{\mbox n}}$ modes as the eigenvector fields of the cavity.
For the eigenvector field on the cross section that couples to these modes, we choose the TM$_{01}$ mode.
Regarding $I_{n1}$ in this case, we expect, as in the preceding subsection, that the magnetic field penetrating the junction becomes stronger as $n$ increases, and increases rapidly above the cutoff frequency.
Thus, $I_{n1}$ is expected to show a trend similar to that considered in the preceding subsection, and the admittance and impedance should resemble those in Figs.~\ref{fig:y11} and \ref{fig:z11}.
As the ``neck'' leading to the junction becomes longer, the coupling becomes weaker and $I_{n1}$ becomes very small. Consequently, the zeros and poles approach one another.

\subsubsection{Very thin resonator}

\begin{figure}[htb]
   \centering
   \includegraphics*[width=20mm]{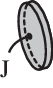}
   \caption{A very ``thin'' cavity.}
   \label{fig:thin}
\end{figure}

Finally, let us consider the input admittance of a very ``thin'' cavity, as shown in Fig.~\ref{fig:thin}.

Since this cavity is simply connected, it has no DC resonant electromagnetic fields.
Moreover, with very few exceptions, its resonance frequencies become extremely high.
The eigenvector fields $\{\bm{H}_n\}$ in $V$ and $\{\bm{E}_{tn}\}$ on $J$ are normalized according to Eqs.~\eqref{eq:hnorm} and \eqref{eq:etnorm}, respectively.
Thus, as the cavity becomes thinner, the magnitude of $\bm{E}_{tn}$ remains unchanged, while that of $\bm{H}_n$ increases.
Consequently, in the limit of vanishing thickness, Eq.~\eqref{eq:ink} gives $|I_{nk}|\rightarrow \infty$.
Substituting this result into Eq.~\eqref{eq:adm1}, we find that $|Y_{11}(j\omega)|\rightarrow \infty$ (or, more generally, $|Y_{lk}(j\omega)|\rightarrow \infty$) for almost all frequencies as the thickness approaches zero.
This means that the input impedance approaches zero, so the boundary behaves as a short-circuited end.
In this limit, the cavity becomes a conducting wall, in agreement with our intuitive understanding.

\section{Matrix equation for coupled modes}
\label{sec:matrix}
In this section, we examine how to determine the resonant frequencies and electromagnetic fields of coupled cavity resonators.
\subsection{Two-resonators case}

\begin{figure}[htb]
   \centering
   \includegraphics*[width=80mm]{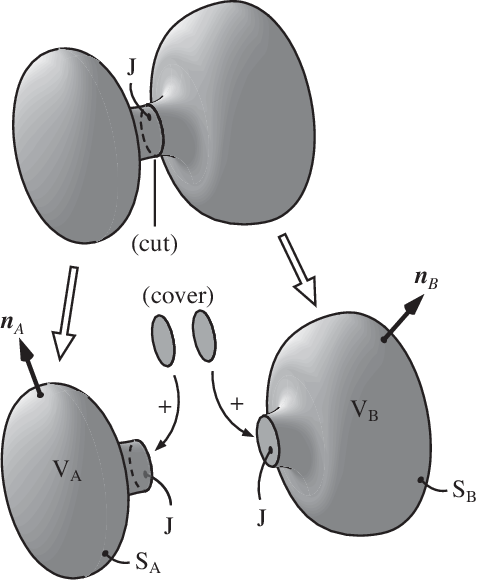}
   \caption{A system of two coupled cavity resonators. Cavities A and B are defined by closing the junction surface $J$ with a perfect conductor.}
   \label{fig:dual}
\end{figure}

First, consider two cavity resonators coupled through a common cross section $J$, as shown in Fig.~\ref{fig:dual}.
If the cross section $J$ is closed by a perfect conductor, two separate cavity resonators, A and B, are formed.
Let S$_A$ and S$_B$ denote the outer walls of cavities A and B excluding $J$, and V$_A$ and V$_B$ their respective interiors.
Let $\bm{n}_A$ and $\bm{n}_B$ be their respective outward unit normal vectors. On $J$, these vectors point in opposite directions.

We seek to express the resonant frequencies and electromagnetic fields of the coupled system in terms of the resonant frequencies and eigenfields of the separate cavities A and B.
The resonant electromagnetic fields of the coupled system must satisfy Maxwell's equations in V$_{\rm A} +$V$_{\rm B}$,
\begin{eqnarray}
\nabla\times\bm{H}&=&j\omega\epsilon\bm{E}\,,\\
\nabla\times\bm{E}&=&-j\omega\mu\bm{H}\,,
\end{eqnarray}
together with the boundary condition on S$_{\rm A} +$S$_{\rm B}$,
\begin{eqnarray}
\bm{n}\times\bm{E}=0\,.
\end{eqnarray}
For $\omega\neq0$, this boundary condition and Maxwell's equations imply that, on S$_{\rm A} +$S$_{\rm B}$,
\begin{eqnarray}
\bm{n}\cdot\bm{H}=0\,.
\end{eqnarray}
Moreover, for $\omega\neq0$, Maxwell's equations imply that
\begin{eqnarray}
\nabla\cdot\bm{E}&=&0\,,\\
\nabla\cdot\bm{H}&=&0\,.
\end{eqnarray}

The method of Sec.~\ref{sub:adm} can be applied to this problem.
The key point is that cavities A and B share the same cross section $J$.
Suppose an electric field represented by \{$V_k$\} is generated on the junction surface $J$.
An electromagnetic field is then generated inside cavity A, inducing on $J$ a magnetic field represented by
\begin{eqnarray}
\label{eq:ila}
 I_{l}^{(A)}=\sum_k Y_{lk}^{(A)}(j\omega)\cdot V_k\,.
\end{eqnarray}
The input admittance of cavity A is defined by
\begin{eqnarray}
 Y_{lk}^{(A)}(j\omega)\coloneqq \sum_n\frac{I_{nl}^{(A)}I_{nk}^{(A)}}
{j \omega_n^{(A)} \mu \left(\frac{\omega}{\omega_n^{(A)}}-\frac{\omega_n^{(A)}}{\omega}\right)}
+\sum_\nu \frac{I_{\nu l}^{(A)}I_{\nu k}^{(A)}}{j \omega \mu}\,.
\end{eqnarray}
The quantities on the right-hand side are calculated from the eigenvector fields in V$_A$ and on $J$.
Likewise, the electric field represented by \{$V_k$\} generates an electromagnetic field inside cavity B, inducing on $J$ a magnetic field represented by
\begin{eqnarray}
\label{eq:ilb}
 I_{l}^{(B)}=\sum_k Y_{lk}^{(B)}(j\omega)\cdot V_k\,.
\end{eqnarray}
The input admittance of cavity B is defined by
\begin{eqnarray}
 Y_{lk}^{(B)}(j\omega)\coloneqq \sum_n\frac{I_{nl}^{(B)}I_{nk}^{(B)}}
{j \omega_n^{(B)} \mu \left(\frac{\omega}{\omega_n^{(B)}}-\frac{\omega_n^{(B)}}{\omega}\right)}
+\sum_\nu \frac{I_{\nu l}^{(B)}I_{\nu k}^{(B)}}{j \omega \mu}\,.
\end{eqnarray}
The quantities on the right-hand side are calculated from the eigenvector fields in V$_B$ and on $J$.

What, then, characterizes a coupled resonant state?
Physically, {\bf the magnetic fields induced on the two sides of the junction surface by the electric field in a resonant state must match}.
In other words, the two magnetic fields calculated above must be continuous across the junction surface.
Taking account of the opposite orientations of the outward normals, this requires, for every $l$,
\begin{eqnarray}
  I_{l}^{(A)}+I_{l}^{(B)}=0\,.
\end{eqnarray}

\begin{figure}[htb]
   \centering
   \includegraphics*[width=100mm]{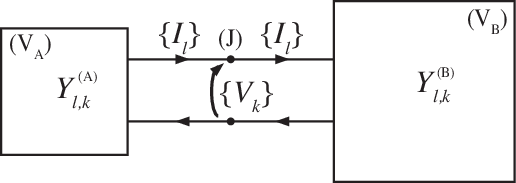}
   \caption{Schematic illustration of a resonant state in a system of two coupled cavities.}
   \label{fig:adm_dual}
\end{figure}

Expressing this condition in terms of the admittances $Y_{lk}$ and the electric-field coefficients $V_k$ on the junction surface gives the homogeneous linear equation
\begin{eqnarray}
\label{eq:homv}
\sum_k \left[Y_{lk}^{(A)}(j\omega)+Y_{lk}^{(B)}(j\omega)\right] V_k=0\,.
\end{eqnarray}
This is one of the main results of this note.
Figure~\ref{fig:adm_dual} illustrates this coupled-resonance condition schematically.

Since $\bm{n}_B=-\bm{n}_A$ on the junction surface, Eq.~\eqref{eq:homv} requires that the tangential magnetic-field components on sides A and B be equal. These components are given, respectively, by
\begin{eqnarray}
\bm{H_{\tiny{/ \! /}}}=\sum_{l}\left(-\bm{n}_A \times\bm{E}_{tl}\right)
\sum_k\left[
\sum_n\frac{I_{nl}^{(A)}I_{nk}^{(A)}}
{j \omega_n^{(A)} \mu \left(\frac{\omega}{\omega_n^{(A)}}-\frac{\omega_n^{(A)}}{\omega}\right)}
+\sum_\nu \frac{I_{\nu l}^{(A)}I_{\nu k}^{(A)}}{j \omega \mu}
\right]V_k\,,
\end{eqnarray}
and
\begin{eqnarray}
\bm{H_{\tiny{/ \! /}}}=\sum_{l}\left(-\bm{n}_B \times\bm{E}_{tl}\right)
\sum_k\left[
\sum_n\frac{I_{nl}^{(B)}I_{nk}^{(B)}}
{j \omega_n^{(B)} \mu \left(\frac{\omega}{\omega_n^{(B)}}-\frac{\omega_n^{(B)}}{\omega}\right)}
+\sum_\nu \frac{I_{\nu l}^{(B)}I_{\nu k}^{(B)}}{j \omega \mu}
\right]V_k\,.
\end{eqnarray}

This expresses Eq.~\eqref{eq:homv} in terms of the electromagnetic fields.

In general, when considering resonant electromagnetic fields near a given frequency, we expect that restricting the expansion on the junction surface to a finite number of eigenvector fields can still give a sufficiently good approximation.
If the number of eigenvector fields on the junction surface is limited to $m$, Eq.~\eqref{eq:homv} takes the finite-dimensional matrix form
\begin{eqnarray}
\label{eq:mat_2cav}
\left(
\begin{array}{cccc}
\mathcal{Y}_{1,1} & \mathcal{Y}_{1,2} & \cdots & \mathcal{Y}_{1,m} \\
\mathcal{Y}_{2,1} & \mathcal{Y}_{2,2} & \cdots & \mathcal{Y}_{2,m} \\
\vdots & \vdots & \ddots & \vdots\\
\mathcal{Y}_{m,1} & \mathcal{Y}_{m,2} & \cdots & \mathcal{Y}_{m,m}
\end{array} 
\right)
\left(\begin{array}{c} V_1\\ V_2 \\ \vdots \\ V_{m}
\end{array}\right)=
\left(\begin{array}{c} 
0 \\ 0 \\ \vdots \\ 0
\end{array}\right)\,,
\end{eqnarray}
or
\begin{eqnarray}
\mathcal{Y}(j\omega)\cdot \bm{V} = 0\,.
\end{eqnarray}
Here, each $\mathcal{Y}_{l,k}$ is a function of $j\omega$, explicitly given by
\begin{eqnarray}
\mathcal{Y}_{l,k}(j\omega)&\coloneqq&
\sum_n\frac{I_{nl}^{(A)}I_{nk}^{(A)}}
{j \omega_n^{(A)} \mu \left(\frac{\omega}{\omega_n^{(A)}}-\frac{\omega_n^{(A)}}{\omega}\right)}
+\sum_\nu \frac{I_{\nu l}^{(A)}I_{\nu k}^{(A)}}{j \omega \mu}\nonumber \\
&+&\sum_n\frac{I_{nl}^{(B)}I_{nk}^{(B)}}
{j \omega_n^{(B)} \mu \left(\frac{\omega}{\omega_n^{(B)}}-\frac{\omega_n^{(B)}}{\omega}\right)}
+\sum_\nu \frac{I_{\nu l}^{(B)}I_{\nu k}^{(B)}}{j \omega \mu}\,.
\end{eqnarray}

The coupled resonant frequencies $\omega=\Omega_r$ are therefore obtained from
\begin{eqnarray}
\det\left[\mathcal{Y} (j \omega)\right] =0\,.
\end{eqnarray}
Here, $r$ labels the solutions.
Furthermore, the eigenvector $\{V_{r,k}\}$ associated with the zero eigenvalue specifies the electric field at the junction surface in the coupled resonant state.
The electromagnetic field of the coupled resonant state is then determined by substituting $\{V_{r,k}\}$ into Eq.~\eqref{eq:exp}.

Note that the matrix $\mathcal{Y}$ is symmetric.
The eigenvector fields introduced in Secs.~\ref{sub:eigen3} and \ref{sub:eigen2} also have an arbitrary overall sign ($\pm 1$).
Since each element of $\mathcal{Y}$ is quadratic in the expansion coefficients, the matrix does not depend on this sign choice.
Both properties also hold for the multi-cavity, multi-aperture systems considered later.

\subsection{Properties of coupled two-resonator systems}

To gain an intuitive understanding of coupled resonant states, let us consider the case $m=1$ in Eq.~\eqref{eq:mat_2cav}, restricting the eigenvector fields on the cross section $J$ to a single mode.
We can then use the properties of the input admittance discussed in Sec.~\ref{sec:single}.
For $m=1$, Eq.~\eqref{eq:mat_2cav} can be written as
\begin{eqnarray}
\label{eq:coupled_dual}
\left[Y_{11}^{(A)}(j\omega)+Y_{11}^{(B)}(j\omega)\right]V_1=0\,.
\end{eqnarray}
Thus, a solution with $V_1\neq 0$ exists only when $\omega$ satisfies
\begin{eqnarray}
\label{eq:coupled_freq_dual}
Y_{11}^{(A)}(j\omega)+Y_{11}^{(B)}(j\omega)=0\,.
\end{eqnarray}
This is the condition for a coupled resonant state.

\begin{figure}[htb]
   \centering
   \includegraphics*[width=100mm]{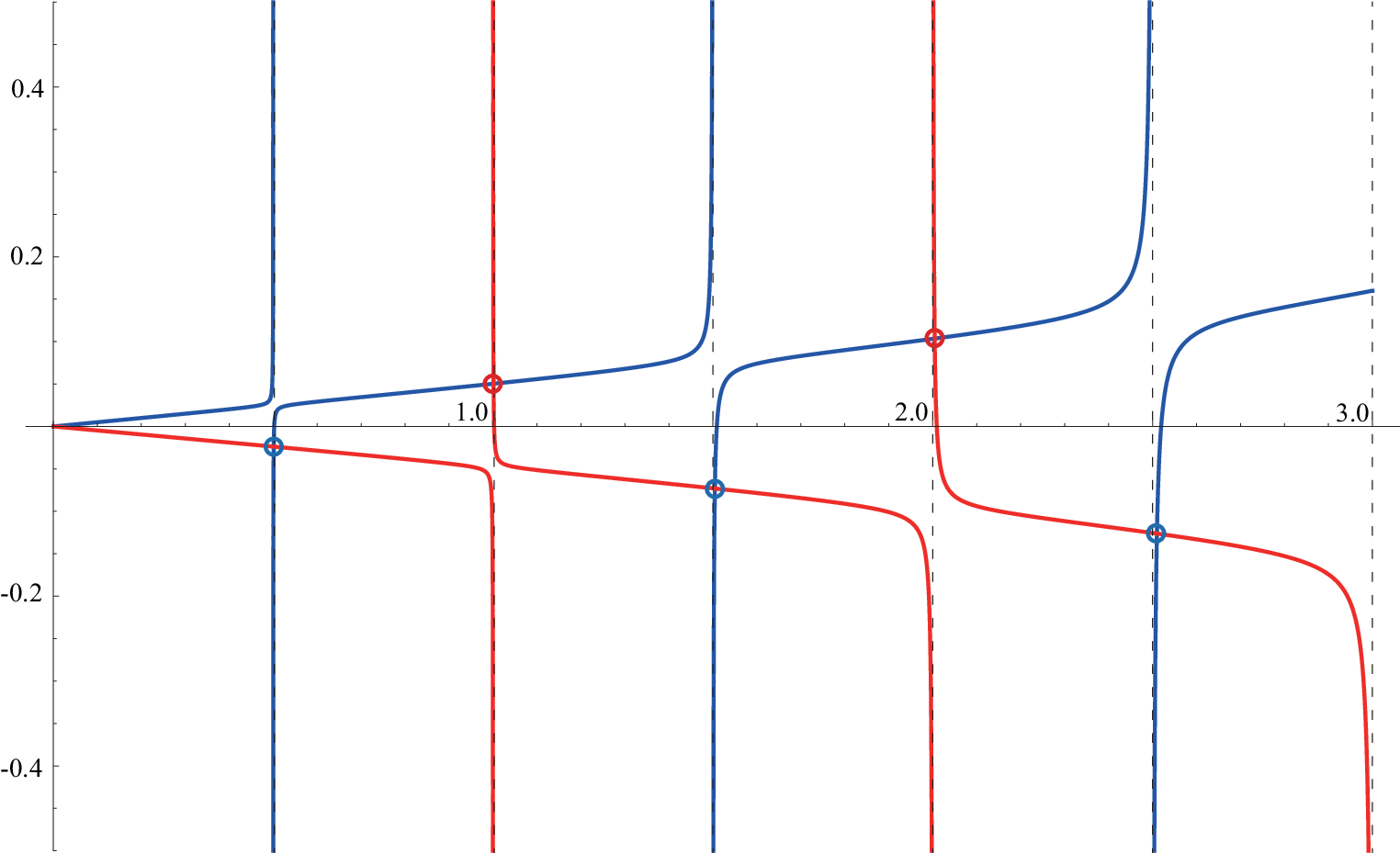}
   \caption{Determination of the resonant frequencies of weakly coupled cavities from their admittances.
The blue curve is the same as in Fig.~\ref{fig:adm_qwr}. The red curve is obtained from Eq.~\eqref{eq:adm1} by setting $\mu=1$, $\omega_n=1.0\times (2n-1),~(n=1,\cdots,20)$ and $I_{n1}=0.015\times n,~(n=1,\cdots,20)$, and then multiplying the result by $-1$.
The intersections of the blue and red curves therefore give the solutions of Eq.~\eqref{eq:coupled_freq_dual}.
The poles of the admittances are indicated by dashed lines.
The intersections occur near the poles of either admittance (the original resonant frequencies of the cavities), at frequencies slightly above the poles.}
   \label{fig:coupled_weak}
\end{figure}

\begin{figure}[htb]
   \centering
   \includegraphics*[width=100mm]{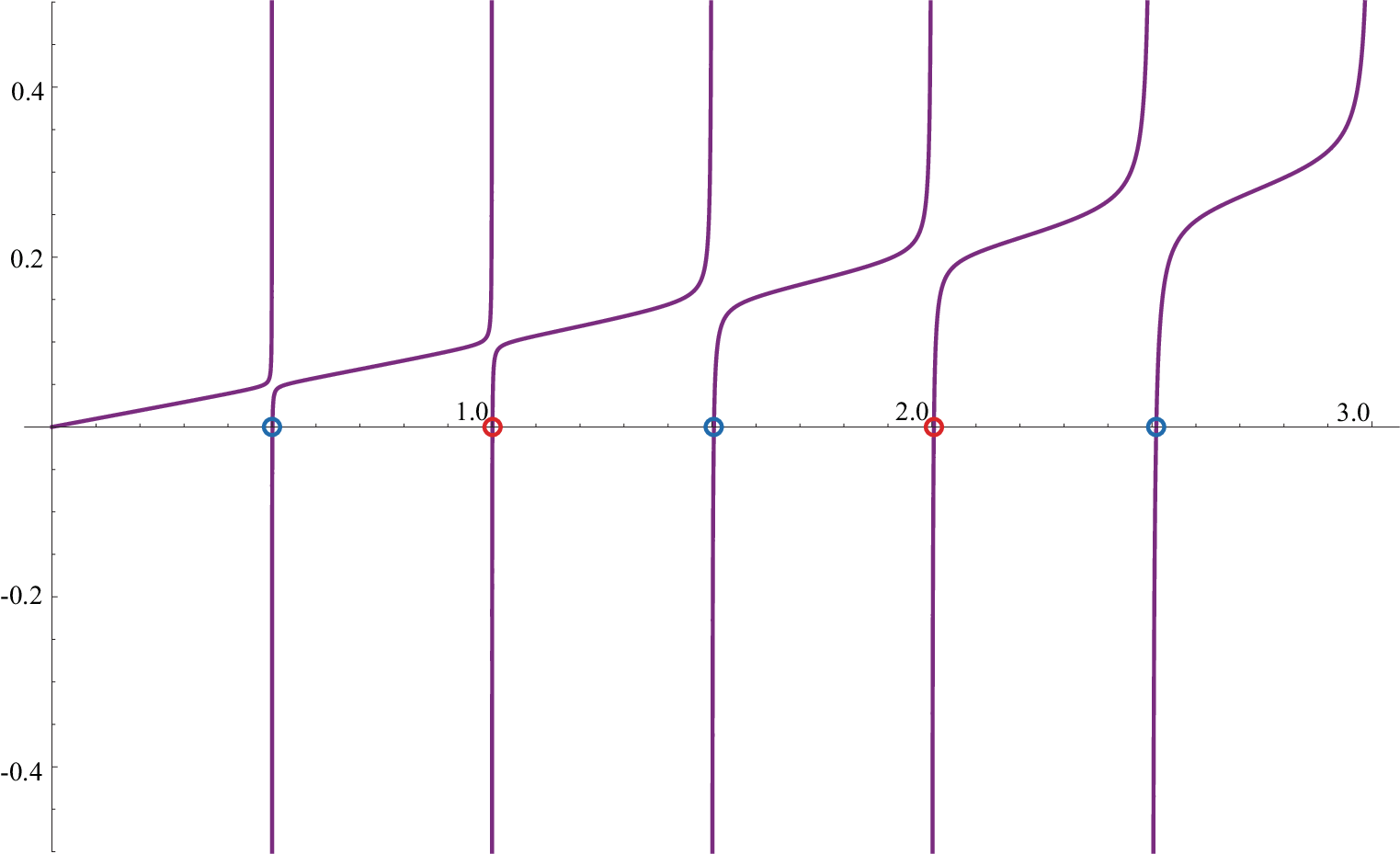}
   \caption{Sum of the input admittances of the two weakly coupled cavities, i.e., the left-hand side of Eq.~\eqref{eq:coupled_freq_dual}. The zeros give the resonant frequencies of the coupled cavities.}
   \label{fig:coupled_total_weak}
\end{figure}

As shown in Figs.~\ref{fig:adm_qwr} and \ref{fig:y11} in Sec.~\ref{sec:single}, when the coupling is weak, the admittance can take a wide range of values near its poles, which correspond to the original resonant frequencies.
Thus, near a pole, one admittance can cancel the other in Eq.~\eqref{eq:coupled_freq_dual}, and the coupled resonant frequencies appear very close to the original resonant frequencies.
Figures~\ref{fig:coupled_weak} and \ref{fig:coupled_total_weak} illustrate how the coupled resonant frequencies are determined for weak coupling.
The intersections of the two curves in Fig.~\ref{fig:coupled_weak} are the solutions of Eq.~\eqref{eq:coupled_freq_dual}.

Let $\Omega_r$ be one of the coupled resonant frequencies determined above.
At this frequency, the ratio of the time-averaged electromagnetic energies stored in the two cavities is given by \cite{bib:mit48}
\begin{eqnarray}
\frac{U^{(A)}}{U^{(B)}}
=\frac{dY_{11}^{(A)}/d\omega | _{\omega=\Omega_r}}{dY_{11}^{(B)}/d\omega | _{\omega=\Omega_r}}\,.
\end{eqnarray}
For weak coupling, as illustrated in Fig.~\ref{fig:adm_qwr}, the slope of one admittance at an intersection is much larger than that of the other. Consequently, the ratio of the energies stored in the two cavities becomes very large.
Thus, although the cavities are coupled, one cavity is scarcely affected by the other. This is the physical meaning of weak coupling.
A closer look at Fig.~\ref{fig:coupled_weak} also shows that the intersections occur at frequencies above the original resonant frequencies (the poles of the admittance).
Thus, the coupled resonant frequencies are higher than those obtained when the junction surface $J$ is covered by a conductor.
For the quarter-wavelength resonator considered in Sec.~\ref{sec:single}, this corresponds to a reduction in the capacitance at the tip when the cavity is coupled, compared with the case in which the junction is covered. The resonance frequency therefore shifts upward.

\begin{figure}[htb]
   \centering
   \includegraphics*[width=100mm]{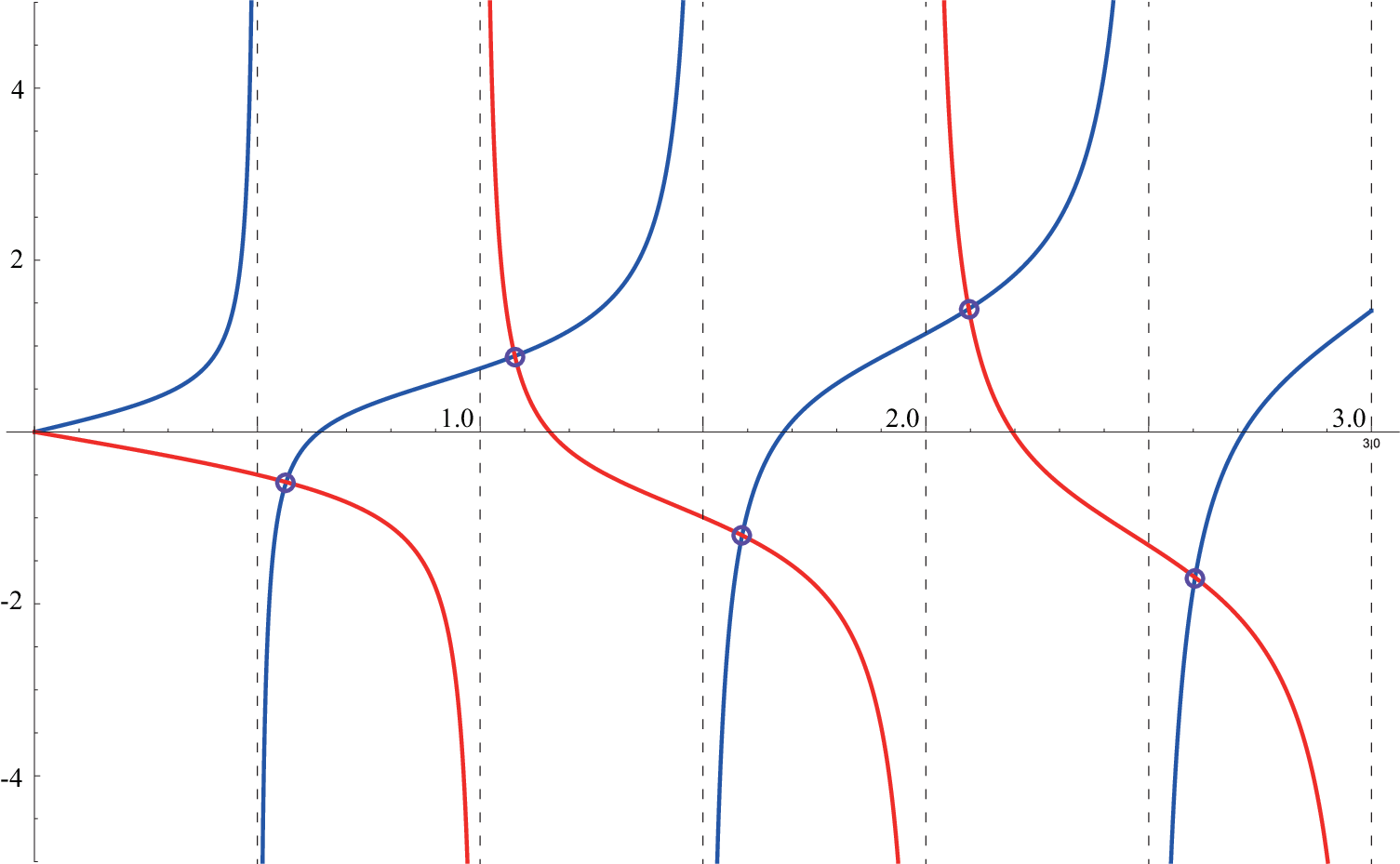}
   \caption{Resonant frequencies of two strongly coupled cavities. The curves are obtained by setting all $I_{n1}$ to 0.5 in the calculation for Fig.~\ref{fig:coupled_weak}. The poles of the admittances are indicated by dashed lines. Note that the intersections are separated from the poles (the original resonant frequencies of the cavities), and the slopes of the two curves at the intersections are comparable.}
   \label{fig:coupled_strong}
\end{figure}

\begin{figure}[htb]
   \centering
   \includegraphics*[width=100mm]{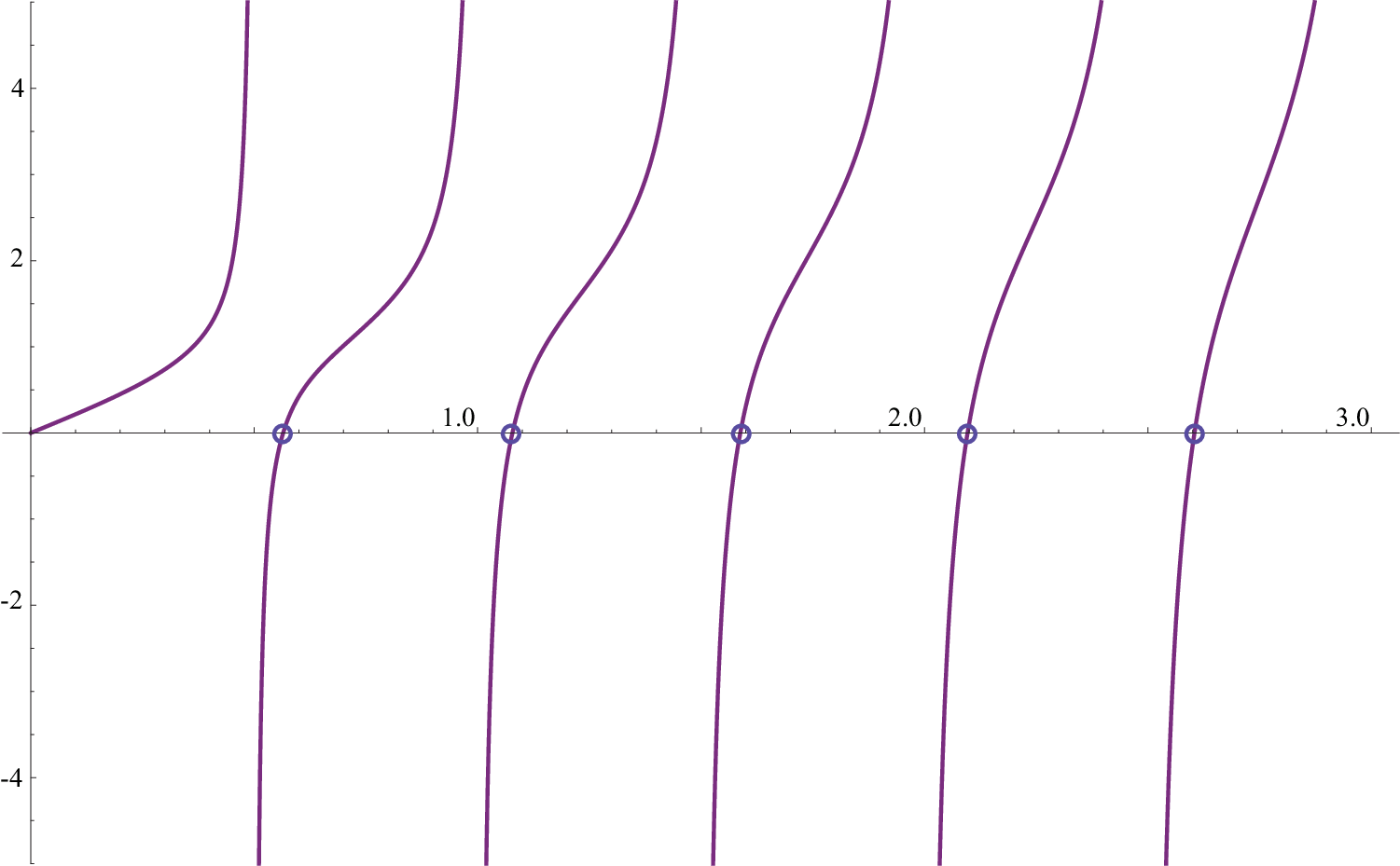}
   \caption{Sum of the input admittances of two strongly coupled cavities, i.e., the left-hand side of Eq.~\eqref{eq:coupled_freq_dual}.}
   \label{fig:coupled_total_strong}
\end{figure}

The situation is different, however, when the input admittance $Y_{11}$ of the original cavity has a form such as that shown in Fig.~\ref{fig:cot}; this corresponds to strong coupling between the resonant modes of the original cavity and the eigenvector field on the cross section $J$.
Figures~\ref{fig:coupled_strong} and \ref{fig:coupled_total_strong} show how the coupled resonant frequencies are determined in this case.
As shown in Fig.~\ref{fig:coupled_strong}, when the coupling is strong, the intersections move away from the original resonant frequencies (the poles).
The slopes of the admittances at the intersections also become comparable, and electromagnetic energy is stored in both cavities.
In other words, a new resonant electromagnetic field is formed, with the two cavities influencing each other.

So far, we have considered resonant states of two coupled cavities, restricting the modes on the junction surface to a single mode.
The frequency splitting that occurs when the two coupled cavities are identical is discussed in Appendix~\ref{ap:split}.

\subsection{General case}

\begin{figure}[htb]
   \centering
   \includegraphics*[width=140mm]{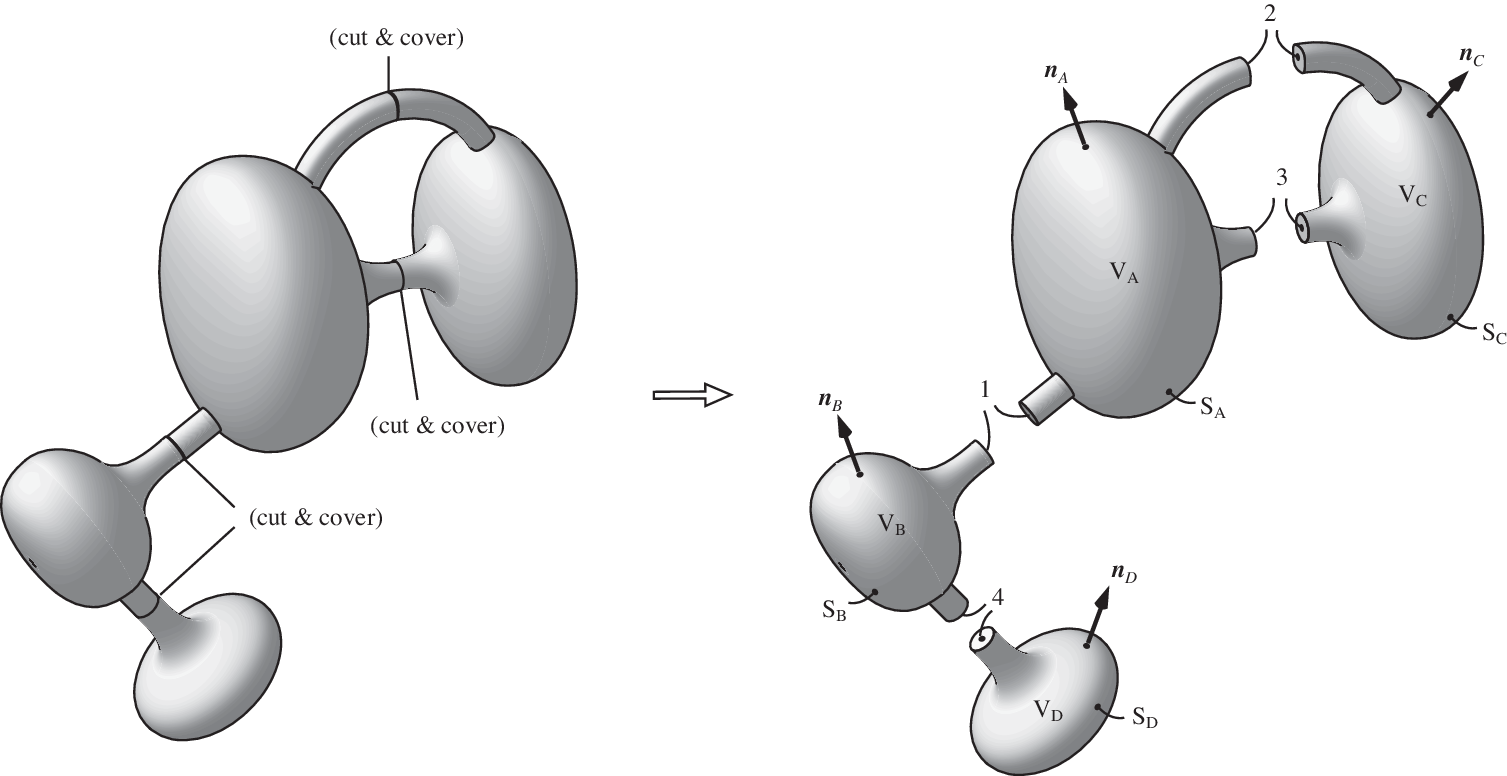}
   \caption{A coupled-cavity system consisting of multiple resonators. The individual cavity resonators are defined by separating the system at the junction surfaces and covering these surfaces with metal.}
   \label{fig:quad}
\end{figure}

We now extend the preceding discussion to the multi-cavity, multi-aperture system shown in Fig.~\ref{fig:quad}.
Although more indices are required, the essential idea remains the same.
We first assume that the numbers of cavities and junction surfaces are finite.
Junction surface 1 is shared by cavities A and B, while junction surfaces 2 and 3 are shared by cavities A and C, and so on.

Consider the magnetic field $\bm{H}$ inside cavity A. As in Eqs.~\eqref{eq:var} and \eqref{eq:varnu},
\begin{eqnarray}
\label{eq:varmul}
\int_{\tiny{\mbox V}_A}\bm{H}_n^{(A)}\cdot\bm{H}dv
&=&-\frac{1}{\left(\frac{\omega_n^{(A)}}{j \omega}+\frac{j \omega}{\omega_n^{(A)}}\right)}\cdot
\frac{1}{\mu\omega_n^{(A)}}
\sum_{J}\int_{\tiny{\mbox S_J}}\bm{n}_A\times\bm{E}\cdot\bm{H}_n^{(A)} dS\,,\\
\label{eq:varnumul}
\int_{\tiny{\mbox V}_A}\bm{H}_\nu^{(A)}\cdot\bm{H}dv
&=&-\frac{1}{j\omega\mu}
\sum_{J}\int_{\tiny{\mbox S_J}}\bm{n}_A\times\bm{E}\cdot\bm{H}_\nu^{(A)} dS\,,
\end{eqnarray}
where the index $J$ runs over all junction surfaces connected to cavity A.

As in the preceding subsection, let $\left\{\bm{E}_{tk}^{(J)}\right\}$ denote the complete set on junction surface $J$. We expand the tangential electric field $\bm{E}_{\tiny{/\! /}}^{(J)}$ on this surface as
\begin{eqnarray}
\bm{E}_{\tiny{/\! /}}^{(J)}\coloneqq\sum_l V_l^{(J)}\bm{E}_{tl}^{(J)}\,.
\end{eqnarray}
Similarly, we write
\begin{eqnarray}
\label{eq:expnj}
\bm{n}_A^{(J)} \times \bm{H}_n^{(A)}=\sum_k I_{nk}^{(AJ)}\bm{E}_{tk}^{(J)}\,,\\
\label{eq:expnuj}
\bm{n}_A^{(J)} \times \bm{H}_\nu^{(A)}=\sum_k I_{\nu k}^{(AJ)}\bm{E}_{tk}^{(J)}\,.
\end{eqnarray}
The surface integrals on the right-hand sides of Eqs.~\eqref{eq:varmul} and \eqref{eq:varnumul} can then be evaluated to give
\begin{eqnarray}
\int_{\tiny{\mbox V}_A}\bm{H}_n^{(A)}\cdot\bm{H}dv
&=&\frac{1}{\left(\frac{\omega_n^{(A)}}{j \omega}+\frac{j \omega}{\omega_n^{(A)}}\right)}\cdot
\frac{1}{\mu\omega_n^{(A)}}
\sum_J\sum_k V_k^{(J)} I_{nk}^{(AJ)}\,,\\
\int_{\tiny{\mbox V}_A}\bm{H}_\nu^{(A)}\cdot\bm{H}dv
&=&\frac{1}{j\omega\mu}
\sum_J\sum_k V_k^{(J)} I_{\nu k}^{(AJ)}\,.
\end{eqnarray}
Thus, when the inputs $\left\{V_k^{(J)}\right\}$ are applied at the boundary surfaces of cavity A, the resulting magnetic field $\bm{H}$ inside A can be expanded in terms of its eigenvector fields as
\begin{eqnarray}
\label{eq:expmul}
\bm{H}=
\sum_n{
\frac{1}{j \left(\frac{\omega}{\omega_n^{(A)}}-\frac{\omega_n^{(A)}}{\omega}\right)}\cdot
\frac{1}{\mu\omega_n^{(A)}}
\sum_J\sum_k V_k^{(J)} I_{nk}^{(AJ)}\bm{H}_n^{(A)}
}+\sum_\nu{
\frac{1}{j \mu\omega}
\sum_J\sum_k V_k^{(J)} I_{\nu k}^{(AJ)}
\bm{H}_\nu^{(A)}
}\,.
\end{eqnarray}
This is the counterpart of Eq.~\eqref{eq:exp}.

Now suppose that junction surface $K$ is shared by cavities A and B.
As in the discussion leading to Eq.~\eqref{eq:kekka1}, we assume that expansion \eqref{eq:expmul} can be applied to the component of the magnetic field parallel to this junction surface.
Substituting expansions \eqref{eq:expnj} and \eqref{eq:expnuj} on junction surface $K$, we obtain
\begin{eqnarray}
\bm{H_{\tiny{/\! /}}}&=&
\sum_n{\frac{1}{j \left(\frac{\omega}{\omega_n^{(A)}}-\frac{\omega_n^{(A)}}{\omega}\right)}\cdot
\frac{1}{\mu\omega_n^{(A)}}
\sum_J\sum_k V_k^{(J)} I_{nk}^{(AJ)}\sum_{l}I_{nl}^{(AK)}
\left(-\bm{n}_A^{(K)} \times\bm{E}_{tl}^{(K)}\right)
}\nonumber \\
&+&\sum_\nu{\frac{1}{j \mu\omega}
\sum_J\sum_k V_k^{(J)} I_{\nu k}^{(AJ)}
\sum_{l}I_{\nu l}^{(AK)}\left(-\bm{n}_A^{(K)} \times\bm{E}_{tl}^{(K)}\right)
}\,.
\end{eqnarray}
Rearranging this expression, we can write the component of the magnetic field parallel to junction surface $K$ as
\begin{eqnarray}
\label{eq:hexpka}
\bm{H_{\tiny{/\! /}}}=\sum_{l}\left(-\bm{n}_A^{(K)} \times\bm{E}_{tl}^{(K)}\right)
\sum_J\sum_k\left[
\sum_n\frac{I_{nl}^{(AK)}I_{nk}^{(AJ)}}
{j \omega_n^{(A)} \mu \left(\frac{\omega}{\omega_n^{(A)}}-\frac{\omega_n^{(A)}}{\omega}\right)}
+\sum_\nu \frac{I_{\nu l}^{(AK)}I_{\nu k}^{(AJ)}}{j \omega \mu}
\right]V_k^{(J)}\,.
\end{eqnarray}
This expression gives the magnetic field on junction surface $K$ generated through Maxwell's equations by the inputs $\left\{V_k^{(J)}\right\}$ applied at the boundary surfaces of cavity A.
Applying the same procedure to cavity B gives
\begin{eqnarray}
\label{eq:hexpkb}
\bm{H_{\tiny{/\! /}}}=\sum_{l}\left(-\bm{n}_B^{(K)} \times\bm{E}_{tl}^{(K)}\right)
\sum_L\sum_k\left[
\sum_n\frac{I_{nl}^{(BK)}I_{nk}^{(BL)}}
{j \omega_n^{(B)} \mu \left(\frac{\omega}{\omega_n^{(B)}}-\frac{\omega_n^{(B)}}{\omega}\right)}
+\sum_\nu \frac{I_{\nu l}^{(BK)}I_{\nu k}^{(BL)}}{j \omega \mu}
\right]V_k^{(L)}\,.
\end{eqnarray}
Here $L$ runs over all junction surfaces connected to cavity B, including $K$.

As discussed in the preceding subsection, in a coupled resonant state the magnetic fields induced by the electric fields at the junction surfaces must match across every junction surface.
For example, at junction surface $K$, we equate Eqs.~\eqref{eq:hexpka} and \eqref{eq:hexpkb}, using $\bm{n}_A^{(K)}=-\bm{n}_B^{(K)}$.
This gives the following homogeneous equation for $V_k^{(J)}$ and $V_k^{(L)}$:
\begin{eqnarray}
&&\sum_J\sum_k\left[
\sum_n\frac{I_{nl}^{(AK)}I_{nk}^{(AJ)}}
{j \omega_n^{(A)} \mu \left(\frac{\omega}{\omega_n^{(A)}}-\frac{\omega_n^{(A)}}{\omega}\right)}
+\sum_\nu \frac{I_{\nu l}^{(AK)}I_{\nu k}^{(AJ)}}{j \omega \mu}
\right]V_k^{(J)}\nonumber \\
&+&\sum_L\sum_k\left[
\sum_n\frac{I_{nl}^{(BK)}I_{nk}^{(BL)}}
{j \omega_n^{(B)} \mu \left(\frac{\omega}{\omega_n^{(B)}}-\frac{\omega_n^{(B)}}{\omega}\right)}
+\sum_\nu \frac{I_{\nu l}^{(BK)}I_{\nu k}^{(BL)}}{j \omega \mu}
\right]V_k^{(L)}\nonumber \\
&=&0\,.
\end{eqnarray}

\begin{figure}[htb]
   \centering
   \includegraphics*[width=150mm]{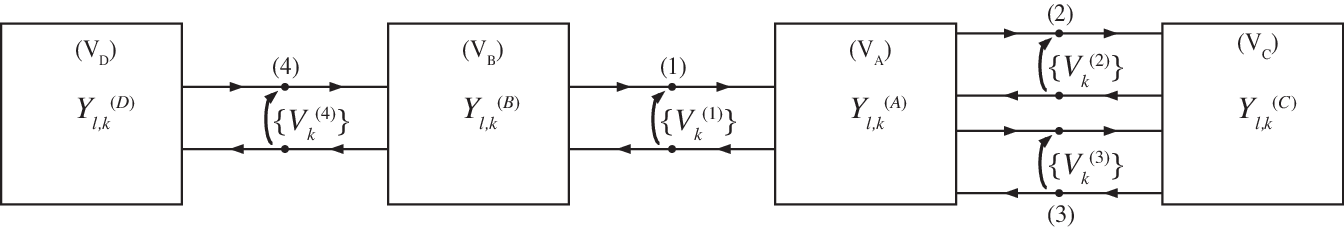}
   \caption{Illustration of the resonant state of coupled cavities in the multi-cavity, multi-aperture case. This corresponds to Fig.~\ref{fig:adm_dual}.}
   \label{fig:adm_multi}
\end{figure}

Applying this procedure to all junction surfaces, we obtain the following linear homogeneous equations describing the resonant state:
\begin{eqnarray}
\label{eq:homv_multi}
\sum_J\sum_k \left[\mathcal{Y}_{l,k}^{(K,J)}(j\omega)\right] V_k^{(J)}=0\,.
\end{eqnarray}
The matrix elements are defined by
\begin{eqnarray}
\mathcal{Y}_{l,k}^{(K,J)}(j \omega)\coloneqq\sum_{A=1}^{N_C}\left[
\sum_n\frac{I_{nl}^{(AK)}I_{nk}^{(AJ)}}
{j \omega_n^{(A)} \mu \left(\frac{\omega}{\omega_n^{(A)}}-\frac{\omega_n^{(A)}}{\omega}\right)}
+\sum_\nu \frac{I_{\nu l}^{(AK)}I_{\nu k}^{(AJ)}}{j \omega \mu}
\right]\,.
\end{eqnarray}
The upper indices of $\mathcal{Y}_{l,k}^{(K,J)}$ refer to the junction surfaces, while the lower indices refer to their eigenvector fields. The index $A$ runs over all resonators, and $N_C$ is the total number of resonators.
By definition, if cavity A does not share junction surface $K$, the coefficients $I_{n l}^{(AK)}$ and $I_{\nu l}^{(AK)}$ vanish.
This situation is illustrated in Fig.~\ref{fig:adm_multi}.

To make the preceding equations easier to handle, we restrict the number of eigenvector fields on each junction surface $J$ to a finite number $m_J$.
In some cases, this restriction is expected to provide a sufficiently accurate approximation.
The homogeneous equations can then be written in matrix form as
\begin{eqnarray}
\left(
\begin{array}{ccccccc}
\mathcal{Y}_{1,1}^{(1,1)} & \mathcal{Y}_{1,2}^{(1,1)} & \cdots & \mathcal{Y}_{1,m_1}^{(1,1)} &  \mathcal{Y}_{1, 1}^{(1, 2)}
&\cdots & \mathcal{Y}_{1, m_{N_J}}^{(1, N_J)} \\
\mathcal{Y}_{2,1}^{(1,1)} & \mathcal{Y}_{2,2}^{(1,1)} & \cdots & \mathcal{Y}_{2,m_1}^{(1,1)} & \mathcal{Y}_{2, 1}^{(1, 2)}
& \cdots &\mathcal{Y}_{2, m_{N_J}}^{(1, N_J)} \\
\vdots & \vdots & \ddots & \vdots& \vdots & \ddots & \vdots \\
\mathcal{Y}_{m_1,1}^{(1,1)} & \mathcal{Y}_{m_1,2}^{(1,1)} & \cdots & \mathcal{Y}_{m_1,m_1}^{(1,1)} & \mathcal{Y}_{m_1, 1}^{(1, 2)}
& \cdots & \mathcal{Y}_{m_1, m_{N_J}}^{(1, N_J)} \\
\mathcal{Y}_{1,1}^{(2,1)} & \mathcal{Y}_{1,2}^{(2,1)} & \cdots & \mathcal{Y}_{1,m_1}^{(2,1)} & \mathcal{Y}_{1, 1}^{(2, 2)}
& \cdots & \mathcal{Y}_{1, m_{N_J}}^{(2, N_J)} \\
\vdots & \vdots & \ddots & \vdots& \vdots & \ddots & \vdots \\
\mathcal{Y}_{m_{N_J},1}^{(N_J,1)} & \mathcal{Y}_{m_{N_J},2}^{(N_J,1)}  & \cdots & 
\mathcal{Y}_{m_{N_J},m_1}^{(N_J,1)} & \mathcal{Y}_{m_{N_J}, 1}^{(N_J, 2)}
& \cdots & \mathcal{Y}_{m_{N_J}, m_{N_J}}^{(N_J, N_J)} \\
\end{array} 
\right)
\left(\begin{array}{c} V_1^{(1)}\\ V_2^{(1)} \\ \vdots \\ V_{m_1}^{(1)} \\ V_1^{(2)} \\ \vdots 
\\V_{ m_{N_J}}^{(N_J)} 
\end{array}\right)=
\left(\begin{array}{c} 
0 \\ 0 \\ \vdots \\ 0 \\ 0 \\ \vdots \\ 0 
\end{array}\right)\,.
\end{eqnarray}
Here $N_J$ denotes the number of junction surfaces.
The matrix $\mathcal{Y}$ on the left-hand side is symmetric and has dimension $(\sum_{J=1}^{N_J} m_J)\times(\sum_{J=1}^{N_J} m_J)$.

The resonant frequencies $\omega=\Omega_r$ of this coupled-resonator system are determined by
\begin{eqnarray}
\det\left[\mathcal{Y}(j \omega)\right] =0\,.
\end{eqnarray}
The eigenvectors corresponding to the zero eigenvalues then determine the resonant states of the coupled system.

We now show that the resonant frequencies determined above are independent of the choice of eigenvector fields on the junction surfaces.
The block corresponding to junction surfaces $K$ and $J$ is an $(m_K\times m_J)$ matrix.
Let $\{O_{ij}^{(m_J)}\}$ and $\{O_{ij}^{(m_K)}\}$ be the orthogonal matrices representing the corresponding basis transformations.
The expansion coefficients transform as
\begin{eqnarray}
I_{nl}^{\prime(AJ)}&=&\sum_{k=1}^{m_J} O_{lk}^{(m_J)}I_{nk}^{(AJ)},\\
I_{nl}^{\prime(AK)}&=&\sum_{k=1}^{m_K} O_{lk}^{(m_K)}I_{nk}^{(AK)}\,,
\end{eqnarray}
so that the corresponding block transforms as
\begin{eqnarray}
\mathcal{Y}^{\prime(K,J)}(j \omega)=O^{(m_K)} \cdot \mathcal{Y}^{(K,J)}(j \omega) \cdot {O^{(m_J)}}^T\,.
\end{eqnarray}
For the full matrix, we obtain
\begin{eqnarray}
\mathcal{Y}^{\prime}(j \omega)=
\left(\begin{array}{cccc}
O^{(m_1)} &&&\\ &O^{(m_2)}&& \\ &&\ddots& \\ &&&O^{(m_{N_J})}
\end{array}\right)
\cdot \mathcal{Y}(j \omega) \cdot 
\left(\begin{array}{cccc} 
{O^{(m_1)}}^T &&&\\ & {O^{(m_2)}}^T&& \\ &&\ddots& \\ &&& {O^{(m_{N_J})}}^T
\end{array}\right)\,.
\end{eqnarray}
Consequently,
\begin{eqnarray}
\det\left[\mathcal{Y}^{\prime}(j \omega)\right] = \det\left[\mathcal{Y}(j \omega)\right]\,.
\end{eqnarray}
Thus, the determinant itself, and hence the resonant frequencies, are independent of the choice of eigenvector fields on the junction surfaces.

\section{Periodic structure}
\label{sec:periodic}

In this chapter, we apply the formulation developed in the previous chapter for coupled resonances in multi-aperture cavities to periodic structures.

\subsection{Formulation}

\begin{figure}[htb]
   \centering
   \includegraphics*[width=140mm]{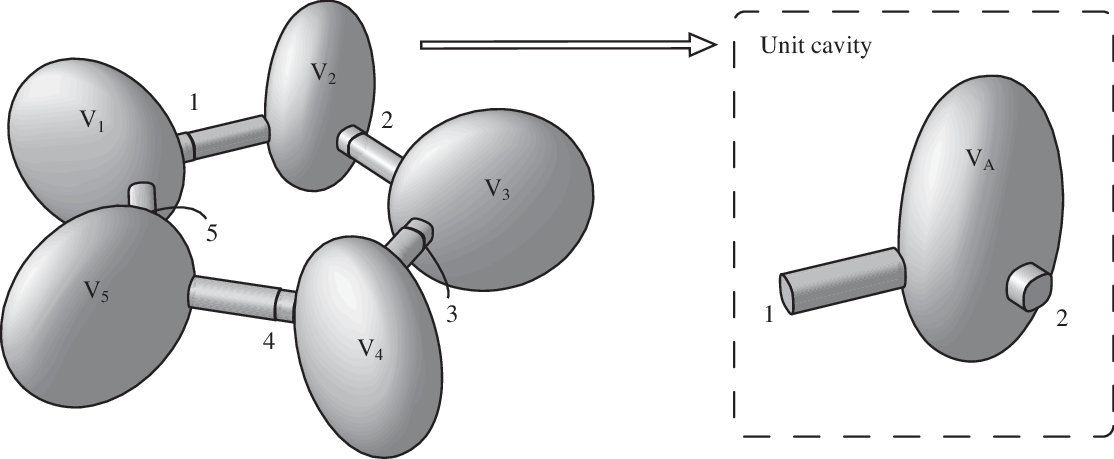}
   \caption{Periodic system of coupled cavity resonators (for $N=5$).
   It consists of identical resonators having two apertures.}
   \label{fig:period}
\end{figure}

First, consider the periodic system shown in Fig.~\ref{fig:period}.
Each unit consists of a cavity with two apertures of the same cross section.
A common set of eigenvector fields can then be used for all cross sections, and a periodic system can be constructed by connecting an arbitrary number $N$ of cavities.

As shown in the preceding section, the coupled resonance frequencies are independent of the choice of eigenvector fields on the cross sections. We may therefore use any convenient basis.
For simplicity, we retain only one eigenvector field on each cross section.
This treatment is appropriate when the frequency under consideration is lower than the cutoff frequencies of the higher-order modes of the cross section.
The case with more than one eigenvector field will be considered in Sec.~\ref{ap:multi}.

In this case, the admittance matrix $\mathcal{Y}(j \omega)$ is an $N\times N$ matrix of the form
\begin{eqnarray}
\label{eq:cycy}
\mathcal{Y}(j \omega)=\left(
\begin{array}{cccccc}
p(j \omega) & q(j \omega) & 0 & \cdots & 0 & q(j \omega) \\
q(j \omega) & p(j \omega) & q(j \omega) & 0 & \cdots & 0\\
0 & q(j \omega) & p(j \omega) & q(j \omega) & \cdots & 0\\
\vdots & \vdots & \vdots & \ddots & \cdots & \vdots\\
0 & 0 & \cdots & q(j \omega) & p(j \omega) & q(j \omega)\\
q(j \omega) & 0 & 0 & \cdots & q(j \omega) & p(j \omega)
\end{array} 
\right)\,.
\end{eqnarray}
Here, $p(j \omega)$ and $q(j \omega)$ are given, respectively, by
\begin{eqnarray}
\label{eq:pk}
p(j \omega)&\coloneqq&\mathcal{Y}_{1,1}^{(1,1)}(j \omega)\nonumber \\
&=&
\sum_n\frac{{I_{n1}^{(1)}}^2+{I_{n1}^{(2)}}^2}
{j \omega_n \mu \left(\frac{\omega}{\omega_n}-\frac{\omega_n}{\omega}\right)}
+\sum_\nu \frac{{I_{\nu 1}^{(1)}}^2+{I_{\nu 1}^{(2)}}^2}{j \omega \mu}\,,
\\
\label{eq:qk}
q(j \omega)&\coloneqq&\mathcal{Y}_{1,1}^{(1,2)}(j \omega)\nonumber \\
&=&
\sum_n\frac{I_{n1}^{(1)}I_{n1}^{(2)}}
{j \omega_n \mu \left(\frac{\omega}{\omega_n}-\frac{\omega_n}{\omega}\right)}
+\sum_\nu \frac{I_{\nu 1}^{(1)}I_{\nu 1}^{(2)}}{j \omega \mu}\,.
\end{eqnarray}
Since the cavities are identical, we have omitted the superscript in the resonance frequency $\omega_n^{(A)}$ of the unit cavity, writing it simply as $\omega_n$.
For the expansion coefficients, $I_{n1}^{(1)}$ refers to the left aperture of the original unit cavity, while $I_{n1}^{(2)}$ refers to its right aperture.

The meaning of these elements can be understood from Fig.~\ref{fig:period}.
Cross section 1 is shared by cavities 1 and 2.
A voltage applied at cross section 1 induces magnetic fields in cavities 1 and 2, producing contributions at that cross section proportional to ${I_{n1}^{(1)}}^2$ and ${I_{n1}^{(2)}}^2$, respectively.
This corresponds to $\mathcal{Y}_{1,1}^{(1,1)}=p(j\omega)$.
On the other hand, cross section 1 is also affected by the voltages applied at cross sections 2 and 5.
These voltages induce magnetic fields in cavities 2 and 1, respectively, which also contribute to the magnetic field at cross section 1.
These contributions are $\mathcal{Y}_{1,1}^{(1,2)}$ and $\mathcal{Y}_{1,1}^{(1,5)}$. Because the cavities are identical, both contributions are proportional to $I_{n1}^{(1)}I_{n1}^{(2)}$ and have the same value $q(j\omega)$.

To express $\mathcal{Y}(j \omega)$, introduce the $N\times N$ identity matrix $I_N$ and the generator $T_N$ of cyclic permutations, defined by
\begin{eqnarray}
T_N=\left(
\begin{array}{cccccc}
0 & 1 & 0 & \cdots & 0 & 0 \\
0 & 0 & 1 & 0 & \cdots & 0\\
0 & 0 & 0 & 1 & \cdots & 0\\
\vdots & \vdots & \vdots & \ddots & \cdots & \vdots\\
0 & 0 & \cdots & 0 & 0 & 1\\
1 & 0 & 0 & \cdots & 0 & 0
\end{array} 
\right)\,.
\end{eqnarray}
We can then write
\begin{eqnarray}
\mathcal{Y}(j \omega)= p(j \omega) I_N +q(j \omega) T_N +q(j \omega) {T_N}^{-1}.
\end{eqnarray}
This shows that $\mathcal{Y}(j \omega)$ is a circulant matrix and satisfies the cyclic symmetry relation
\begin{eqnarray}
T_N\cdot \mathcal{Y}(j \omega)=\mathcal{Y}(j \omega) \cdot T_N\,.
\end{eqnarray}
It can therefore be diagonalized using the symmetry of the cyclic group.

To carry out the diagonalization, first note that
\begin{eqnarray}
(T_N)^N=I_N,
\end{eqnarray}
so the eigenvalues of $T_N$ are $\lambda^k\quad (k=0,1,\ldots,N-1)$, where $\lambda=\exp(2j\pi/N)$.
An eigenvector corresponding to $\lambda^k$ is
\begin{eqnarray}
\label{eq:ev}
\bm{\xi}_k&\coloneqq
\frac{1}{\sqrt{N}}{ {(1, \lambda^k, \lambda^{2k},...,\lambda^{(N-1)k})}^T}\,.
\end{eqnarray}
Now introduce the unitary matrix $U_N$:
\begin{eqnarray}
\label{eq:un}
U_N&\coloneqq&(\bm{\xi}_0,\cdots,\bm{\xi}_{N-1})\nonumber \\
&=&\frac{1}{\sqrt{N}}\left(
\begin{array}{ccccc}
1 & 1 & 1 & \cdots & 1 \\
1 & e^{2j\pi/N} & e^{4j\pi/N} &  \cdots & e^{2(N-1)j\pi/N}\\
1 & e^{4j\pi/N} & e^{8j\pi/N} &  \cdots & e^{4(N-1)j\pi/N}\\
\vdots & \vdots & \vdots & \ddots & \vdots\\
1 & e^{2(N-1)j\pi/N} & e^{4(N-1)j\pi/N} & \cdots & e^{2(N-1)^2j\pi/N}
\end{array} 
\right)\,.
\end{eqnarray}
Then $\mathcal{Y}(j \omega)$ is diagonalized as
\begin{eqnarray}
{U_N}^{-1}\cdot \mathcal{Y}(j \omega) \cdot U_N=
\left(
\begin{array}{ccccc}
f_0(j \omega) &  &  &  &  \\
& f_1(j \omega)  &  &  &  \\
&  &  f_2(j \omega) &  &  \\
&  &  & \ddots & \\
&  &  &  & f_{N-1}(j \omega)
\end{array} 
\right)\,,
\end{eqnarray}
where
\begin{eqnarray}
\label{eq:fk}
f_k(j \omega)&\coloneqq&p(j \omega)+2 q(j \omega)\cos\left(\frac{2 k \pi}{N}\right) \nonumber\\
&=&
\sum_n\frac{{I_{n1}^{(1)}}^2+{I_{n1}^{(2)}}^2+2 I_{n1}^{(1)}I_{n1}^{(2)}\cos\left(\frac{2 k \pi}{N}\right)}
{j \omega_n \mu \left(\frac{\omega}{\omega_n}-\frac{\omega_n}{\omega}\right)}
+\sum_\nu \frac{{I_{\nu 1}^{(1)}}^2+{I_{\nu 1}^{(2)}}^2
+2 I_{\nu1}^{(1)}I_{\nu1}^{(2)}\cos\left(\frac{2 k \pi}{N}\right)}
{j \omega \mu}\,.
\end{eqnarray}
Note that each $f_k(j \omega)$ is purely imaginary.

The coupled resonance frequencies are given by the zeros of the functions $f_k(j \omega)$.
When one of the $f_k$ vanishes at $\omega=\Omega_{k,m}$, that is, when $f_k(j \Omega_{k,m})=0$, the eigenvector corresponding to the zero eigenvalue is $\bm{\xi}_k$.
Thus, for a given $k$, the voltage distributions on the boundary surfaces share the same translational symmetry.
However, as can be seen by substituting $\omega=\Omega_{k,m}$ into Eq.~\eqref{eq:expmul}, the amplitudes of the $\bm{H}_n^{(A)}$ in the superposition depend on the resonance frequency.
A specific example will be considered in the next section.

The eigenvector $\bm{\xi}_k$ can be regarded as representing a ``traveling wave.''
We now express the coupled resonant electromagnetic fields in terms of ``standing waves.''
Suppose that $k$ is not zero and, when $N$ is even, is not $N/2$.
Equation~\eqref{eq:fk} then gives
\begin{eqnarray}
\label{eq:fknk}
f_k(j \omega)=f_{N-k}(j \omega)\,.
\end{eqnarray}
Therefore, when $k\neq 0,N/2$, the eigenvalues of $\mathcal{Y}(j \omega)$ are degenerate, and
\begin{eqnarray}
\mathcal{Y}(j\omega)&\cdot&(\bm{\xi}_k,\bm{\xi}_{N-k})
=(\bm{\xi}_k,\bm{\xi}_{N-k}) \cdot
\left(
\begin{array}{cc}
f_k(j\omega) & 0 \\
0 & f_k(j\omega) \\
\end{array} 
\right)
\end{eqnarray}
holds. Here, noting from Eq.~\eqref{eq:ev} that
\begin{eqnarray}
\bm{\xi}_{N-k}=\overline{\bm{\xi}_k}\,,
\end{eqnarray}
we find that $\bm{u}_k$ and $\bm{v}_k$, defined by the unitary transformation
\begin{eqnarray}
\label{eq:u2}
(\bm{u}_k,\bm{v}_k) \coloneqq 
(\bm{\xi}_k,\bm{\xi}_{N-k})\cdot
\left(
\begin{array}{cc}
1/\sqrt{2} & -j/\sqrt{2}\\
1/\sqrt{2} & j/\sqrt{2} \\
\end{array} 
\right)
\end{eqnarray}
are real, and
\begin{eqnarray}
\mathcal{Y}(j\omega)&\cdot&(\bm{u}_k,\bm{v}_k)
=(\bm{u}_k,\bm{v}_k) \cdot
\left(
\begin{array}{cc}
f_k(j\omega) & 0\\
0 & f_k(j\omega) \\
\end{array} 
\right)
\end{eqnarray}
holds.
Thus, $\mathcal{Y}(j\omega)$ has been diagonalized in a real basis.
These real basis vectors correspond to standing waves.
Since $\mathcal{Y}(j\omega)$ is a purely imaginary symmetric matrix, it can also be diagonalized by an orthogonal matrix, that is, a real unitary matrix. The procedure above provides an explicit construction.

Finally, replacing $j\omega$ with $s$ in $f_k(j\omega)$ gives
\begin{eqnarray}
f_k(s)&=&
\sum_n\frac{s}{\mu}\cdot
\frac{{I_{n1}^{(1)}}^2+{I_{n1}^{(2)}}^2+2 I_{n1}^{(1)}I_{n1}^{(2)}\cos\left(\frac{2 k \pi}{N}\right)}
{s^2+{\omega_n}^2}
+\sum_\nu \frac{1}{\mu}\cdot
\frac{{I_{\nu 1}^{(1)}}^2+{I_{\nu 1}^{(2)}}^2
+2 I_{\nu1}^{(1)}I_{\nu1}^{(2)}\cos\left(\frac{2 k \pi}{N}\right)}{s}\nonumber \\
&=&
\sum_n\frac{1}{2 \mu}\cdot
\frac{{I_{n1}^{(1)}}^2+{I_{n1}^{(2)}}^2+2 I_{n1}^{(1)}I_{n1}^{(2)}\cos\left(\frac{2 k \pi}{N}\right)}
{s+j\omega_n}
+\sum_n\frac{1}{2 \mu}\cdot
\frac{{I_{n1}^{(1)}}^2+{I_{n1}^{(2)}}^2+2 I_{n1}^{(1)}I_{n1}^{(2)}\cos\left(\frac{2 k \pi}{N}\right)}
{s-j\omega_n}
\nonumber \\
&+&\sum_\nu \frac{1}{\mu}\cdot
\frac{{I_{\nu 1}^{(1)}}^2+{I_{\nu 1}^{(2)}}^2
+2 I_{\nu1}^{(1)}I_{\nu1}^{(2)}\cos\left(\frac{2 k \pi}{N}\right)}{s}\,.\nonumber \\
\end{eqnarray}
The residues at all the poles $s=0,\pm j\omega_n$ satisfy
\begin{eqnarray}
0\leq \left[ |I_{n1}^{(1)}|- |I_{n1}^{(2)}|\right]^2\leq
{I_{n1}^{(1)}}^2+{I_{n1}^{(2)}}^2+2 I_{n1}^{(1)}I_{n1}^{(2)}\cos\left(\frac{2 k \pi}{N}\right)
\leq\left[ |I_{n1}^{(1)}|+ |I_{n1}^{(2)}|\right]^2\,,
\end{eqnarray}
and are therefore nonnegative; in general, they are positive.
Consequently, each $f_k(s)$ is a positive-real function, and the above expression is a Foster expansion.
Thus, if the resonance frequencies $\left\{\omega_n \right\}$ of the original unit cavity are not degenerate, there is exactly one zero of each $f_k(j\omega)$ between $\omega_n$ and $\omega_{n+1}$.

\subsection{Solvable example}

\begin{figure}[htb]
   \centering
   \includegraphics*[width=140mm]{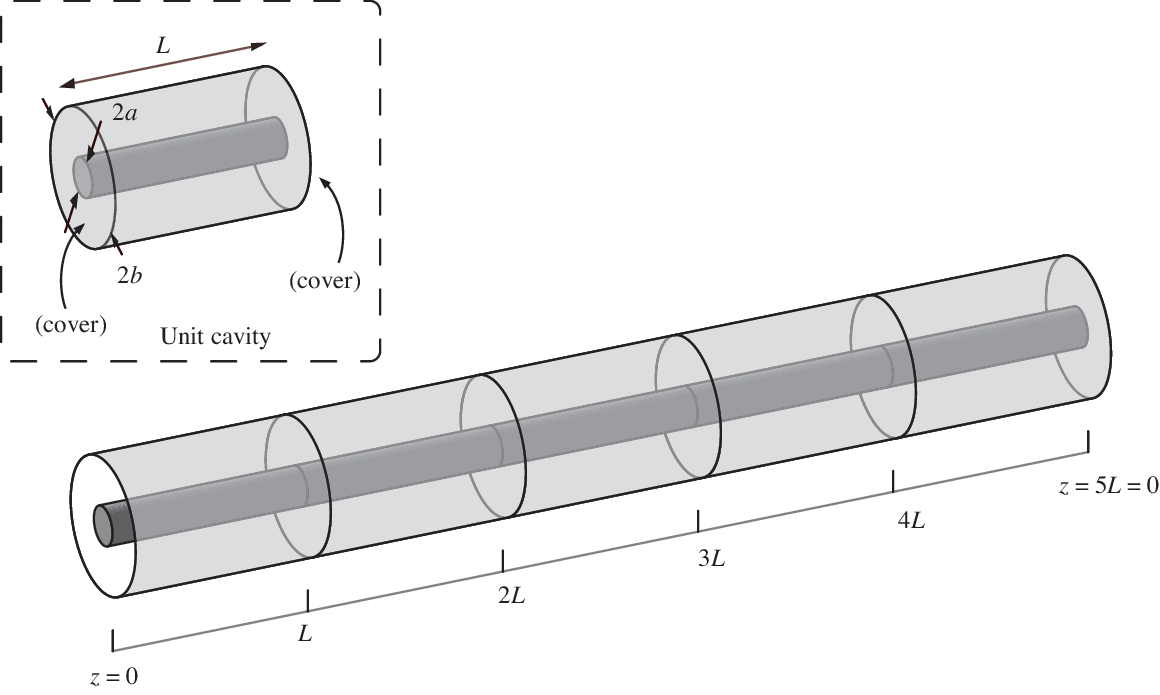}
   \caption{Periodic system of coupled resonators consisting of coaxial resonators (for $N=5$).}
   \label{fig:coax}
\end{figure}

As an exactly solvable example, consider the periodic system shown in Fig.~\ref{fig:coax}, consisting of coupled half-wavelength coaxial resonators of length $L$, inner-conductor radius $a$, and outer-conductor radius $b$. We apply the formulation of the preceding section to this system.
The eigenvector fields and eigenvalues of the unit cavity are given by Eqs.~\eqref{eq:hnpol} and \eqref{eq:eigentem}, respectively. The lower-order eigenvector fields have already been illustrated in Fig.~\ref{fig:unitcell}.
We also retain only the cross-sectional eigenvector field given by Eq.~\eqref{eq:dammen}.

\begin{figure}[htb]
   \centering
   \includegraphics*[width=120mm]{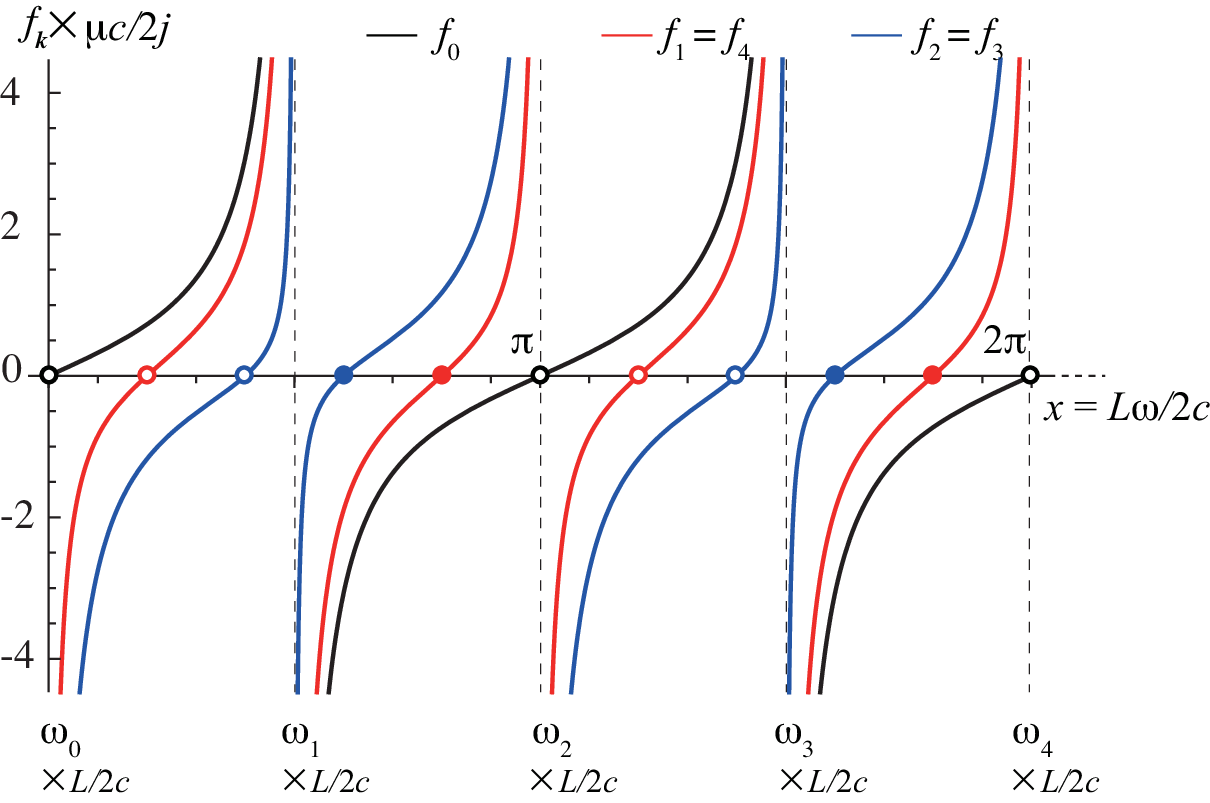}
   \caption{Examples of $f_k(j\omega)$. These correspond to Fig.~\ref{fig:coax}, with $N=5$.
   The vertical axis is $f_k\times\mu c/2j$, and the horizontal axis is $\omega L/2c=x$.
   The positions of the resonance frequencies $\omega_n$ of the unit cavity are also shown.}
   \label{fig:fk}
\end{figure}

Using these fields and Eq.~\eqref{eq:ink}, we obtain the expansion coefficients on the left junction surface of the unit cavity:
\begin{eqnarray}
\label{eq:coeff1}
I_{n1}^{(1)} &=&\int_{z=0}\bm{E}_{t1}\cdot\left(-\bm{e}_z \times\bm{H}_n\right)dS
\nonumber\\
&=&\frac{2}{\sqrt{2L\left(1+\delta_{n0}\right)}}\,.
\end{eqnarray}
Similarly, those on the right junction surface are
\begin{eqnarray}
\label{eq:coeff2}
I_{n1}^{(2)} &=&\int_{z=L}\bm{E}_{t1}\cdot\left(\bm{e}_z \times\bm{H}_n\right)dS
\nonumber\\
&=&(-)^{n+1}\cdot\frac{2}{\sqrt{2L\left(1+\delta_{n0}\right)}}\,.
\end{eqnarray}
Substituting these results into Eq.~\eqref{eq:fk} and rearranging, we obtain
\begin{eqnarray}
j \mu L \cdot f_k(j \omega)
=\frac{2-2\cos\left(\frac{2 k \pi}{N}\right)}{\omega}
+\sum_{n=1}^{\infty}
\frac{4+4\cdot(-)^{n+1}\cos\left(\frac{2 k \pi}{N}\right)}
{\omega_n \left(\frac{\omega}{\omega_n}-\frac{\omega_n}{\omega}\right)}\,.
\end{eqnarray}
Separating the terms with even and odd $n$, we obtain
\begin{eqnarray}
\frac{j \mu L}{4} \cdot f_k(j \omega)
=\sin^2\left(\frac{k \pi}{N}\right)
\left[\frac{1}{\omega}+
\sum_{m=1}^{\infty}
\frac{2\omega}{\omega^2-{\omega_{2m}}^2}
\right]
+2\cos^2\left(\frac{k \pi}{N}\right)
\sum_{m=0}^{\infty}
\frac{\omega}{\omega^2-{\omega_{2m+1}}^2}\,.
\end{eqnarray}
Next, using Eq.~\eqref{eq:eigentem} and introducing the variable
\begin{eqnarray}
x\coloneqq\frac{L}{2c}\omega\,,
\end{eqnarray}
we can rewrite this expression as
\begin{eqnarray}
\frac{j \mu c}{2} \cdot f_k
=\sin^2\left(\frac{k \pi}{N}\right)
\left[\frac{1}{x}+
\sum_{m=1}^{\infty}
\frac{2x}{x^2-m^2\pi^2}
\right]
+\cos^2\left(\frac{k \pi}{N}\right)
\sum_{m=0}^{\infty}
\frac{8x}{4x^2-(2m+1)^2\pi^2}\,.
\end{eqnarray}
Using the partial-fraction expansions of $\tan$ and $\cot$ in Eqs.~\eqref{eq:cot} and \eqref{eq:tan},
we obtain
\begin{eqnarray}
\label{eq:findroot}
f_k(j \omega)
=\frac{2}{j \mu c}
\left[\sin^2\left(\frac{k \pi}{N}\right)
\cot\left(\frac{L \omega}{2 c}\right)
-\cos^2\left(\frac{k \pi}{N}\right)
\tan\left(\frac{L \omega}{2 c}\right)
\right]\,.
\end{eqnarray}
Figure~\ref{fig:fk} illustrates the behavior of $f_k(j\omega)$.
Note also that Eq.~\eqref{eq:fknk}, that is, $f_k(j\omega)=f_{N-k}(j\omega)$, holds.

\begin{figure}[htb]
   \centering
   \includegraphics*[width=140mm]{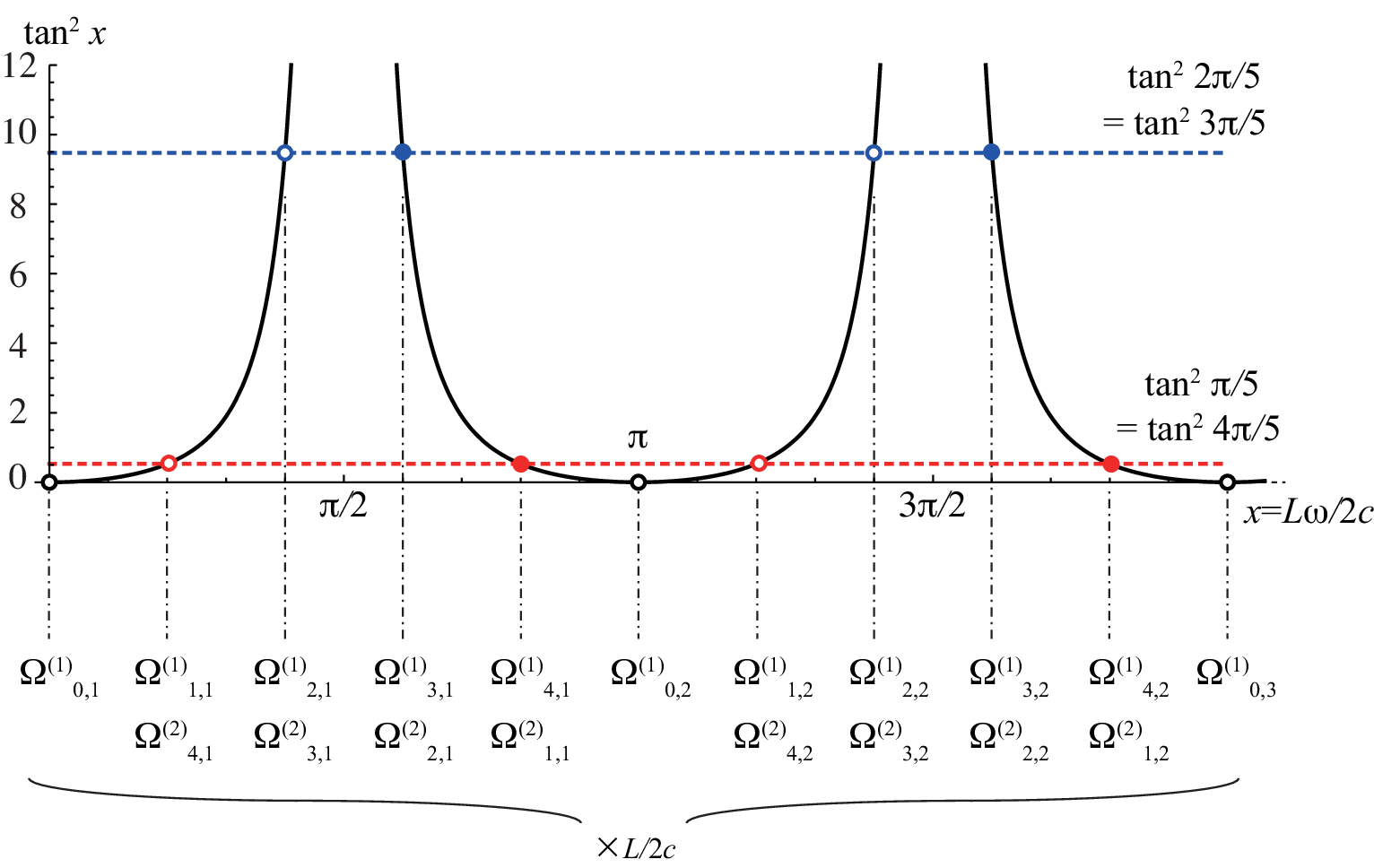}
   \caption{How the roots $\Omega^{(p)}_{k,m}$ of Eq.~\eqref{eq:tan2x} are determined, for $N=5$.}
   \label{fig:tan2x}
\end{figure}

\begin{figure}[htb]
   \centering
   \includegraphics*[width=120mm]{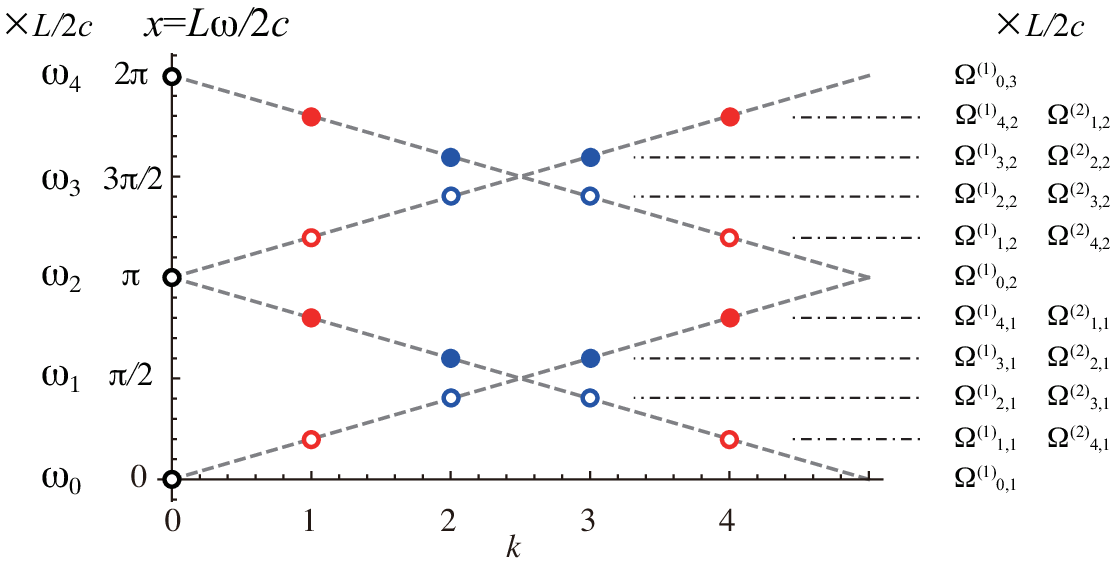}
   \caption{Dispersion relation corresponding to Fig.~\ref{fig:coax}, with $N=5$.
   The horizontal axis is the cyclic-symmetry index $k$, and the vertical axis is the frequency of the coupled resonant electromagnetic field.
   The resonance frequencies $\omega_n$ of the unit cavity are also indicated.}
   \label{fig:disp_exact}
\end{figure}

From Eq.~\eqref{eq:findroot}, the roots of $f_k(j\omega)$ are solutions of
\begin{eqnarray}
\tan^2\left(\frac{L \omega}{2 c}\right)=\tan^2\left(\frac{k \pi}{N}\right), \qquad k=0,1,\cdots, N-1\,.
\label{eq:tan2x}
\end{eqnarray}
Figure~\ref{fig:tan2x} illustrates the roots of Eq.~\eqref{eq:tan2x}.
For convenience in the following calculations, we divide the roots into two series:
\begin{eqnarray}
\Omega^{(1)}_{k,m}
\coloneqq\frac{2c}{L}\cdot\left[\frac{k \pi}{N}+\left(m-1\right)\pi\right], \qquad m=1,2,\cdots\,,
\end{eqnarray}
and
\begin{eqnarray}
\Omega^{(2)}_{k,m}
\coloneqq\frac{2c}{L}\cdot\left(m \pi-\frac{k \pi}{N}\right), \qquad m=1,2,\cdots\,.
\end{eqnarray}
The index $k$ in $\Omega^{(1)}_{k,m}$ denotes the cyclic-symmetry index.
On the other hand, as shown in Fig.~\ref{fig:tan2x}, $m$ does not represent the order of the roots.
For example, $\Omega^{(1)}_{3,1}$ is the second root of $f_3$, while the first root of $f_3$, $\Omega^{(2)}_{3,1}$, coincides with $\Omega^{(1)}_{2,1}$, as shown in Fig.~\ref{fig:tan2x}.
In general, the solutions of the two series satisfy
\begin{eqnarray}
\label{eq:omgknk}
\Omega^{(1)}_{k,m}=\Omega^{(2)}_{N-k,m}\,.
\end{eqnarray}
Thus, although the two-series description is redundant, it is convenient for expressing the coupled resonant electromagnetic fields considered below.
The dispersion relation in this case, namely, the change in the frequencies of the coupled resonant electromagnetic fields with the cyclic-symmetry index $k$, is shown in Fig.~\ref{fig:disp_exact}.

Having obtained the coupled resonance frequencies, we now consider the electromagnetic fields.
First, consider the electromagnetic field in the first resonator ($0<z<L$) in Fig.~\ref{fig:coax}.
The index $p=1,2$ distinguishes the two series of roots.
As discussed in the previous section, when $\omega=\Omega^{(p)}_{k,m}$, the eigenvector is $\bm{\xi}_k$ in Eq.~\eqref{eq:ev}.
From this and the expansion coefficients in Eqs.~\eqref{eq:coeff1} and \eqref{eq:coeff2}, we obtain
\begin{eqnarray}
V^{(1)}I_{n1}^{(1)}+V^{(2)}I_{n1}^{(2)}=\frac{2}{\sqrt{2 N L\left(1+\delta_{n0}\right)}}
\left[1+(-)^{n+1}\lambda^k\right]\,.
\end{eqnarray}
Using this result and the eigenvector field in Eq.~\eqref{eq:hnpol}, the left-hand side of Eq.~\eqref{eq:expmul} is calculated as
\begin{eqnarray}
\label{eq:hpkm0}
\bm{H}^{(p)}_{k,m}=
\frac{\bm{e}_\theta}{r}\cdot
\frac{1}{j \mu L \sqrt{2 N \pi \log\frac{b}{a}}}
\sum_{n=0}^\infty
{
\cos\left(\frac{\omega_n}{c}z\right)\cdot\frac{2}{1+\delta_{n0}}\cdot
\frac{\Omega^{(p)}_{k,m}}{{\Omega^{(p)}_{k,m}}^2-{\omega_n}^2}\cdot
\left[1+(-)^{n+1}\lambda^k\right]
}\,.
\end{eqnarray}
Thus, the electromagnetic field of a coupled resonance is a superposition of eigenvector fields $\bm{H}_n^{(A)}$ of the original unit cavity.
For a given $k$, the voltage distributions on the boundary surfaces share the same translational symmetry, whereas the coefficients of the individual $\bm{H}_n^{(A)}$ depend on the coupled resonance frequency $\Omega^{(p)}_{k,m}$.

We now continue the calculation.
First, note that $\lambda^k=\cos\left(\frac{2k\pi}{N}\right)+j\sin\left(\frac{2k\pi}{N}\right)$.
We also introduce
\begin{eqnarray}
\label{eq:rhodef}
\rho^{(p)}_{k,m}\coloneqq\frac{L}{c\pi}\Omega^{(p)}_{k,m}.
\end{eqnarray}
With this definition, we have
\begin{eqnarray}
\label{eq:rho1}
\cos\rho^{(1)}_{k,m}\pi&=&\cos\rho^{(2)}_{k,m}\pi=\cos\left(\frac{2 k \pi}{N}\right)\,,\\
\label{eq:rho2}
\sin\rho^{(1)}_{k,m}\pi&=&-\sin\rho^{(2)}_{k,m}\pi=\sin\left(\frac{2 k \pi}{N}\right)\,.
\end{eqnarray}
Substituting Eqs.~\eqref{eq:rhodef}, \eqref{eq:rho1}, \eqref{eq:rho2}, and \eqref{eq:eigentem} into Eq.~\eqref{eq:hpkm0} and carrying out the calculation, we obtain
\begin{eqnarray}
\label{eq:hpkm1}
\bm{H}^{(1)}_{k,m}&=&
\frac{\bm{e}_\theta}{r}\cdot
\frac{1}{j \mu c \sqrt{2 N \pi \log\frac{b}{a}}}
\sum_{n=0}^\infty
\cos\left(\frac{n\pi z}{L}\right)\nonumber \\
&\times&
\frac{2}{1+\delta_{n0}}\cdot
\frac{\rho^{(1)}_{k,m}}{\pi\left({\rho^{(1)}_{k,m}}^2-n^2\right)}\cdot
\left[1-(-)^{n}\cos\rho^{(1)}_{k,m}\pi-(-)^{n}j\sin\rho^{(1)}_{k,m}\pi\right]
\nonumber \\
&=&
\frac{\bm{e}_\theta}{r}\cdot
\frac{1}{j \mu c \sqrt{2 N \pi \log\frac{b}{a}}}
\left[\sin\left(\frac{\rho^{(1)}_{k,m}\pi z}{L}\right)-j\cos\left(\frac{\rho^{(1)}_{k,m}\pi z}{L}\right)\right]
\nonumber \\
&=&
-\frac{\bm{e}_\theta}{r}\cdot
\frac{1}{\mu c \sqrt{2 N \pi \log\frac{b}{a}}}
\exp\left(j\frac{\rho^{(1)}_{k,m}\pi z}{L}\right)\,.
\end{eqnarray}
Here we have used the Fourier expansion formulas
\begin{eqnarray}
\sin\left(\frac{\rho\pi z}{L}\right)&=&\frac{1-\cos\rho\pi}{\rho\pi}+
\sum_{m=1}^\infty\cos\left(\frac{m\pi z}{L}\right)\cdot\frac{2}{\pi}\cdot
\frac{\rho\left[1-(-)^{m}\cos\rho\pi\right]}{\rho^2-m^2},\nonumber \\
\cos\left(\frac{\rho\pi z}{L}\right)&=&\frac{\sin\rho\pi}{\rho\pi}+
\sum_{m=1}^\infty\cos\left(\frac{m\pi z}{L}\right)\cdot(-)^m\cdot\frac{2}{\pi}\cdot
\frac{\rho\sin\rho\pi}{\rho^2-m^2}\,.
\end{eqnarray}
Similarly, we obtain
\begin{eqnarray}
\label{eq:hpkm2}
\bm{H}^{(2)}_{k,m}=
-\frac{\bm{e}_\theta}{r}\cdot
\frac{1}{\mu c \sqrt{2 N \pi \log\frac{b}{a}}}
\exp\left(-j\frac{\rho^{(2)}_{k,m}\pi z}{L}\right)\,.
\end{eqnarray}
This is the electromagnetic field of the coupled resonance in the first resonator ($0<z<L$).

In the $n$th resonator ($(n-1)L<z<nL$), if the $z$ coordinate is chosen as in Fig.~\ref{fig:coax}, the expression for the eigenvector $\bm{\xi}_k$ leads, through the same procedure as above, to
\begin{eqnarray}
\bm{H}^{(1)}_{k,m}=
-\frac{\bm{e}_\theta}{r}\cdot
\frac{1}{\mu c \sqrt{2 N \pi \log\frac{b}{a}}}
\exp\left[j\frac{\rho^{(1)}_{k,m}\pi}{L}\left\{z-(n-1)L\right\}\right]
\exp\left[j(n-1)\frac{2\pi k}{N}\right]\,.
\end{eqnarray}
Using Eqs.~\eqref{eq:rho1} and \eqref{eq:rho2}, however, we readily find that
\begin{eqnarray}
\bm{H}^{(1)}_{k,m}=
-\frac{\bm{e}_\theta}{r}\cdot
\frac{1}{\mu c \sqrt{2 N \pi \log\frac{b}{a}}}
\exp\left(j\frac{\rho^{(1)}_{k,m}\pi z}{L}\right)\,.
\end{eqnarray}
Thus, the same expression describes the coupled resonant electromagnetic field in every resonator, not only in the first.
Using Eq.~\eqref{eq:rhodef} to express the result in terms of $\Omega^{(1)}_{k,m}$, we obtain
\begin{eqnarray}
\bm{H}^{(1)}_{k,m}=
-\frac{\bm{e}_\theta}{r}\cdot
\frac{1}{\mu c \sqrt{2 N \pi \log\frac{b}{a}}}
\exp\left(j\frac{\Omega^{(1)}_{k,m}}{c}z\right)\,.
\end{eqnarray}
Similarly, for $\bm{H}^{(2)}_{k,m}$, we obtain
\begin{eqnarray}
\bm{H}^{(2)}_{k,m}=
-\frac{\bm{e}_\theta}{r}\cdot
\frac{1}{\mu c \sqrt{2 N \pi \log\frac{b}{a}}}
\exp\left(-j\frac{\Omega^{(2)}_{k,m}}{c}z\right)\,.
\end{eqnarray}
Corresponding to Eq.~\eqref{eq:omgknk}, we also have
\begin{eqnarray}
\label{eq:hknk}
\bm{H}^{(2)}_{N-k,m}=\overline{\bm{H}^{(1)}_{k,m}}\,.
\end{eqnarray}

The coupled resonant electromagnetic fields obtained above correspond to traveling waves.
When expressed in a real basis using Eq.~\eqref{eq:u2}, they give the standing waves
\begin{eqnarray}
\frac{1}{\sqrt{2}}\left(\bm{H}^{(1)}_{k,m}+\bm{H}^{(2)}_{N-k,m}\right)&=&
\frac{1}{\sqrt{2}}\left(\bm{H}^{(1)}_{k,m}+\overline{\bm{H}^{(1)}_{k,m}}\right)\nonumber\\
&=&
-\frac{\bm{e}_\theta}{r}\cdot
\frac{1}{\mu c \sqrt{N \pi \log\frac{b}{a}}}
\cos\left(\frac{\Omega^{(1)}_{k,m}}{c}z\right),\\
\frac{1}{j\sqrt{2}}\left(\bm{H}^{(1)}_{k,m}-\bm{H}^{(2)}_{N-k,m}\right)&=&
\frac{1}{j\sqrt{2}}\left(\bm{H}^{(1)}_{k,m}-\overline{\bm{H}^{(1)}_{k,m}}\right)\nonumber\\
&=&
-\frac{\bm{e}_\theta}{r}\cdot
\frac{1}{\mu c \sqrt{N \pi \log\frac{b}{a}}}
\sin\left(\frac{\Omega^{(1)}_{k,m}}{c}z\right)\,.
\end{eqnarray}
Equivalently,
\begin{eqnarray}
\frac{1}{\sqrt{2}}\left(\bm{H}^{(2)}_{k,m}+\bm{H}^{(1)}_{N-k,m}\right)&=&
\frac{1}{\sqrt{2}}\left(\bm{H}^{(2)}_{k,m}+\overline{\bm{H}^{(2)}_{k,m}}\right)\nonumber\\
&=&
-\frac{\bm{e}_\theta}{r}\cdot
\frac{1}{\mu c \sqrt{N \pi \log\frac{b}{a}}}
\cos\left(\frac{\Omega^{(2)}_{k,m}}{c}z\right),\\
\frac{1}{j\sqrt{2}}\left(\bm{H}^{(2)}_{k,m}-\bm{H}^{(1)}_{N-k,m}\right)&=&
\frac{1}{j\sqrt{2}}\left(\bm{H}^{(2)}_{k,m}-\overline{\bm{H}^{(2)}_{k,m}}\right)\nonumber\\
&=&
\frac{\bm{e}_\theta}{r}\cdot
\frac{1}{\mu c \sqrt{N \pi \log\frac{b}{a}}}
\sin\left(\frac{\Omega^{(2)}_{k,m}}{c}z\right)\,.
\end{eqnarray}

\begin{figure}[htb]
   \centering
   \includegraphics*[width=100mm]{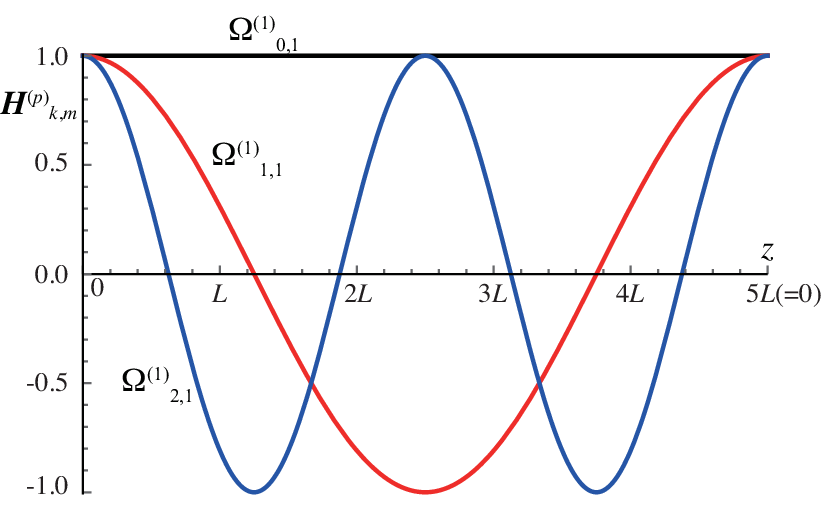}
   \caption{Examples of the magnetic fields of coupled resonances for $N=5$.
   The corresponding $\Omega^{(p)}_{k,m}$ are also indicated.
   Coupled resonances with resonance frequencies below the original resonance frequency $\omega_1$ appear correctly, and they satisfy the periodic boundary conditions.}
   \label{fig:hk}
\end{figure}

Examples of the resulting magnetic fields in the coupled cavities are shown in Fig.~\ref{fig:hk}.
Note that coupled resonances appear whose resonance frequencies are lower than the resonance frequency $\omega_1$ of the original cavity.
Moreover, only eigenvector fields satisfying the periodic boundary conditions appear.
This result is physically reasonable and supports the validity of the formulation developed in this note.

\subsection{General dispersion relation}
We now examine the dispersion relation for a general periodic system, restricting the eigenvector fields on each cross section to a single mode. In other words, we examine how the zeros of $f_k(j\omega)$ depend on $k$.
The starting point is Eq.~\eqref{eq:fk}.
Since $f_k(j\omega)$ and $q(j\omega)$ are purely imaginary for real $\omega$, we introduce the real-valued functions $g_k(\omega)$ and $r(\omega)$:
\begin{eqnarray}
g_k(\omega)\coloneqq f_k(j \omega)/j, \qquad r(\omega)\coloneqq q(j \omega)/j\,.
\end{eqnarray}
The zeros of $r(\omega)$, if they exist, are called ``transmission zeros'' \cite{bib:ant97}.
As we shall see, the presence or absence of transmission zeros leads to two types of dispersion relations.

\subsubsection{Dispersion relation without transmission zero}

\begin{figure}[htb]
   \centering
   \includegraphics*[width=100mm]{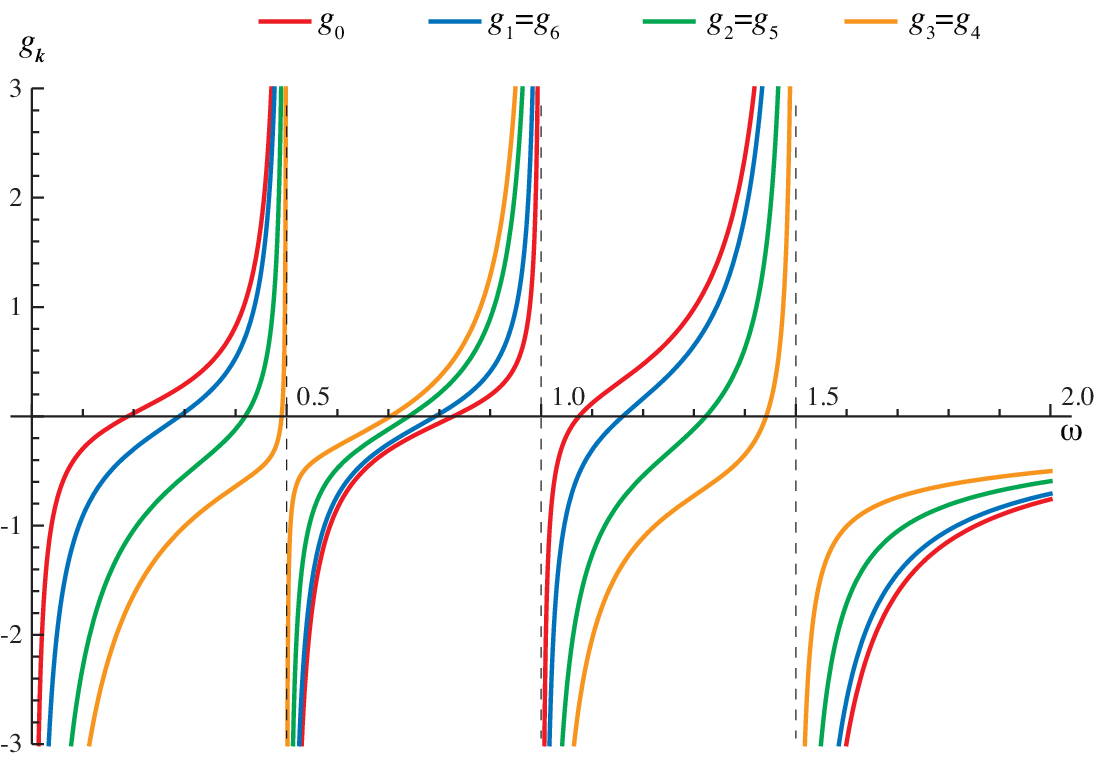}
   \caption{Plots of $g_k(\omega)$ calculated using $\omega_n$ and $I_{nk}$ in Table \ref{tab:cos}. We set $N=7$.}
   \label{fig:diag1}
\end{figure}

\begin{figure}[htb]
   \centering
   \includegraphics*[width=100mm]{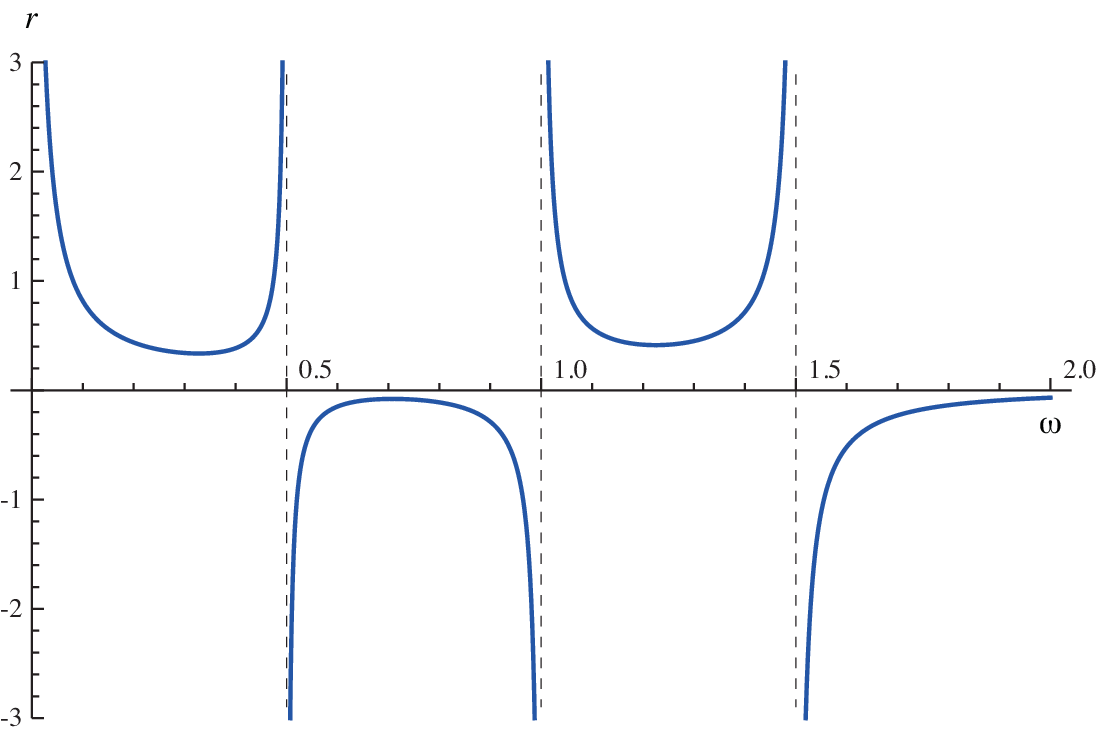}
   \caption{Plot of $r(\omega)$ calculated using $\omega_n$ and $I_{nk}$ in Table \ref{tab:cos}.}
   \label{fig:offdiag1}
\end{figure}

We choose the frequencies $\omega_n$ and coefficients $I_{nk}$ as listed in Table~\ref{tab:cos}.
For weak coupling between cavities, the coupled resonant frequencies cluster near the poles and become difficult to distinguish. We therefore consider relatively strong coupling here.
The characteristics of the dispersion relations discussed below, however, do not depend on the coupling strength.
The functions $g_k(\omega)$ and $r(\omega)$ obtained from Eqs.~\eqref{eq:pk}, \eqref{eq:qk}, and \eqref{eq:fk} are shown in Figs.~\ref{fig:diag1} and \ref{fig:offdiag1}.
These numerical results suggest that, when no transmission zero is present, the residues $I_{n1}^{(1)}\cdot I_{n1}^{(2)}$ at the poles $s=j\omega_n$ alternate in sign as $n$ increases.

\begin{table}[htb]
\caption{Values of $\omega_n$ and coefficients $I_{nk}$ used to obtain Figs. \ref{fig:diag1} and \ref{fig:offdiag1}}\label{tab:cos}
\begin{tabular}{c|cccc}
 $n$&0&1&2&3\\
\tableline
$\omega_n$&0.0&0.5&1.0&1.5\\ 
$I_{n1}^{(1)}$&0.2&0.3&0.4&0.4\\  
$I_{n1}^{(2)}$&-0.4&0.3&-0.2&0.2\\ 
\tableline
\end{tabular}
\end{table}


\begin{figure}[htb]
   \centering
   \includegraphics*[width=100mm]{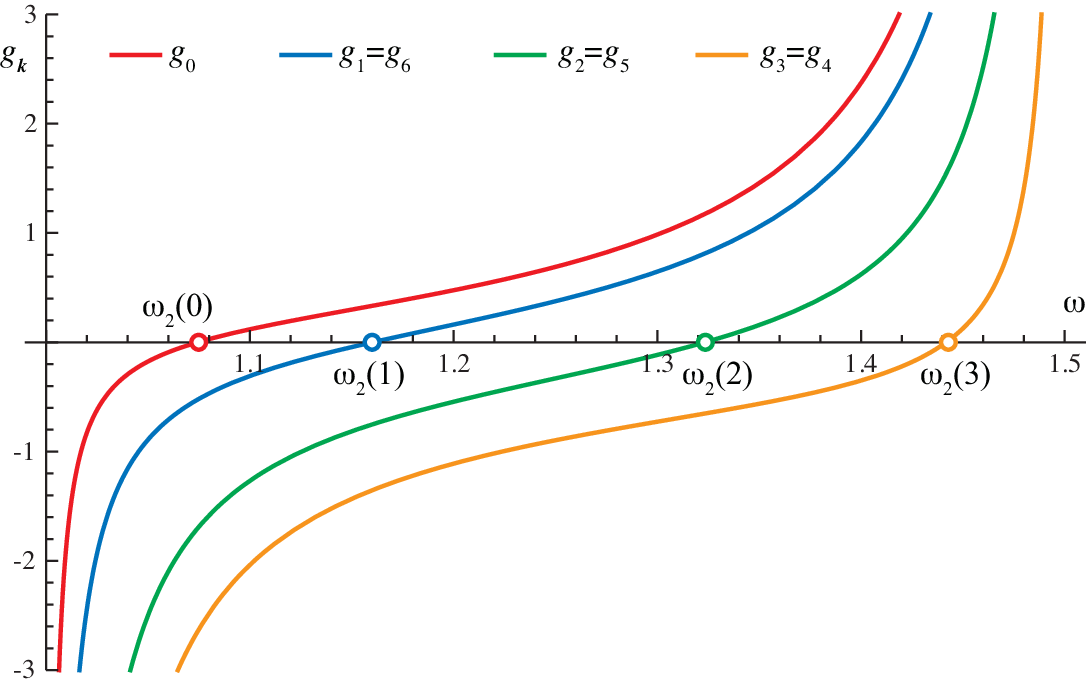}
   \caption{Plots of $g_k(\omega)$, showing an enlargement of the interval
   [$\omega_2, \omega_3$] in Fig. \ref{fig:diag1}.
   The positions of the zeros $\omega_2(k)$ of $g_k(\omega)$ are indicated by open circles.}
   \label{fig:diagzoom1}
\end{figure}
\begin{figure}[htb]
   \centering
   \includegraphics*[width=100mm]{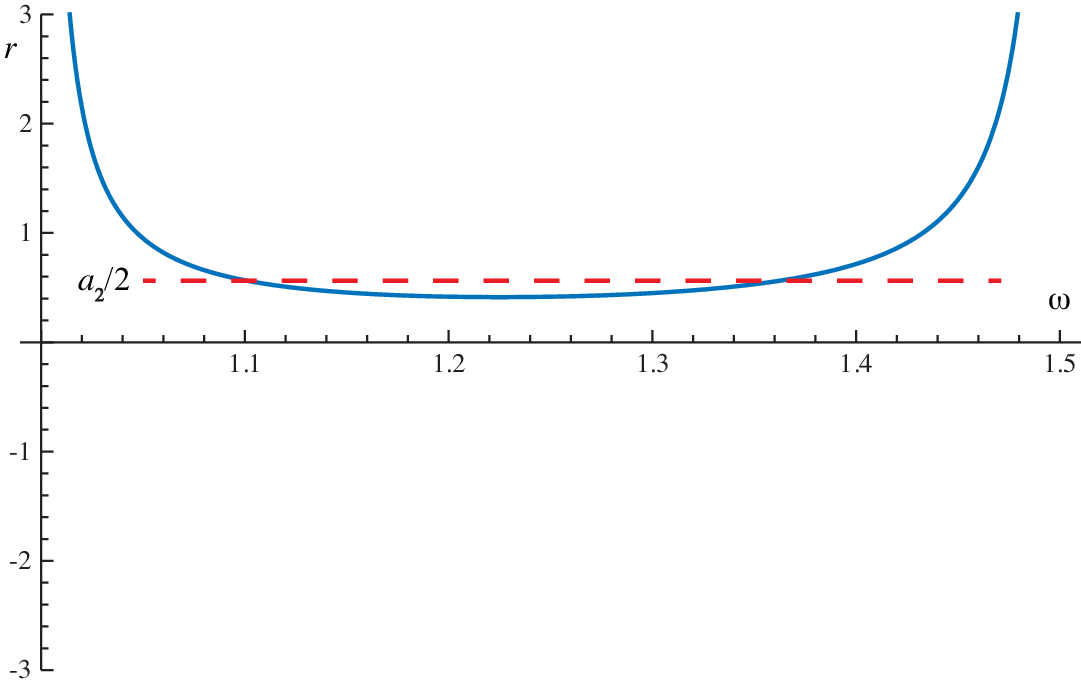}
   \caption{Plot of $r(\omega)$, showing an enlargement of the interval
   [$\omega_2, \omega_3$] in Fig. \ref{fig:offdiag1}.
   The dotted line represents the approximate value $a_2/2$ of $r(\omega)$ in this interval.}
   \label{fig:offdiagzoom1}
\end{figure}
\begin{figure}[htb]
   \centering
   \includegraphics*[width=100mm]{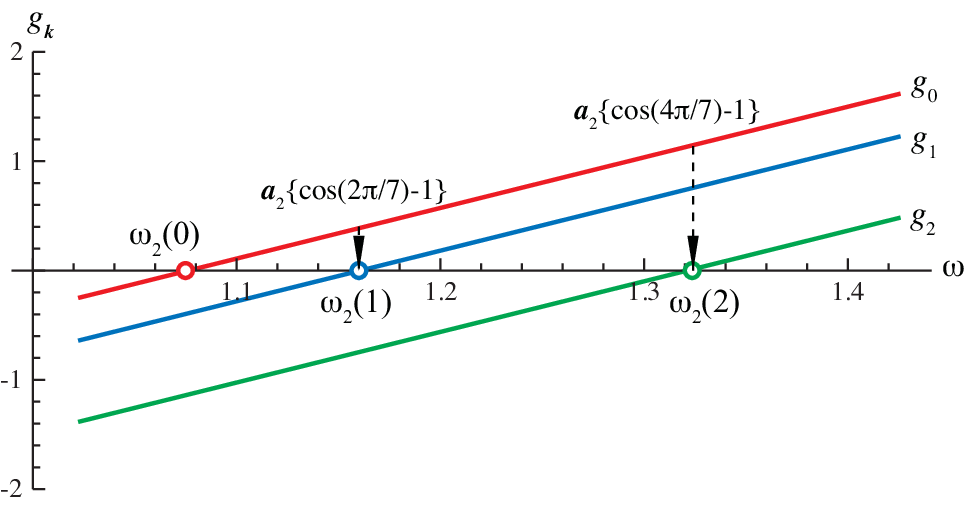}
   \caption{Linear approximations to $g_k(\omega)$ in Fig. \ref{fig:diagzoom1}.}
   \label{fig:approx1}
\end{figure}

Let us focus on the interval [$\omega_n, \omega_{n+1}$].
Taking $n=2$, we show enlarged views of this interval in Figs. \ref{fig:diagzoom1} and \ref{fig:offdiagzoom1}.
Figure~\ref{fig:offdiagzoom1} shows that $r(\omega)$ is nearly constant over most of this interval.
Denoting its approximate value by $a_n/2$, we obtain from Eq.~\eqref{eq:fk} the approximation
\begin{eqnarray}
g_k(\omega)\approx g_0(\omega)+a_n\left\{\cos\left(\frac{2 k \pi}{N}\right)-1\right\}\,.
\end{eqnarray}
Thus, $g_k(\omega)$ can be regarded as a vertical translation of $g_0(\omega)$.
Let $\omega_n(k)$ denote the root of $g_k(\omega)$ in this interval. Approximating each $g_k(\omega)$ by a straight line, as shown in Fig.~\ref{fig:approx1}, gives
\begin{eqnarray}
\omega_n(k)-\omega_n(0)\approx b_n\left\{\cos\left(\frac{2 k \pi}{N}\right)-1\right\}\,,
\end{eqnarray}
where $b_n$ is a constant.
Using this approximation, we obtain the dispersion relation
\begin{eqnarray}
\label{eq:disp1}
\omega_n(k)^2-\omega_n(0)^2&=&\left\{\omega_n(k)+\omega_n(0)\right\}
\left\{\omega_n(k)-\omega_n(0)\right\}\nonumber \\
&\approx& 2 \omega_n(0) b_n\left\{\cos\left(\frac{2 k \pi}{N}\right)-1\right\}\,.
\end{eqnarray}
Here we have used $\omega_n(k)+\omega_n(0)\approx 2\omega_n(0)$.
This approximation is sufficiently accurate when $\left|\omega_n(k)/\omega_n(0)-1\right| \ll 1$.

\begin{figure}[htb]
   \centering
   \includegraphics*[width=150mm]{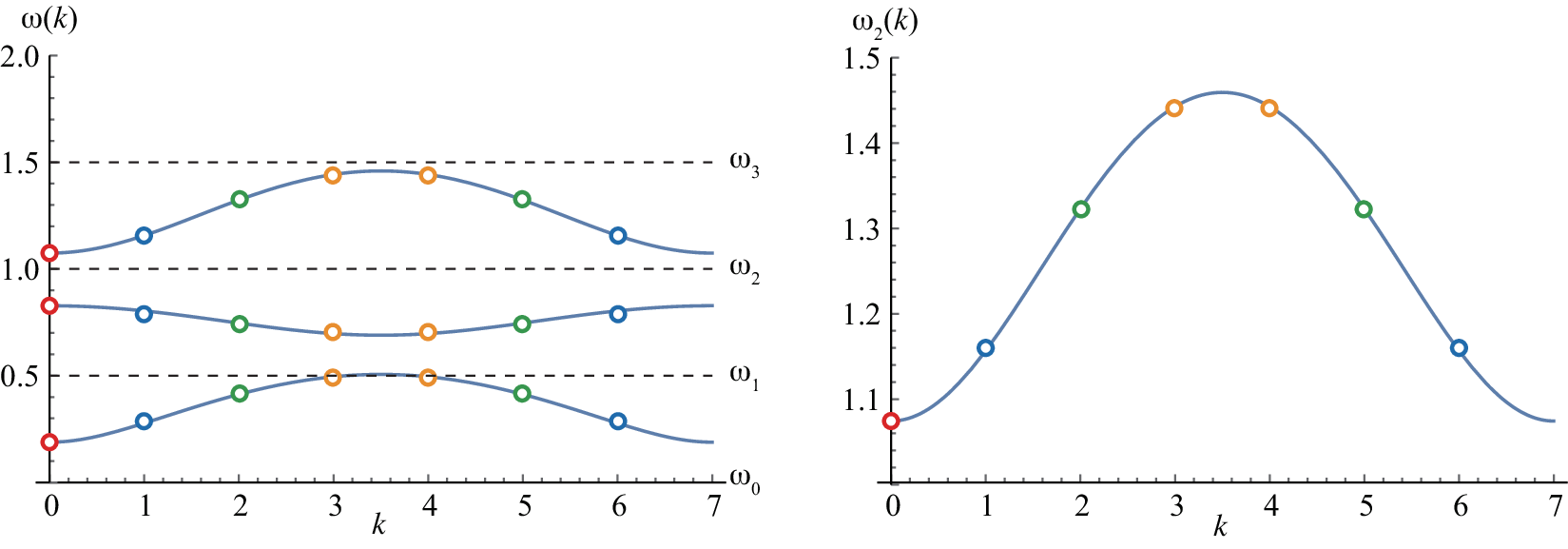}
   \caption{Open circles represent numerical calculations of the dispersion relation $\omega(k)$, and solid lines show the results of fitting with \eqref{eq:disp1}.
   The only free parameter in the fitting is $b_n$.
   The left panel shows three branches, and the right panel is an enlargement of one of them ($\omega_2(k)$).}
   \label{fig:disp1}
\end{figure}

Figure \ref{fig:disp1} compares the dispersion relation
$\omega_n(k)$ obtained by numerical calculation with
a fit using \eqref{eq:disp1}.
It can be seen that the function of the form
\eqref{eq:disp1} gives a good approximation.
The alternating signs of the residues noted above
also explain why upward-convex and downward-convex
dispersion curves appear alternately.

\subsubsection{Dispersion relation with transmission zeros}

\begin{figure}[htb]
   \centering
   \includegraphics*[width=100mm]{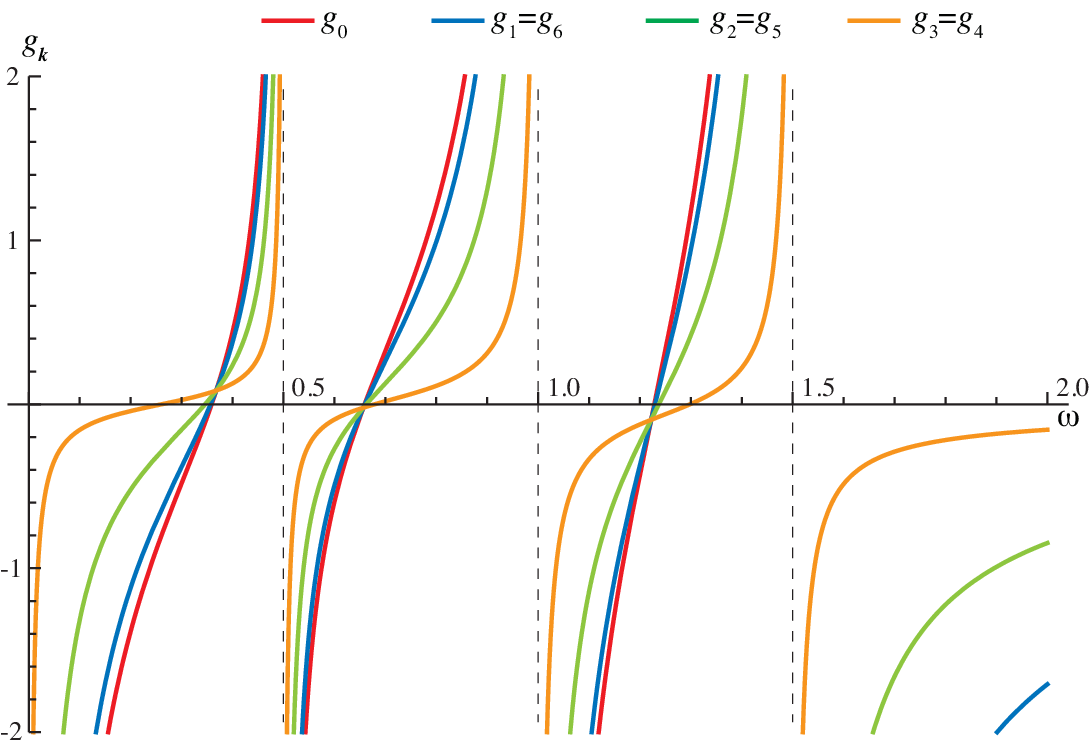}
   \caption{Plots of $g_k(\omega)$ calculated using $\omega_n$ and $I_{nk}$ in Table \ref{tab:new}. We set $N=7$.}
   \label{fig:diag2}
\end{figure}
\begin{figure}[htb]
   \centering
   \includegraphics*[width=100mm]{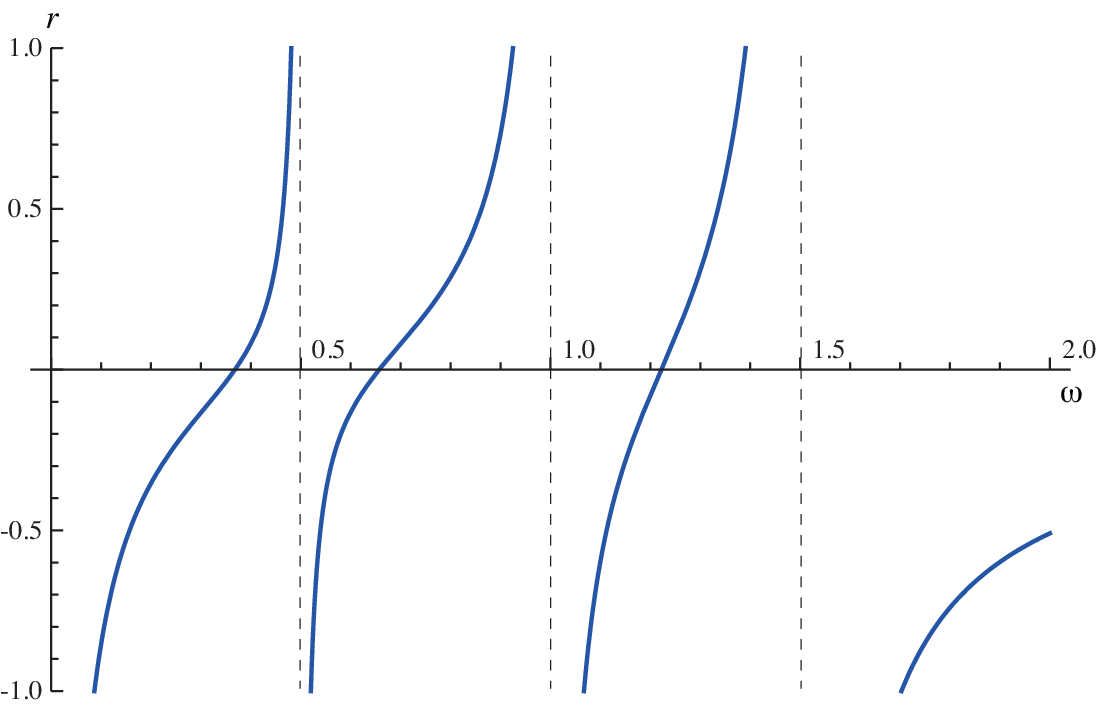}
   \caption{Plot of $r(\omega)$ calculated using $\omega_n$ and $I_{nk}$ in Table \ref{tab:new}.}
   \label{fig:offdiag2}
\end{figure}

Next, we consider the case in which transmission zeros are present.
Figures~\ref{fig:diag2} and \ref{fig:offdiag2} show $g_k(\omega)$ and $r(\omega)$ for this case. The frequencies $\omega_n$ and coefficients $I_{nk}$ used in the calculations are listed in Table~\ref{tab:new}.
These numerical results suggest that, when a transmission zero is present, the residues $I_{n1}^{(1)}\cdot I_{n1}^{(2)}$ have the same sign at least one pair of adjacent poles $s=j\omega_n$.

\begin{table}
\caption{Values of $\omega_n$ and coefficients $I_{nk}$ used to obtain Figs. \ref{fig:diag2} and \ref{fig:offdiag2}}\label{tab:new}
\begin{tabular}{c|cccc}
 $n$&0&1&2&3\\
\tableline
$\omega_n$&0.0&0.5&1.0&1.5\\ 
$I_{n1}^{(1)}$&0.3&0.2&0.3&0.4\\  
$I_{n1}^{(2)}$&0.3&0.3&0.5&0.5\\ 
\tableline
\end{tabular}
\end{table}

\begin{figure}[htb]
   \centering
   \includegraphics*[width=100mm]{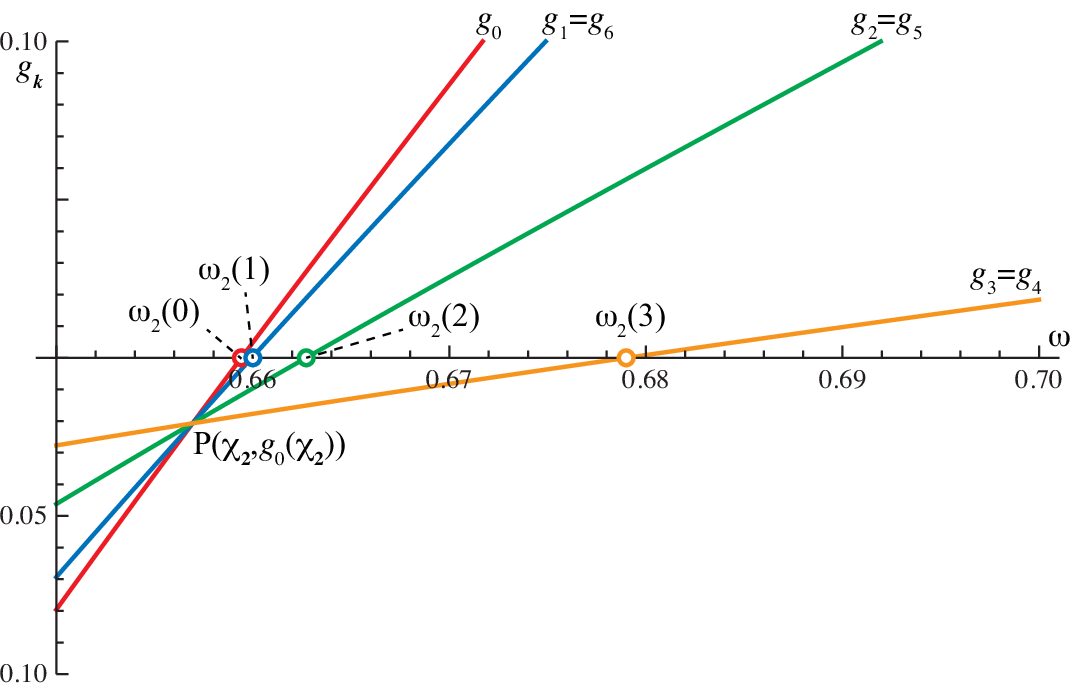}
   \caption{Plots of $g_k(\omega)$, showing an enlargement of the interval
   [$\omega_2, \omega_3$] in Fig. \ref{fig:diag2}.
   The positions of the zeros $\omega_2(k)$ are indicated by open circles.
   All $g_k$ pass through point P in the figure.}
   \label{fig:diagzoom2}
\end{figure}
\begin{figure}[htb]
   \centering
   \includegraphics*[width=100mm]{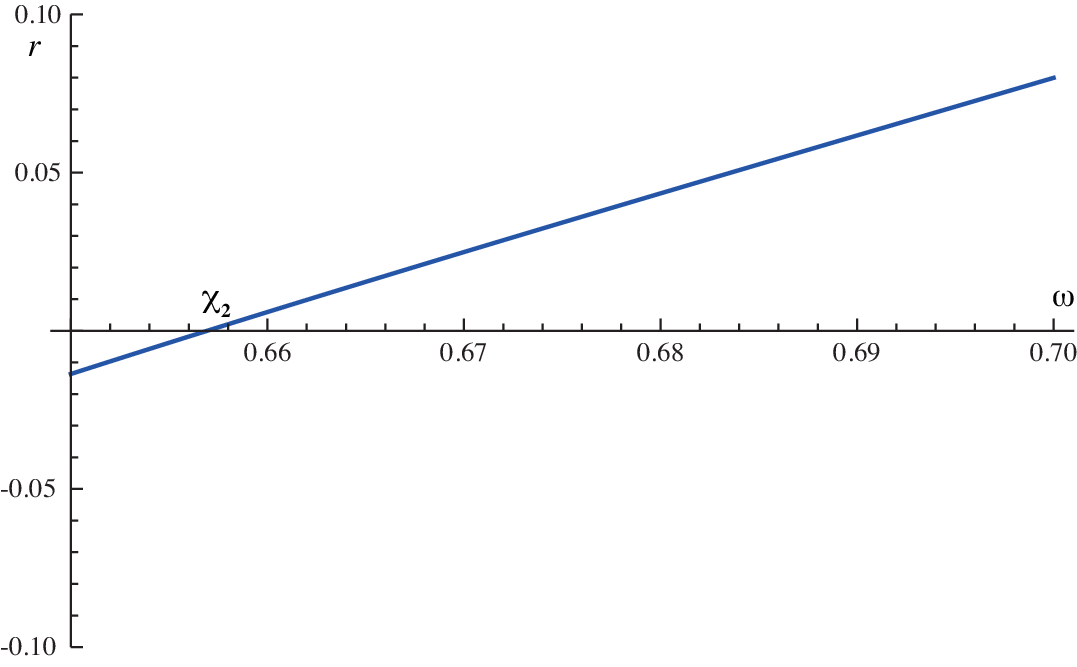}
   \caption{Plot of $r(\omega)$, showing an enlargement of the interval
   [$\omega_2, \omega_3$] in Fig. \ref{fig:offdiag2}.
   The transmission zero is denoted by $\chi_2$, corresponding to $n=2$.}
   \label{fig:offdiagzoom2}
\end{figure}
\begin{figure}[htb]
   \centering
   \includegraphics*[width=100mm]{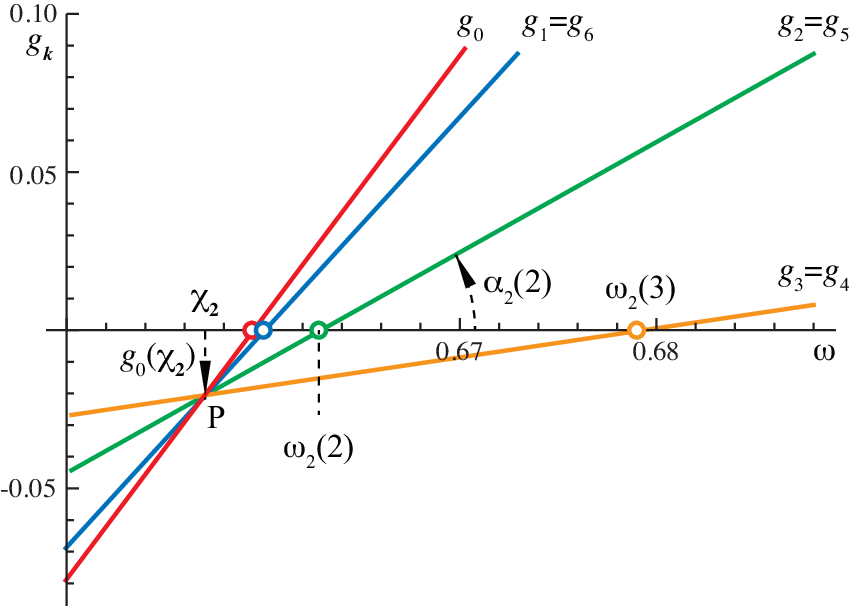}
   \caption{Linear approximations to $g_k(\omega)$.
   Equation \eqref{eq:katamuki} is obtained geometrically.}
   \label{fig:approx2}
\end{figure}

Consider an interval [$\omega_n, \omega_{n+1}$] containing a transmission zero.
Again taking $n=2$, we show enlarged views of the interval in Figs. \ref{fig:diagzoom2} and \ref{fig:offdiagzoom2}.
Let $\omega=\chi_n$ denote the transmission zero
($\chi_2$ in Figs. \ref{fig:diagzoom2} and \ref{fig:offdiagzoom2}).
Since $r(\chi_n)=0$, Eq.~\eqref{eq:fk} shows that all the curves $g_k(\omega)$ pass through the point P at ($\chi_n,g_0(\chi_n)$).
To examine the behavior of $g_k(\omega)$ near point P, we retain only the contribution from a single $\omega_n$.
This approximation need not preserve the exact property that all the curves pass through P.
Thus,
\begin{eqnarray}
g_k(\omega)&\approx&
-\frac{{I_{n1}^{(1)}}^2+{I_{n1}^{(2)}}^2+2 I_{n1}^{(1)}I_{n1}^{(2)}\cos\left(\frac{2 k \pi}{N}\right)}
{\omega_n \mu \left(\frac{\omega}{\omega_n}-\frac{\omega_n}{\omega}\right)}\nonumber\\
&=&-\frac{1}{\mu}\cdot\frac{\omega}{\omega^2-{\omega_n}^2}
\cdot\left[{I_{n1}^{(1)}}^2+{I_{n1}^{(2)}}^2+2 I_{n1}^{(1)}I_{n1}^{(2)}\cos\left(\frac{2 k \pi}{N}\right)\right]\,.
\end{eqnarray}
As illustrated in Fig.~\ref{fig:approx2}, we approximate each $g_k$ by a straight line near P, taking the common point of the original curves as the reference point.
Estimating the slope at P from the single-term approximation gives
\begin{eqnarray}
\frac{dg_k}{d\omega}(\chi_n)&\approx&
\frac{1}{\mu}\cdot\frac{{\chi_n}^2+{\omega_n}^2}{({\chi_n}^2-{\omega_n}^2)^2}
\cdot\left[{I_{n1}^{(1)}}^2+{I_{n1}^{(2)}}^2+2 I_{n1}^{(1)}I_{n1}^{(2)}\cos\left(\frac{2 k \pi}{N}\right)\right]
\nonumber\\
&\coloneqq&\alpha_n(k)\,.
\end{eqnarray}
Here, $\alpha_n(k)$ is nonnegative and, in general, positive.
In the following, we assume that $\alpha_n(k)>0$ for the modes under consideration.

\begin{figure}[htb]
   \centering
   \includegraphics*[width=150mm]{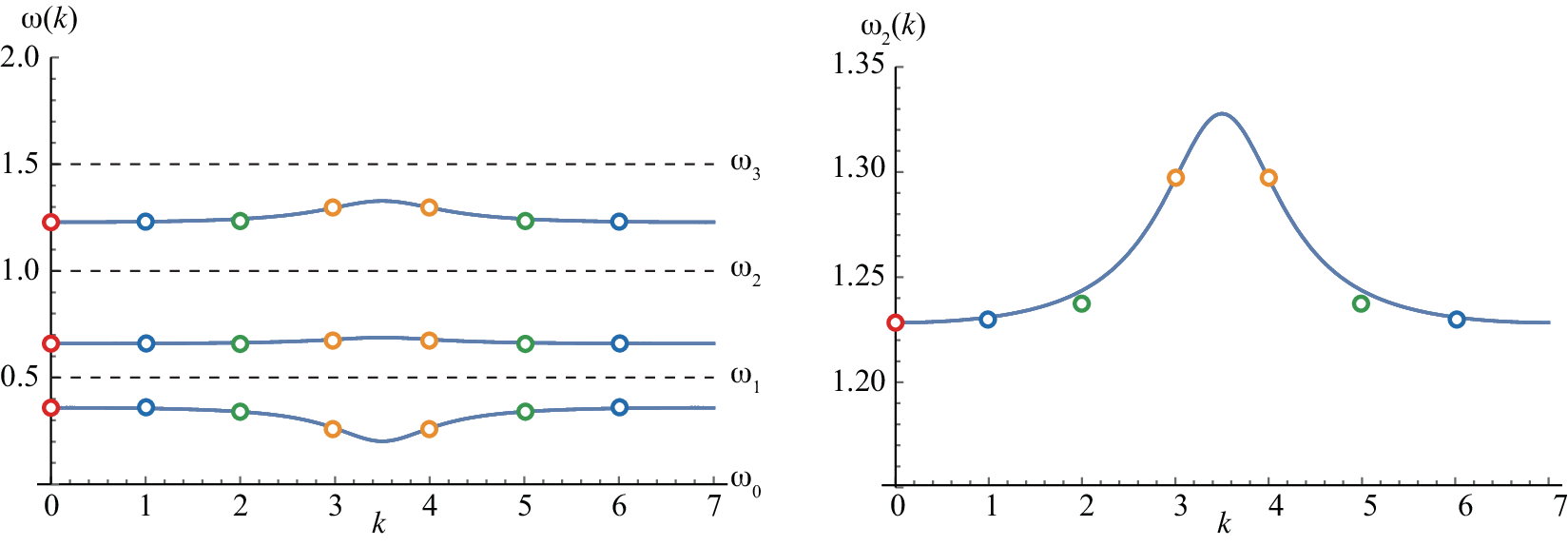}
   \caption{
   Open circles represent numerical calculations of the dispersion relation $\omega(k)$, and solid lines show the results of fitting with \eqref{eq:disp2}.
   The only free parameter in the fitting is $c_n$.
   The left panel shows three branches, and the right panel is an enlargement of one of them ($\omega_2(k)$).}
   \label{fig:disp2}
\end{figure}

Under the linear approximation shown in Fig.~\ref{fig:approx2}, the root $\omega_n(k)$ and the slope $\alpha_n(k)$ are related by
\begin{eqnarray}
\label{eq:katamuki}
\omega_n(k)-\chi_n\approx -\frac{g_0(\chi_n)}{\alpha_n(k)}\,,
\end{eqnarray}
or
\begin{eqnarray}
\omega_n(k)\approx \chi_n-\frac{g_0(\chi_n)}{\alpha_n(k)}\,.
\end{eqnarray}
We then obtain the dispersion relation
\begin{eqnarray}
\label{eq:disp2}
\omega_n(k)^2-\omega_n(0)^2&=&\left\{\omega_n(k)+\omega_n(0)\right\}
\left\{\omega_n(k)-\omega_n(0)\right\}\nonumber \\
&\approx& 2 \omega_n(0) g_0(\chi_n)\left\{\frac{1}{\alpha_n(0)}-\frac{1}{\alpha_n(k)}\right\}\nonumber \\
&=&2 \omega_n(0) g_0(\chi_n)\left\{\frac{\alpha_n(k)-\alpha_n(0)}{\alpha_n(0)\alpha_n(k)}\right\}
\nonumber\\
&=&-\frac{2 \omega_n(0)\chi_n\left({\chi_n}^2-{\omega_n}^2\right)}{{\chi_n}^2+{\omega_n}^2}
\cdot\frac{2I_{n1}^{(1)}I_{n1}^{(2)}\left\{\cos\left(\frac{2 k \pi}{N}\right)-1\right\}}
{\left\{{I_{n1}^{(1)}}^2+{I_{n1}^{(2)}}^2+2 I_{n1}^{(1)}I_{n1}^{(2)}
\cos\left(\frac{2 k \pi}{N}\right)\right\}}\nonumber \\
&=&c_n\frac{K\left\{1-\cos\left(\frac{2 k \pi}{N}\right)\right\}}
{1-K\left\{1-\cos\left(\frac{2 k \pi}{N}\right)\right\}}\,,
\end{eqnarray}
where we have introduced
\begin{eqnarray}
\label{eq:cn}
c_n\coloneqq\frac{2 \omega_n(0)\chi_n\left({\chi_n}^2-{\omega_n}^2\right)}{{\chi_n}^2+{\omega_n}^2}\,,
\end{eqnarray}
and
\begin{eqnarray}
K\coloneqq\frac{2I_{n1}^{(1)}I_{n1}^{(2)}}{\left(I_{n1}^{(1)}+I_{n1}^{(2)}\right)^2}\,.
\end{eqnarray}
In the second line, we also used $\omega_n(k)+\omega_n(0)\approx 2 \omega_n(0)$.
This approximation is expected to be sufficiently accurate when $\left|\omega_n(k)/\omega_n(0)-1\right| \ll 1$.
When $\left|\chi_n/\omega_n-1\right| \ll 1$, Eq. \eqref{eq:cn} simplifies to
\begin{eqnarray}
c_n\approx 2 \omega_n(0)\left(\chi_n-\omega_n\right)\,.
\end{eqnarray}
Figure~\ref{fig:disp2} compares the numerically calculated dispersion relation $\omega_n(k)$ with a fit based on Eq.~\eqref{eq:disp2}.
A function of the form in Eq.~\eqref{eq:disp2} again gives a good approximation.

\subsection{Multidimensional case\label{ap:multi}}

Let us consider a periodic system consisting of $N_c$ resonators, with $N_m$ eigenvector fields on each junction surface (Fig.~\ref{fig:period}).
The admittance matrix $Y(s)$ is then an $(N_c\cdot N_m)\times (N_c\cdot N_m)$ matrix of the following form:
\begin{eqnarray}
Y(s)=\left(
\begin{array}{cccccc}
P(s) & Q(s) & 0 & \cdots & 0 & {Q(s)}^T \\
{Q(s)}^T & P(s) & Q(s) & 0 & \cdots & 0\\
0 & {Q(s)}^T & P(s) & Q(s) & \cdots & 0\\
\vdots & \vdots & \vdots & \ddots & \cdots & \vdots\\
0 & 0 & \cdots & {Q(s)}^T & P(s) & Q(s)\\
Q(s) & 0 & 0 & \cdots & {Q(s)}^T & P(s)
\end{array} 
\right)\,.
\end{eqnarray}
Here, both $P(s)$ and $Q(s)$ are $N_m \times N_m$ matrices.
In terms of Kronecker products, $Y(s)$ becomes
\begin{eqnarray}
Y(s) = I_{N_c} \otimes P(s) +T_{N_c} \otimes Q(s) +{T_{N_c}}^{-1} \otimes {{Q(s)}^T}\,.
\end{eqnarray}
The matrices $P(s)$ and $Q(s)$ are explicitly given by
\begin{eqnarray}
P(s)=
\left(
\begin{array}{cccc}
Y_{1,1}^{(1,1)}(s) & Y_{1,2}^{(1,1)}(s) & \cdots & Y_{1,N_m}^{(1,1)}(s) \\
Y_{2,1}^{(1,1)}(s) & Y_{2,2}^{(1,1)}(s) & \cdots & Y_{2,N_m}^{(1,1)}(s) \\
\vdots & \vdots & \ddots & \vdots\\
Y_{N_m,1}^{(1,1)}(s) & Y_{N_m,2}^{(1,1)}(s) & \cdots & Y_{N_m,N_m}^{(1,1)}(s)
\end{array} 
\right)\,,
\end{eqnarray}
\begin{eqnarray}
Q(s)=
\left(
\begin{array}{cccc}
Y_{1,1}^{(1,2)}(s) & Y_{1,2}^{(1,2)}(s) & \cdots & Y_{1,N_m}^{(1,2)}(s) \\
Y_{2,1}^{(1,2)}(s) & Y_{2,2}^{(1,2)}(s) & \cdots & Y_{2,N_m}^{(1,2)}(s) \\
\vdots & \vdots & \ddots & \vdots\\
Y_{N_m,1}^{(1,2)}(s) & Y_{N_m,2}^{(1,2)}(s) & \cdots & Y_{N_m,N_m}^{(1,2)}(s)
\end{array} 
\right)\,.
\end{eqnarray}
Their components are given as functions of $s$ by
\begin{eqnarray}
Y_{l,k}^{(1,1)}(s)&=&
\sum_n\frac{I_{nl}^{(1)}I_{nk}^{(1)}+I_{nl}^{(2)}I_{nk}^{(2)}}
{\mu \omega_n \left(\frac{s}{\omega_n}+\frac{\omega_n}{s}\right)}
+\sum_\nu \frac{I_{\nu l}^{(1)}I_{\nu k}^{(1)}+I_{\nu l}^{(2)}I_{\nu k}^{(2)}}{\mu s}\,,
\\
Y_{l,k}^{(1,2)}(s)&=&
\sum_n\frac{I_{nl}^{(1)}I_{nk}^{(2)}}
{\mu \omega_n \left(\frac{s}{\omega_n}+\frac{\omega_n}{s}\right)}
+\sum_\nu \frac{I_{\nu l}^{(1)}I_{\nu k}^{(2)}}{\mu s}.
\end{eqnarray}

Separating the dependence on $s$, we may write $Y(s)$ as
\begin{eqnarray}
\label{eq:prmat}
Y(s)=
\sum_n\frac{1}
{\mu \omega_n \left(\frac{s}{\omega_n}+\frac{\omega_n}{s}\right)}R_n
+\sum_\nu \frac{1}{\mu s}R_{\nu}\,.
\end{eqnarray}
Here, $R_n$ is the real symmetric matrix
\begin{eqnarray}
\label{eq:residue}
R_n &=& I_{N_c} \otimes V_n +T_{N_c} \otimes W_n +{T_{N_c}}^{-1} \otimes {{W_n}^T} \\
&=&\left(
\begin{array}{cccccc}
V_n & W_n & 0 & \cdots & 0 & {W_n}^T \\
{W_n}^T & V_n & W_n & 0 & \cdots & 0\\
0 & {W_n}^T & V_n & W_n & \cdots & 0\\
\vdots & \vdots & \vdots & \ddots & \cdots & \vdots\\
0 & 0 & \cdots & {W_n}^T & V_n & W_n\\
W_n & 0 & 0 & \cdots & {W_n}^T & V_n
\end{array} 
\right)\,.
\end{eqnarray}
Here, the matrices $V_n$ and $W_n$ are given by
\begin{eqnarray}
V_n=
\left(
\begin{array}{ccc}
I_{n1}^{(1)}I_{n1}^{(1)}+I_{n1}^{(2)}I_{n1}^{(2)} 
& \cdots & I_{n1}^{(1)}I_{nN_m}^{(1)}+I_{n1}^{(2)}I_{nN_m}^{(2)} \\
\vdots &  \ddots & \vdots\\
I_{nN_m}^{(1)}I_{n1}^{(1)}+I_{nN_m}^{(2)}I_{n1}^{(2)}
& \cdots &  I_{nN_m}^{(1)}I_{nN_m}^{(1)}+I_{nN_m}^{(2)}I_{nN_m}^{(2)}
\end{array} 
\right)\,,
\end{eqnarray}
\begin{eqnarray}
W_n=
\left(
\begin{array}{ccc}
I_{n1}^{(1)}I_{n1}^{(2)} 
& \cdots & I_{n1}^{(1)}I_{nN_m}^{(2)} \\
\vdots &  \ddots & \vdots\\
I_{nN_m}^{(1)}I_{n1}^{(2)}  
& \cdots &  I_{nN_m}^{(1)}I_{nN_m}^{(2)}
\end{array} 
\right)\,.
\end{eqnarray}
The matrix $R_\nu$ can be expressed similarly.
Equation~\eqref{eq:prmat} is the standard form of a lossless positive-real matrix, and $R_n$ and $R_\nu$ are called residue matrices
\footnote{
A matrix-valued function $\bm{Z}(s)$ of the complex frequency $s=\sigma+j\omega$ is called a positive-real matrix if each component is analytic in the right half-plane $\mathrm{Re}\,s>0$, if $\bm{Z}(s^*)=\bm{Z}(s)^*$ holds, and if, furthermore,
\[
\frac{\bm{Z}(s)+\bm{Z}(s)^\dagger}{2}
\succeq 0
\]
holds.
Here, $\dagger$ denotes the Hermitian conjugate, and $\succeq 0$ denotes positive semidefiniteness.
The impedance matrix of a passive multiport circuit is a positive-real matrix.
In particular, in the lossless case, $\bm{Z}(j\omega)+\bm{Z}(j\omega)^\dagger=0$ holds except at poles on the imaginary axis.
For a lossless real-rational positive-real matrix, all finite poles are simple poles on the imaginary axis, and the residue matrix at each pole is Hermitian positive semidefinite~\cite{bib:and73}.
}.

Importantly, all the residue matrices have translational symmetry.
More precisely, defining
\begin{eqnarray}
{\widetilde{T}}_{N_c}\coloneqq ({T_{N_c\cdot N_m}})^{N_m}=T_{N_c} \otimes I_{N_m}\,,
\end{eqnarray}
we have
\begin{eqnarray}
\label{eq:trans}
{\widetilde{T}}_{N_c}\cdot R_n=R_n \cdot {\widetilde{T}}_{N_c}\,, \qquad
{\widetilde{T}}_{N_c}\cdot R_{\nu}=R_{\nu} \cdot {\widetilde{T}}_{N_c}\,.
\end{eqnarray}
This property allows $Y(s)$ to be decomposed into blocks.
We first note that
\begin{eqnarray}
({\widetilde{T}}_{N_c})^{N_c}=I_{N_c\cdot N_m}\,.
\end{eqnarray}
Consequently, defining
\begin{eqnarray}
\label{eq:lambda_def}
\lambda\coloneqq\exp(2j\pi/N_c)\,,
\end{eqnarray}
we find that the eigenvalues of ${\widetilde{T}}_{N_c}$ are $\lambda^k$ $(k=0,1,\ldots,N_c-1)$.
Each eigenvalue is $N_m$-fold degenerate, and the eigenvectors corresponding to $\lambda^k$ can be written using a Kronecker product as follows.
\begin{eqnarray}
\label{eq:lev}
\bm{\xi}_{(k,i)}&\coloneqq&\frac{1}{\sqrt{N_c}}
{(\underbrace{\overbrace{0,...,0,1}^{i},0,...,0}_{N_m},
\underbrace{\overbrace{0,...,0, \lambda^k}^{i},0,...,0}_{N_m},
\cdots,
\underbrace{\overbrace{0,...,0, \lambda^{(N_c-1)k}}^{i},0,...,0}_{N_m})^T} \nonumber \\
&=& 
\frac{1}{\sqrt{N_c}}{(1, \lambda^k, \lambda^{2k},...,\lambda^{(N_c-1)k})^T}
\otimes
{(\underbrace{\overbrace{0,...,0,1}^{i},0,...,0}_{N_m})^T}, \qquad (i=1,...,N_m)\,.
\end{eqnarray}
These vectors form an orthonormal basis. Denoting the Hermitian inner product by $\langle \bm{u},\bm{v}\rangle$, we have
\begin{eqnarray}
\langle \bm{\xi}_{(k,i)}, \bm{\xi}_{(m,j)} \rangle = \delta_{km}\cdot \delta_{ij}\,.
\end{eqnarray}
By Eq.~\eqref{eq:trans}, we also have
\begin{eqnarray}
\langle \bm{\xi}_{(k,i)}, {\widetilde{T}}_{N_c}\cdot R_n\bm{\xi}_{(m,j)} \rangle
=\langle \bm{\xi}_{(k,i)}, R_n \cdot {\widetilde{T}}_{N_c} \bm{\xi}_{(m,j)} \rangle\,.
\end{eqnarray}
The right-hand side is
\begin{eqnarray}
\langle \bm{\xi}_{(k,i)}, R_n \cdot {\widetilde{T}}_{N_c} \bm{\xi}_{(m,j)} \rangle
=\lambda^m \langle \bm{\xi}_{(k,i)}, R_n \bm{\xi}_{(m,j)} \rangle\,,
\end{eqnarray}
while, since ${\widetilde{T}}_{N_c}$ is real, the left-hand side becomes
\begin{eqnarray}
\langle \bm{\xi}_{(k,i)}, {\widetilde{T}}_{N_c}\cdot R_n \bm{\xi}_{(m,j)} \rangle
&=&\langle ({\widetilde{T}}_{N_c})^T \bm{\xi}_{(k,i)}, R_n \bm{\xi}_{(m,j)} \rangle \nonumber \\
&=&\langle \lambda^{-k} \bm{\xi}_{(k,i)}, R_n \bm{\xi}_{(m,j)} \rangle \nonumber \\
&=&\lambda^k \langle \bm{\xi}_{(k,i)}, R_n \bm{\xi}_{(m,j)} \rangle\,.
\end{eqnarray}
It follows that the residue matrices are block diagonal in this basis:
\begin{eqnarray}
\langle \bm{\xi}_{(k,i)}, R_n  \bm{\xi}_{(m,j)} \rangle
=\delta_{km} \langle \bm{\xi}_{(k,i)}, R_n \bm{\xi}_{(m,j)} \rangle\,.
\end{eqnarray}
To express this explicitly, we use the unitary matrix $U_{N_c}$ of Eq.~\eqref{eq:un} to define
\begin{eqnarray}
{\widetilde{U}}_{N_c}\coloneqq U_{N_c} \otimes I_{N_m}\,.
\end{eqnarray}
Then, for each term in Eq.~\eqref{eq:residue},
\begin{eqnarray}
\left( I_{N_c} \otimes V_n \right) \cdot \widetilde{U}_{N_c} =
\left( I_{N_c} \otimes V_n \right) \cdot \left( U_{N_c} \otimes I_{N_m} \right)=
U_{N_c} \otimes V_n\,,
\end{eqnarray}
\begin{eqnarray}
\left( T_{N_c} \otimes W_n \right) \cdot \widetilde{U}_{N_c} =
\left( T_{N_c} \otimes W_n \right) \cdot \left( U_{N_c} \otimes I_{N_m} \right)=
\left(U_{N_c} \cdot \Lambda \right) \otimes W_n\,,
\end{eqnarray}
\begin{eqnarray}
\left( {T_{N_c}}^{-1} \otimes {{W_n}^T} \right) \cdot \widetilde{U}_{N_c} =
\left( {T_{N_c}}^{-1} \otimes {{W_n}^T} \right) \cdot \left( U_{N_c} \otimes I_{N_m} \right)=
\left(U_{N_c} \cdot \Lambda^{-1} \right) \otimes {{W_n}^T}
\end{eqnarray}
hold, where $\Lambda$ is the diagonal matrix
\begin{eqnarray}
\Lambda=
\left(
\begin{array}{ccccc}
1 &  &  &  &  \\
& \lambda  &  &  &  \\
&  &  \lambda^2 &  &  \\
&  &  & \ddots & \\
&  &  &  & \lambda^{N_c-1}
\end{array} 
\right)\,.
\end{eqnarray}
Therefore,
\begin{eqnarray}
\label{eq:hn}
(\widetilde{U}_{N_c})^{-1}\cdot R_n \cdot {\widetilde{U}}_{N_c}&=&
I_{N_c} \otimes V_n + \Lambda \otimes W_n + {\Lambda}^{-1} \otimes {{W_n}^T}\nonumber \\
&=&\left(
\begin{array}{ccccc}
H_{n,0} &  &  &  &  \\
& H_{n,1}  &  &  &  \\
&  &  H_{n,2}  & \\
&  &  & \ddots & \\
&  &  &  & H_{n,N_c-1}
\end{array} 
\right)
\end{eqnarray}
is block diagonal, where
\begin{eqnarray}
\label{eq:hnk_def}
H_{n,k}\coloneqq V_n+\lambda^k W_n+\bar{\lambda}^k {{W_n}^T}\,,
\end{eqnarray}
and each of these matrices is Hermitian.
The admittance matrix $Y(s)$ is likewise block diagonalized:
\begin{eqnarray}
(\widetilde{U}_{N_c})^{-1}\cdot Y(s) \cdot {\widetilde{U}}_{N_c}&=&
I_{N_c} \otimes P(s) + \Lambda \otimes Q(s) + {\Lambda}^{-1} \otimes {{Q(s)}^T}\nonumber \\
&=&\left(
\begin{array}{ccccc}
F_0(s) &  &  &  &  \\
& F_1(s)  &  &  &  \\
&  &  F_2(s)  & \\
&  &  & \ddots & \\
&  &  &  & F_{N_c-1}(s)
\end{array} 
\right)\,,
\end{eqnarray}
and
\begin{eqnarray}
\label{eq:fks}
\label{eq:yblk}
F_k(s)=
\sum_n\frac{1}
{\mu \omega_n \left(\frac{s}{\omega_n}+\frac{\omega_n}{s}\right)}H_{n,k}
+\sum_\nu \frac{1}{\mu s}H_{\nu,k}\,.
\end{eqnarray}
Since $H_{n,k}$ is a Hermitian matrix, $F_k(s)$ is also in the standard form of a lossless positive-real matrix~\cite{bib:and73}.

We next examine the properties of the residue matrix $H_{n,k}$.
Expanding Eq.~\eqref{eq:hnk_def} gives a simple expression for $H_{n,k}$.
Define the vectors $\bm{p}$ and $\bm{q}$ by
\begin{eqnarray}
 \bm{p}=\begin{pmatrix} I_{n1}^{(1)}\\ I_{n2}^{(1)}\\\vdots\\ I_{nN_m}^{(1)}\end{pmatrix}\,,\qquad
 \bm{q}=\begin{pmatrix} I_{n1}^{(2)}\\ I_{n2}^{(2)}\\\vdots\\ I_{nN_m}^{(2)}\end{pmatrix}\,,
\end{eqnarray}
and define
\begin{eqnarray}
 \bm{u}=\bm{p}+e^{-j k\phi}\bm{q}\,.
\end{eqnarray}
Then $H_{n,k}$ can be expressed as a Kronecker product:
\begin{eqnarray}
H_{n,k}=\bm{u}\otimes {\bm{u}^{\ast}}^T\,,
\end{eqnarray}
where we have used Eq.~\eqref{eq:lambda_def} and set $\phi\coloneqq 2\pi/N_c$.
Such a matrix has rank one provided $\bm{u}\neq\bm{0}$\footnote{
Let us see this.
Consider the $N$-dimensional column vectors
\begin{eqnarray*}
 \bm{u}=\begin{pmatrix}u_1\\u_2\\\vdots\\u_N\end{pmatrix}\,,\qquad
 \bm{v}=\begin{pmatrix}v_1\\v_2\\\vdots\\v_N\end{pmatrix}\,,
\end{eqnarray*}
and define the matrix formed from their Kronecker product as
\begin{eqnarray*}
 M=\bm{u}\otimes {\bm{v}}^T
 =\begin{pmatrix}
 u_1v_1&u_1v_2&\cdots&u_1v_N\\
 u_2v_1&u_2v_2&\cdots&u_2v_N\\
 \vdots&\vdots&\ddots&\vdots\\
 u_Nv_1&u_Nv_2&\cdots&u_Nv_N
 \end{pmatrix}\,.
\end{eqnarray*}
The $j$th column of $M$ is
\begin{eqnarray*}
 \begin{pmatrix}u_1v_j\\u_2v_j\\\vdots\\u_Nv_j\end{pmatrix}
 =v_j\bm{u}\,.
\end{eqnarray*}
Therefore,
\begin{eqnarray*}
 M=\begin{pmatrix}v_1\bm{u}&v_2\bm{u}&\cdots&v_N\bm{u}\end{pmatrix}\,.
\end{eqnarray*}
Thus every column is a scalar multiple of the same vector $\bm{u}$, so that there is at most one linearly independent column vector.
Furthermore, if $\bm{u}\neq\bm{0}$ and $\bm{v}\neq\bm{0}$, at least one $v_j$ is nonzero and $M$ has a nonzero column.
In this case,
\begin{eqnarray*}
{\rm rank}(M)= 1\,.
\end{eqnarray*}
This result also follows from the identity relating the Kronecker product and rank,
\begin{eqnarray*}
{\rm rank}(A\otimes B)= {\rm rank}(A) {\rm rank}(B)\,.
\end{eqnarray*}
}.
Moreover,
\begin{eqnarray}
H_{n,k}\cdot\bm{u}=\bm{u}\otimes{\bm{u}^{\ast}}^T\cdot\bm{u}=\bm{u}||\bm{u}||^2\,,
\end{eqnarray}
and hence the eigenvalue is
\begin{eqnarray}
||\bm{u}||^2=\sum_{l=1}^{N_m}\left[{I_{nl}^{(1)}}^2+{I_{nl}^{(2)}}^2+
2I_{nl}^{(1)}I_{nl}^{(2)}\cos\left(\frac{2k\pi}{N_c}\right)\right]\,,
\end{eqnarray}
which is nonnegative (and, in general, positive).
This is consistent with the positive-real property of $F_k(s)$.

Returning to the coupled system, its resonant frequencies are determined, as in the one-dimensional case, by the vanishing of the determinant of each block:
\begin{eqnarray}
\det\left[F_k(j\omega)\right]=0\,.
\end{eqnarray}
With a single eigenvector field, each block was
a scalar function, whereas here $F_k(j\omega)$
is an $N_m\times N_m$ matrix.
At first sight, this might suggest that increasing
the number of eigenvector fields on the junction
surface also increases the number of roots.
As we shall see, however, the rank-one structure
of the residue matrices ensures that at most one
eigenvalue branch diverges at each pole.

\begin{table}
\caption{Values of $\omega_n$ and coefficients $I_{nk}$ used for Fig.~\ref{fig:multi}}\label{tab:multi}
\begin{tabular}{c|cccc}
 $n$&0&1&2&3\\
\tableline
$\omega_n$&0.0&0.5&1.0&1.5\\ 
$I_{n1}^{(1)}$&0.1&0.2&0.3&0.4\\  
$I_{n2}^{(1)}$&0.1&0.4&0.5&0.3\\  
$I_{n3}^{(1)}$&0.2&0.5&0.6&0.4\\  
$I_{n1}^{(2)}$&0.2&0.3&0.4&0.3\\ 
$I_{n2}^{(2)}$&0.3&0.3&0.5&0.3\\ 
$I_{n3}^{(2)}$&0.4&0.4&0.6&0.4\\ 
\tableline
\end{tabular}
\end{table}

\begin{figure}[h]
   \centering
   \includegraphics*[width=85mm]{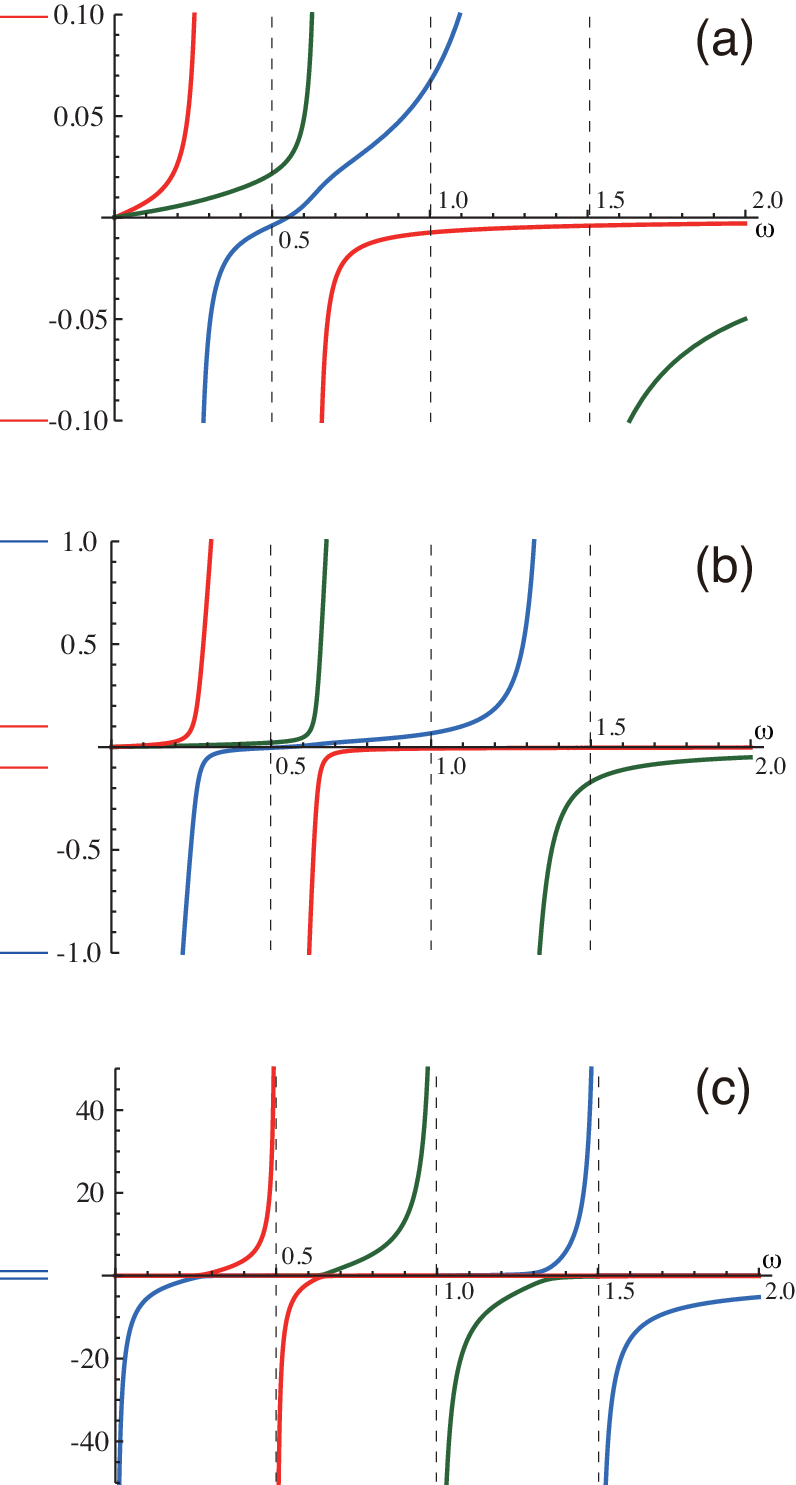}
   \caption{An example of the behavior of the coupled resonant frequencies when the number of eigenvector fields on the cross section is increased.
   With $N_c=4$ and $N_m=3$, the eigenvalues of $F_0(j \omega)$ were calculated as functions of $\omega$ using the coefficients $I_{nk}$ given in Table~\ref{tab:multi}, and plotted with different full-scale ranges on the vertical axis.
   The full-scale ranges are (a): $\pm 0.1$, (b): $\pm 1$, and (c): $\pm 50$.
   In (a), some branches can be seen to ``pass over'' the poles.
   When the scale is reduced to that of (c), it becomes clear that at most one branch diverges at a given pole.}
   \label{fig:multi}
\end{figure}

Figure~\ref{fig:multi} shows the behavior of the eigenvalues of $F_k(j\omega)$.
The eigenvalues generally form $N_m$ branches. However, not every branch diverges at each pole: at most one branch is seen to diverge at a given pole.
This is due to the fact that {\bf the residue matrix $H_{n,k}$ of $F_k(j\omega)$ has rank one}.
This property is discussed in Refs.~\cite{bib:oon50, bib:ced57} in connection with causality as embodied in positive-real matrices.

Finally, as in the single-eigenvector-field case, we examine whether diagonalization in a real basis is possible.
Let $k$ be an integer different from zero and, when $N_c$ is even, also different from $N_c/2$.
Then, for the eigenvalues, clearly,
\begin{eqnarray}
\lambda^k=\overline{\lambda^{N_c-k}}\,,
\end{eqnarray}
and, from Eq.~\eqref{eq:lev}, for the eigenvectors,
\begin{eqnarray}
\label{eq:xi}
\bm{\xi}_{(k,i)}=\overline{\bm{\xi}_{(N_c-k,i)}}
\end{eqnarray}
hold.
Consequently, the $k$th and $(N_c-k)$th blocks in Eq.~\eqref{eq:hn} are complex conjugates,
\begin{eqnarray}
H_{n,k}=\overline{H_{n,N_c-k}}\,.
\end{eqnarray}
Using this in Eq.~\eqref{eq:fks}, we obtain, on the imaginary axis $s=j\omega$,
\begin{eqnarray}
\label{eq:lfk}
F_k(j\omega)/j=\overline{F_{N_c-k}(j\omega)/j}\,.
\end{eqnarray}
The matrix $F_k(j\omega)/j$ is an $N_m \times N_m$ Hermitian matrix and can be diagonalized by a unitary matrix $U_{(k)}(j \omega)$ of the same size. Namely,
\begin{eqnarray}
\label{eq:fkdiag}
U_{(k)}(j \omega)^{-1}\cdot F_k(j\omega)/j\cdot U_{(k)}(j \omega)=
\left(
\begin{array}{ccccc}
g_{k,1}(\omega) &  &  &    \\
& g_{k,2}(\omega)  &  &    \\
&  &   \ddots & \\
&  &   & g_{k,N_m}(\omega)
\end{array} 
\right)\,,
\end{eqnarray}
where all the diagonal elements are real functions of $\omega$.
Using the orthonormal eigenvectors of $F_k(j\omega)/j$, defined by
\begin{eqnarray}
\label{eq:eta}
(\bm{\eta}_{(k,1)}(\omega),\cdots,\bm{\eta}_{(k,N_m)}(\omega)) \coloneqq 
(\bm{\xi}_{(k,1)},\cdots,\bm{\xi}_{(k,N_m)})\cdot U_{(k)}(j \omega)\,,
\end{eqnarray}
we can write
\begin{eqnarray}
Y(j\omega)/j&\cdot&(\bm{\eta}_{(k,1)}(\omega),\cdots,\bm{\eta}_{(k,N_m)}(\omega)) \nonumber\\
&=&(\bm{\eta}_{(k,1)}(\omega),\cdots,\bm{\eta}_{(k,N_m)}(\omega)) \cdot
\left(
\begin{array}{ccccc}
g_{k,1}(\omega) &  &  &    \\
& g_{k,2}(\omega)  &  &    \\
&  &   \ddots & \\
&  &   & g_{k,N_m}(\omega)
\end{array} 
\right)\,.
\end{eqnarray}
On the other hand, taking the complex conjugate of Eq.~\eqref{eq:fkdiag} and using Eq.~\eqref{eq:lfk}, we obtain
\begin{eqnarray}
\label{eq:nck}
\overline{U_{(k)}(j \omega)}^{-1}\cdot \overline{F_k(j\omega)/j}\cdot 
\overline{U_{(k)}(j \omega)}&=&
\overline{U_{(k)}(j \omega)}^{-1}\cdot F_{N_c-k}(j\omega)/j\cdot 
\overline{U_{(k)}(j \omega)}\nonumber \\
&=&\left(
\begin{array}{ccccc}
g_{k,1}(\omega) &  &  &  &  \\
& g_{k,2}(\omega)  &  &  &  \\
&  &  g_{k,3}(\omega)  & \\
&  &  & \ddots & \\
&  &  &  & g_{k,N_m}(\omega)
\end{array} 
\right)\,.
\end{eqnarray}
Here, using Eqs.~\eqref{eq:xi} and \eqref{eq:eta}, we have
\begin{eqnarray}
\label{eq:eta1}
(\bm{\xi}_{(N_c-k,1)},\cdots,\bm{\xi}_{(N_c-k,N_m)})\cdot \overline{U_{(k)}(j \omega)}
&=&
(\overline{\bm{\xi}_{(k,1)}},\cdots,\overline{\bm{\xi}_{(k,N_m)}})\cdot \overline{U_{(k)}(j \omega)}\nonumber \\
&=&(\overline{\bm{\eta}_{(k,1)}(\omega)},\cdots,\overline{\bm{\eta}_{(k,N_m)}(\omega)}).
\end{eqnarray}
Thus, Eq.~\eqref{eq:nck} implies
\begin{eqnarray}
Y(j\omega)/j&\cdot&(\overline{\bm{\eta}_{(k,1)}(\omega)},\cdots,\overline{\bm{\eta}_{(k,N_m)}(\omega)})
\nonumber\\
&=&(\overline{\bm{\eta}_{(k,1)}(\omega)},\cdots,\overline{\bm{\eta}_{(k,N_m)}(\omega)}) \cdot
\left(
\begin{array}{ccccc}
g_{k,1}(\omega) &  &  &    \\
& g_{k,2}(\omega)  &  &    \\
&  &   \ddots & \\
&  &   & g_{k,N_m}(\omega)
\end{array} 
\right)\,.
\end{eqnarray}
Therefore, rearranging the order of the basis vectors, we obtain
\begin{eqnarray}
Y(j\omega)/j&\cdot&(\bm{\eta}_{(k,l)}(\omega),\overline{\bm{\eta}_{(k,l)}(\omega)})
\nonumber\\
&=&(\bm{\eta}_{(k,l)}(\omega),\overline{\bm{\eta}_{(k,l)}(\omega)}) \cdot
\left(
\begin{array}{cc}
g_{k,l}(\omega) & \\
& g_{k,l}(\omega) \\
\end{array} 
\right)\,.
\end{eqnarray}
Furthermore, applying the following unitary transformation to the basis,
\begin{eqnarray}
\label{eq:eta2}
(\bm{u}_{(k,l)}(\omega),\bm{v}_{(k,l)}(\omega)) \coloneqq 
(\bm{\eta}_{(k,l)}(\omega),\overline{\bm{\eta}_{(k,l)}(\omega)})\cdot
\left(
\begin{array}{cc}
1/\sqrt{2} & -j/\sqrt{2}\\
1/\sqrt{2} & j/\sqrt{2} \\
\end{array} 
\right)\,,
\end{eqnarray}
the vectors $\bm{u}_{(k,l)}(\omega)$ and $\bm{v}_{(k,l)}(\omega)$ are real, and
\begin{eqnarray}
Y(j\omega)/j&\cdot&(\bm{u}_{(k,l)}(\omega),\bm{v}_{(k,l)}(\omega))
\nonumber\\
&=&(\bm{u}_{(k,l)}(\omega),\bm{v}_{(k,l)}(\omega)) \cdot
\left(
\begin{array}{cc}
g_{k,l}(\omega) & \\
& g_{k,l}(\omega) \\
\end{array} 
\right)\,.
\end{eqnarray}
Thus, $Y(j\omega)/j$ has been diagonalized in a real basis.
Indeed, for $s=j\omega$, the matrix $Y(j\omega)/j$ is real symmetric and can therefore be diagonalized by an orthogonal (real unitary) matrix. The construction above makes this diagonalization explicit.

\section*{Acknowledgment}
I thank ChatGPT for help with English editing and for suggestions that improved the exposition.
I learned about RF accelerator systems, their underlying physics and mathematics, and many technical aspects from Dr. Yoshiaki Chiba of RIKEN.
In particular, Dr. Chiba introduced me to Professor Kurokawa's textbook \cite{bib:kur69}.
I am sincerely grateful for his guidance.
For many years, I have worked toward common goals with members of the RIKEN RI Beam Factory, especially those in the RF group.
I would like to express my gratitude to them here.

\appendix
\section{Variational description of coupled modes\label{ap:var}}
In this appendix, we show that Eq.~\eqref{eq:var} can be derived from a variational formulation.
Consider again two coupled cavity resonators, as shown in Fig.~\ref{fig:dual}.
We divide the system at the junction plane, referring to the left and right regions as A and B, respectively. We denote the junction surface by S.
We neglect losses in the outer walls of the resonators and treat them as perfect conductors.
The boundary of resonator A consists of S$_A$ and S, while that of resonator B consists of S$_B$ and S.
On S, the outward normal vectors $\bm{n}_A$ and $\bm{n}_B$ point in opposite directions.
Consider the extremum problem for the following functional in the combined region:
\begin{eqnarray}
k^2\left(\bm{E}\right)=\frac{\int_{\tiny {\mbox V_A}+{\mbox V_B}}
[\left(\nabla\times\bm{E}\right)^2+
\left(\nabla\cdot\bm{E}\right)^2]dv-2\int_{\tiny {\mbox S_A}+{\mbox S_B}}
\bm{n}\times\bm{E}\cdot\nabla\times\bm{E}dS}
{\int_{\tiny {\mbox V_A}+{\mbox V_B}}\bm{E}^2dv}\,.
\end{eqnarray}
We consider a solution of type (B), as discussed in Sec.~\ref{sub:eigen3}.
If the extremal value of this functional is $k^2$ and the corresponding vector field is $\bm{E}$, Eqs.~\eqref{eq:efun1}--\eqref{eq:efun3} give
\begin{eqnarray}
\label{eq:e1}
\nabla\times\nabla\times\bm{E}
-k^2\bm{E}&=&0\qquad \quad {\mbox {in V}}\,,\\
\nabla\cdot\bm{E}&=&0\qquad \quad{\mbox {in V}}\,,\\
\bm{n}\times\bm{E}&=&0\qquad \quad{\mbox {on S}}\,,
\end{eqnarray}
and
\begin{eqnarray}
\label{eq:k2}
k^2=\frac{\int_{\tiny {\mbox V_A}+{\mbox V_B}}
\left(\nabla\times\bm{E}\right)^2dv}
{\int_{\tiny {\mbox V_A}+{\mbox V_B}}\bm{E}^2dv}\,.
\end{eqnarray}
For the field $\bm{E}$ obtained in this way, we define $\bm{H}$ by the following relation, with $\omega=ck$:
\begin{eqnarray}
\bm{H}\coloneqq-\frac{1}{j\omega \mu}\nabla\times\bm{E}\,.
\end{eqnarray}
It follows from Eq.~\eqref{eq:e1} that $\bm{H}$ satisfies
\begin{eqnarray}
\label{eq:maxh}
\nabla\times\bm{H}=j\omega\epsilon\bm{E}\,,
\end{eqnarray}
together with the boundary condition
\begin{eqnarray}
\bm{n}\cdot\bm{H}=0\qquad \quad{\mbox {on S}}\,.
\end{eqnarray}
Thus, ($\omega$, $\bm{E}$, $\bm{H}$) constitutes a solution of the Maxwell eigenvalue problem in the combined region.

We assume that complete systems of the vector fields discussed in Sec.~\ref{sub:eigen3} are available in both regions A and B.
We denote these systems by $\{\bm{E}_n^{(A)}\}$, $\{\bm{H}_n^{(A)}\}$, $\{\bm{E}_n^{(B)}\}$, and $\{\bm{H}_n^{(B)}\}$.
We expand the denominator of Eq.~\eqref{eq:k2} using $\{\bm{E}_n^{(A)}\}$ and $\{\bm{E}_n^{(B)}\}$, and the numerator using $\{\bm{H}_n^{(A)}\}$ and $\{\bm{H}_n^{(B)}\}$.
By the orthonormality of these bases, Eq.~\eqref{eq:k2} becomes
\begin{eqnarray}
k^2=\left(\sum_{n}{A'_n}^2+\sum_{m}{B'_m}^2\right)/
\left(\sum_{n}{A_n}^2+\sum_{m}{B_m}^2\right)\,,
\end{eqnarray}
where
\begin{eqnarray}
\label{eq:an}
A_n&=&\int_{\tiny {\mbox V_A}}\bm{E}_n^{(A)} \cdot \bm{E}dv,\\
\label{eq:bm}
B_m&=&\int_{\tiny {\mbox V_B}}\bm{E}_m^{(B)} \cdot \bm{E}dv,\\
\label{eq:and}
A'_n&=&\int_{\tiny {\mbox V_A}}\bm{H}_n^{(A)} \cdot \nabla\times\bm{E}dv,\\
\label{eq:bmd}
B'_m&=&\int_{\tiny {\mbox V_B}}\bm{H}_m^{(B)} \cdot \nabla\times\bm{E}dv.
\end{eqnarray}
The coefficient $A'_n$ can be rewritten as follows:
\begin{eqnarray}
A'_n&=&\int_{\tiny {\mbox V_A}}\bm{H}_n^{(A)} \cdot \nabla\times\bm{E}dv\nonumber\\
&=&k_n^{(A)}\int_{\tiny {\mbox V_A}}\bm{E}_n^{(A)} \cdot \bm{E}dv
-\int_{\tiny{\mbox S}_A\cup\tiny{\mbox S}}\bm{H}_n^{(A)}\times\bm{E}\cdot\bm{n}_A dS
\nonumber\\
&=&k_n^{(A)} A_n -\int_{\tiny{\mbox S}}\bm{H}_n^{(A)}\times\bm{E}\cdot\bm{n}_A dS\,.
\end{eqnarray}

At the stationary value $k^2$, the first variation vanishes under an infinitesimal change of $\bm{E}$ proportional to any $\bm{E}_n^{(A)}$.
Since $\bm{E}_n^{(A)}\times\bm{n}_A=0$, such a variation does not affect the boundary term in $A'_n$.
We may therefore write
\begin{eqnarray}
\frac{\partial k^2}{\partial A_n}=0\,.
\end{eqnarray}
Substituting Eqs.~\eqref{eq:an}--\eqref{eq:bmd} and evaluating the derivative gives
\begin{eqnarray}
\frac{\partial k^2}{\partial A_n}=2 \left[k_n^{(A)}\left(k_n^{(A)}A_n
-\int_{\tiny{\mbox S}}\bm{H}_n^{(A)}\times\bm{E}\cdot\bm{n}_A dS\right)
-k^2 A_n\right]/
\left(\sum_{n}{A_n}^2+\sum_{m}{B_m}^2\right)\,.
\end{eqnarray}
Equivalently,
\begin{eqnarray}
k^2 A_n={k_n^{(A)}}^2A_n-k_n^{(A)}
\int_{\tiny{\mbox S}}\bm{H}_n^{(A)}\times\bm{E}\cdot\bm{n}_A dS\,.
\end{eqnarray}
Multiplying both sides by $c^2$ and writing $A_n$ explicitly, we obtain
\begin{eqnarray}
\omega^2 \int_{\tiny {\mbox V_A}}\bm{E}_n^{(A)} \cdot \bm{E}dv
={\omega_n^{(A)}}^2\int_{\tiny {\mbox V_A}}\bm{E}_n^{(A)} \cdot \bm{E}dv+c\omega_n^{(A)}
\int_{\tiny{\mbox S}}\bm{n}_A\times\bm{E}\cdot\bm{H}_n^{(A)} dS\,.
\end{eqnarray}
Therefore,
\begin{eqnarray}
\label{eq:anresult}
\left(\omega^2-{\omega_n^{(A)}}^2\right) \int_{\tiny {\mbox V_A}}\bm{E}_n^{(A)} \cdot \bm{E}dv
=c\omega_n^{(A)}\int_{\tiny{\mbox S}}\bm{n}_A\times\bm{E}\cdot\bm{H}_n^{(A)} dS\,.
\end{eqnarray}
On the other hand, taking the inner product of Eq.~\eqref{eq:maxh} with $\bm{E}_n^{(A)}$ and integrating over region A, we obtain, using the properties of $\{\bm{E}_n^{(A)}\}$,
\begin{eqnarray}
\int_{\tiny {\mbox V_A}}\bm{E}_n^{(A)} \cdot \bm{E}dv
=\frac{k_n^{(A)}}{j\omega\epsilon}
\int_{\tiny {\mbox V_A}}\bm{H}_n^{(A)} \cdot \bm{H}dv\,.
\end{eqnarray}
Substituting this result into Eq.~\eqref{eq:anresult} gives
\begin{eqnarray}
\int_{\tiny {\mbox V_A}}\bm{H}_n^{(A)} \cdot \bm{H}dv
&=&\frac{jc^2\epsilon\omega}
{\left(\omega^2-{\omega_n^{(A)}}^2\right)}
\int_{\tiny{\mbox S}}\bm{n}_A\times\bm{E}\cdot\bm{H}_n^{(A)} dS\nonumber \\
&=&-\frac{1}{\left(\frac{\omega_n^{(A)}}{j \omega}+\frac{j \omega}{\omega_n^{(A)}}\right)}\cdot
\frac{1}{\mu\omega_n^{(A)}}
\int_{\tiny{\mbox S}}\bm{n}_A\times\bm{E}\cdot\bm{H}_n^{(A)} dS\,.
\end{eqnarray}
We have thus derived Eq.~\eqref{eq:var}.

\section{Frequency splitting of coupled identical two-resonator systems\label{ap:split}}

Consider the coupled two-cavity system discussed in Sec.~\ref{sec:matrix} when the uncoupled cavities have equal resonance frequencies.
This system exhibits phenomena familiar in accelerator RF systems, including level repulsion and the splitting into symmetric and antisymmetric modes.
We examine these phenomena in terms of input admittances.

For simplicity, we retain only one mode on the junction cross section.
The basic equation is then Eq.~\eqref{eq:coupled_dual}, namely,
\begin{eqnarray*}
\left[Y_{11}^{(A)}(j\omega)+Y_{11}^{(B)}(j\omega)\right]V_1=0\,.
\end{eqnarray*}
Here $V_1$ denotes the voltage on the junction surface J in the coupled resonant state.
Omitting the mode index, we obtain
\begin{eqnarray}
\label{eq:coupled_simp}
\left[Y_A(j\omega)+Y_B(j\omega)\right]V=0\,.
\end{eqnarray}
Thus, as discussed in Sec.~\ref{sec:matrix}, coupled modes with a nonzero junction voltage $V\neq0$ are ordinarily determined by $Y_A+Y_B=0$.
However, in the limit where the two cavities become identical ($Y_A=Y_B$), symmetry gives rise to a mode for which the junction voltage $V$ is exactly zero.
This mode is no longer represented by a zero of $Y_A+Y_B$. As shown below, however, its eigenvalue persists in the characteristic matrix and, in the identical-cavity limit, coincides with a pole of the individual-cavity admittance $Y_A(=Y_B)$.

Let the resonance frequencies of the two individual cavities be $\omega_1$ and $\omega_2$, and define
\begin{eqnarray}
  x=\omega^2\,,
  \qquad a=\omega_1^2\,,
  \qquad b=\omega_2^2\,.
\end{eqnarray}
Assume that $\omega_1$ and $\omega_2$ are sufficiently close, and consider the admittance near these frequencies.
Using Eq.~\eqref{eq:adm1}, we collect the Foster terms other than those associated with the two nearby poles $\omega_1$ and $\omega_2$ into a background term \(A\), assumed to vary slowly over the frequency range of interest. We then obtain
\begin{eqnarray}
  \frac{Y_1+Y_2}{j\omega}
  \simeq
  F(x)
  :=A+\frac{K_1}{a-x}+\frac{K_2}{b-x}
  \label{eq:Ftwopoles}\,.
\end{eqnarray}
The function $F(x)$ is real, and coupled modes with nonzero junction voltage are determined by \(F(x)=0\).
Multiplying by the denominators gives
\begin{eqnarray}
  D(x)
  :=A(a-x)(b-x)+K_1(b-x)+K_2(a-x)=0
  \label{eq:characteristicD}\,.
\end{eqnarray}
Assuming \(A\neq0\), define
\begin{eqnarray}
  \alpha_1=\frac{K_1}{A}\,,
  \qquad
  \alpha_2=\frac{K_2}{A}\,.
\end{eqnarray}
The equation then becomes
\begin{eqnarray}
  (a-x)(b-x)+\alpha_1(b-x)+\alpha_2(a-x)=0\,.
\end{eqnarray}
The two solutions are
\begin{eqnarray}
  x_\pm=\frac12\left[
  a+b+\alpha_1+\alpha_2
  \pm
  \sqrt{
  (a-b+\alpha_1-\alpha_2)^2+4\alpha_1\alpha_2
  }
  \right]
  \label{eq:generalroots}\,,
\end{eqnarray}
and the frequencies of the coupled system are
\begin{eqnarray}
  \omega_\pm=\sqrt{x_\pm}\,.
\end{eqnarray}
In the present case, Eq.~\eqref{eq:adm1} implies \(\alpha_1\alpha_2>0\).
Equation~\eqref{eq:generalroots} can therefore be expressed as
\begin{eqnarray}
  \det\left[
  \begin{pmatrix}
    a+\alpha_1 & \sqrt{\alpha_1\alpha_2}\\
    \sqrt{\alpha_1\alpha_2} & b+\alpha_2
  \end{pmatrix}
  -xI
  \right]=0
  \label{eq:coupledmatrix_general}\,.
\end{eqnarray}

To describe two slightly different cavities, we write
\begin{eqnarray}
  a=\omega_0^2-d\,,
  \qquad
  b=\omega_0^2+d\,,
  \qquad
  K_1=K_2=K
  \label{eq:symmetricparam}\,.
\end{eqnarray}
We also define
\begin{eqnarray}
  \alpha=\frac{K}{A}\,.
\end{eqnarray}
Then Eq.~\eqref{eq:generalroots} becomes
\begin{eqnarray}
  x_\pm
  =\omega_0^2+\alpha
  \pm\sqrt{d^2+\alpha^2}
  \label{eq:avoidedroots}\,.
\end{eqnarray}
The two branches given by Eq.~\eqref{eq:avoidedroots} vary continuously with \(d\) and do not cross for nonzero coupling, \(\alpha\neq0\).
At \(d=0\), the minimum splitting in \(x=\omega^2\) is
\begin{eqnarray}
  x_+-x_-=2|\alpha|\,.
\end{eqnarray}
When \(|\alpha|\ll\omega_0^2\), the corresponding frequency splitting is approximately
\begin{eqnarray}
  |\omega_+-\omega_-|
  \simeq\frac{|\alpha|}{\omega_0}
  \label{eq:freqsplitting}\,.
\end{eqnarray}

The eigenvalues in Eq.~\eqref{eq:avoidedroots} are those of the matrix
\begin{eqnarray}
  M(d)=
  \begin{pmatrix}
    \omega_0^2-d+\alpha & \alpha\\
    \alpha & \omega_0^2+d+\alpha
  \end{pmatrix}
  \label{eq:Mdetuning}\,.
\end{eqnarray}
For \(|d|\gg|\alpha|\), the eigenvectors are approximately
\begin{eqnarray}
  \begin{pmatrix}1\\0\end{pmatrix}\,,
  \qquad
  \begin{pmatrix}0\\1\end{pmatrix}\,.
\end{eqnarray}
Each mode is then localized predominantly in one of the cavities.
As \(|d|\) decreases, the two states mix continuously. At \(d=0\), they become
\begin{eqnarray}
  \bm q_{\rm S}=\frac{1}{\sqrt2}
  \begin{pmatrix}1\\1\end{pmatrix}\,,
  \qquad
  \bm q_{\rm A}=\frac{1}{\sqrt2}
  \begin{pmatrix}1\\-1\end{pmatrix}
  \label{eq:brightdarkvectors}\,.
\end{eqnarray}
These are the symmetric and antisymmetric modes, respectively.
Transforming to the basis \(\{\bm q_{\rm S},\bm q_{\rm A}\}\), Eq.~\eqref{eq:Mdetuning} becomes
\begin{eqnarray}
  M_{\rm SA}(d)=
  \begin{pmatrix}
    \omega_0^2+2\alpha & -d\\
    -d & \omega_0^2
  \end{pmatrix}
  \label{eq:Mbrightdark}\,.
\end{eqnarray}
Thus, for identical cavities, \(d=0\), the symmetric and antisymmetric modes are decoupled, whereas a small asymmetry \(d\neq0\) mixes them.

For \(d\neq0\), \(F(x)=0\) ordinarily has two zeros.
Taking \(d\to0\), Eq.~\eqref{eq:avoidedroots} gives
\begin{eqnarray}
  x\longrightarrow
  \omega_0^2\,,
  \qquad
  \omega_0^2+2\alpha
  \label{eq:identicallimits}\,.
\end{eqnarray}
On the other hand, if we set \(a=b=\omega_0^2\) exactly, then
\begin{eqnarray}
  F(x)=A+\frac{2K}{\omega_0^2-x}
  =\frac{A(\omega_0^2-x)+2K}{\omega_0^2-x}
  \label{eq:Fidentical}\,.
\end{eqnarray}
This expression has only one ordinary zero,
\begin{eqnarray}
  x=\omega_0^2+\frac{2K}{A}
  =\omega_0^2+2\alpha\,.
\end{eqnarray}
By contrast, clearing the denominator gives the characteristic polynomial
\begin{eqnarray}
  D(x)
  =(\omega_0^2-x)
  \left[A(\omega_0^2-x)+2K\right]
  \label{eq:Dfactorized}\,,
\end{eqnarray}
which retains the two roots
\begin{eqnarray}
  x=\omega_0^2,
  \qquad
  x=\omega_0^2+2\alpha
  \label{eq:twoidenticalmodes}\,.
\end{eqnarray}

The factor $\omega_0^2-x$ cancels against the denominator in Eq.~\eqref{eq:Fidentical}.
Thus, at the special point $a=b$, one factor of $D(x)$ is canceled by the denominator of $F(x)$.
As the degenerate point is approached, one of the zeros approaches a pole. At degeneracy, the common factor in the numerator and denominator cancels, and the corresponding zero is no longer visible as a zero of $F(x)$.
The sum of positive-real functions in Eq.~\eqref{eq:adm1s} is itself a positive-real function, and Foster's theorem allows only first-order zeros or first-order poles on the imaginary axis.
The cancellation described above is consistent with this property of positive-real functions.
Physically, this mode may be regarded as an internal mode that is not captured by $Y=I/V$, since $V=0$.

Thus, even when the two cavities are identical, the coupled system generally has two eigenmodes with distinct frequencies, provided the coupling is nonzero.
In the simplified model of this note,
\begin{eqnarray}
  \omega_{\rm A}^2=\omega_0^2,
  \qquad
  \omega_{\rm S}^2=\omega_0^2+2\alpha\,.
\end{eqnarray}
One frequency appears to remain at the individual-cavity resonance frequency, while the other is shifted.
This behavior can be visualized, for example, by considering how the pole of the red curve approaches that of the blue curve in Fig.~\ref{fig:coupled_weak}.

\bibliography{manubib}

@article{bib:koh04,
  title = {Flattop acceleration system in the RIKEN AVF cyclotron},
  author = {Kohara, S. and Goto, A. and Sakamoto, N. and Kamigaito, O. and Watanabe, S. and Teranishi, T. and Katayama, T. and Chiba, Y. and Kase, M. and Yano, Y.},
  journal = {Nuclear Instruments and Methods in Physics Research Section A: Accelerators, Spectrometers, Detectors and Associated Equipment},
  volume = {526},
  number = {3},
  pages = {230--238},
  year = {2004}
}

@book{bib:kur69,
  title     = {An Introduction to the Theory of Microwave Circuits},
  author    = {Kurokawa, Kaneyuki},
  series    = {Electrical Science Series},
  year      = {1969},
  publisher = {Academic Press},
  address   = {New York},
  chapter={4}
}

@article{bib:kur58,
  author    = {Kurokawa, Kaneyuki},
  title     = {The Expansions of Electromagnetic Fields in Cavities},
  journal   = {IRE Transactions on Microwave Theory and Techniques},
  volume    = {6},
  number    = {2},
  pages     = {178--187},
  year      = {1958},
  publisher = {IEEE}
}

@book{bib:sob99,
  author     = {Sobenin, N. P. and Zverev, B. V.},
  title      = {Electrodynamic Characteristics of Accelerating Cavities},
  year       = {1999},
  publisher = {CRC Press},
  address    = {London},
  chapter={2}
}

@article{bib:bev61,
  title     = {Periodic electromagnetic and quantum systems},
  author    = {Bevensee, Robert M.},
  journal   = {Annals of Physics},
  volume    = {12},
  number    = {2},
  pages     = {222--263},
  year      = {1961},
  publisher = {Elsevier},
}

@book{bib:bev64,
  title     = {Electromagnetic Slow Wave Systems},
  author    = {Bevensee, R. M.},
  year      = {1964},
  publisher = {John Wiley \& Sons},
  address   = {New York},
  chapter={4, 5}
}

@article{bib:nis66,
  title={Dispersion relation and frequency characteristics of alternating periodic structure for linear accelerators},
  author={Nishikawa, T. and Giordano, S. and Carter, D.},
  journal={Review of Scientific Instruments},
  volume={37},
  number={5},
  pages={652--661},
  year={1966},
  publisher={AIP Publishing}
}

@article{bib:nag67,
  title={Coupled Resonator Model for Standing Wave Accelerator Tanks},
  author={Nagle, D. E. and Knapp, E. A. and Knapp, B. C.},
  journal={Review of Scientific Instruments},
  volume={38},
  number={11},
  pages={1583--1587},
  year={1967},
  publisher={AIP Publishing}
}

@inproceedings{bib:dom92,
  author    = {D{\^o}me, Georges},
  title     = {Basic RF theory, waveguides and cavities},
  booktitle = {Proceedings of CERN Accelerator School: Course on RF engineering for particle accelerators},
  editor    = {Turner, S.},
  year      = {1992},
  pages     = {1--96},
  address   = {Oxford, UK},
  publisher = {CERN},
}

@book{bib:cou53,
  title={Methods of Mathematical Physics},
  author={Courant, Richard and Hilbert, David},
  volume={1},
  year={1953},
  publisher={Interscience Publishers, John Wiley \& Sons},
  address={New York},
  chapter={V, VI, VII}
}

@book{bib:wad06,
  author    = {Wadhwa, C. L.},
  title     = {Network Synthesis},
  booktitle = {Network Analysis and Synthesis},
  edition   = {2nd},
  chapter   = {10},
  year      = {2006},
  publisher = {New Age International Publishers},
  address   = {New Delhi}
}

@book{bib:mit48,
  author    = {Montgomery, Carol G. and Dicke, Robert H. and Purcell, Edward M.},
  title     = {Elements of Network Theory},
  booktitle = {Principles of Microwave Circuits},
  series    = {MIT Radiation Laboratory Series},
  volume    = {8},
  chapter   = {5},
  year      = {1948},
  publisher = {McGraw-Hill},
  address   = {New York}
}

@book{bib:ant97,
  author    = {Antsaklis, Panos J. and Michel, Anthony N.},
  title     = {Linear Systems},
  booktitle = {Linear Systems},
  pages     = {564},
  year      = {1997},
  publisher = {McGraw-Hill},
  address   = {New York},
  series    = {McGraw-Hill Series in Electrical and Computer Engineering}
}

@book{bib:and73,
  author    = {Anderson, Brian D. O. and
               Vongpanitlerd, Sumeth},
  title     = {Network Analysis and Synthesis:
               A Modern Systems Theory Approach},
  publisher = {Prentice-Hall},
  address   = {Englewood Cliffs, NJ},
  year      = {1973}
}

@article{bib:oon50,
  author    = {Oono, Yosiro},
  title     = {Synthesis of a finite $2n$-terminal network by a group of networks each of which contains only one ohmic resistance},
  journal   = {Journal of Mathematics and Physics},
  volume    = {29},
  number    = {1},
  pages     = {13--26},
  year      = {1950},
  publisher = {Massachusetts Institute of Technology}
}

@article{bib:ced57,
  author    = {Cederbaum, Israel},
  title     = {On matrices of residues of the impedance or admittance matrices of $n$-ports},
  journal   = {IRE Transactions on Circuit Theory},
  volume    = {4},
  number    = {1},
  pages     = {20--21},
  year      = {1957},
  publisher = {IEEE}
}

@book{bib:wan08,
  author    = {Wangler, Thomas P.},
  title     = {Radio-Frequency Linear Accelerators},
  booktitle = {RF Linear Accelerators},
  edition   = {2nd},
  chapter   = {3},
  year      = {2008},
  publisher = {Wiley-VCH},
  address   = {Weinheim}
}

@inproceedings{bib:kag97,
  author     = {T. Kageyama and K. Akai and N. Akasaka and E. Ezura and H. Mizuno and F. Naito and H. Nakanishi and H. Sakai and Y. Takeuchi and Y. Yamazaki and T. Kobayashi},
  title      = {Development of High-Power ARES Cavities},
  booktitle  = {Proceedings of the 1997 Particle Accelerator Conference (PAC 97)},
  pages      = {2902--2904},
  year       = {1997},
  address    = {Vancouver, B.C., Canada}
}

@article{bib:shi94,
  author    = {Shintake, Tsumoru},
  title     = {Proposal of Accelerating RF-cavity Coupled with an Energy Storage Cavity for Heavy Beam Loading Accelerators},
  journal   = {Particle Accelerators},
  volume    = {44},
  pages     = {131},
  year      = {1994}
}

@book{bib:tes01,
  author       = {{TESLA Collaboration} and Richard, F. and Schneider, J. R. and Trines, D. and Wagner, A.},
  title        = {{TESLA Technical Design Report, Part I-VI: The superconducting electron-positron linear collider with an integrated X-ray laser laboratory}},
  publisher    = {{DESY}},
  address      = {{Hamburg}},
  year         = {2001},
  reportNumber = {DESY-2001-011, ECFA-2001-209}
}

@book{bib:van60,
  author    = {Van Valkenburg, M. E.},
  title     = {Introduction to Modern Network Synthesis},
  publisher = {John Wiley \& Sons},
  address   = {New York},
  year      = {1960}
}

@book{bib:mon03,
  author    = {Monk, Peter},
  title     = {Finite Element Methods for Maxwell's Equations},
  publisher = {Oxford University Press},
  year      = {2003},
}

\end{document}